\documentclass{article}
\usepackage{amssymb,amsmath}
\usepackage{amssymb,amsthm}
\usepackage{enumerate}
\usepackage{graphicx}
\usepackage{float}
\usepackage{epstopdf}

\usepackage[colorlinks=true,citecolor=blue]{hyperref}
\usepackage{cite}

\theoremstyle{definition}

\numberwithin{Theorem}{section}
\numberwithin{Lemma}{section}

\numberwithin{Corollary}{section}
\numberwithin{Example}{section}
\numberwithin{Remark}{section}
\date{}
\usepackage{parskip}
\usepackage{amsmath}
\usepackage{graphicx}
\usepackage[numbers]{natbib}
\usepackage{float}
\usepackage[small,labelsep=space]{caption}
\usepackage[labelformat=simple]{subcaption}

\usepackage{url}
\usepackage{enumitem}
\usepackage{booktabs}
\usepackage{appendix}
\usepackage{xcolor}
\usepackage{pifont}   
\newcommand{\tick}{\ding{52}}
\newcommand{\cross}{\ding{55}}
\usepackage{numprint}
\npthousandsep{,}\npthousandthpartsep{}\npdecimalsign{.}  
\usepackage{siunitx}
\usepackage{afterpage}   

\usepackage{hyperref}
\hypersetup{linktoc=all,
            unicode=true,
            colorlinks=true,
            linkcolor=blue,
            citecolor=blue,
            urlcolor=blue,
            pdfstartview=,
            pdfnewwindow=true}

\usepackage{listings}
\lstdefinestyle{custommatlab}{
  belowcaptionskip=1\baselineskip,
  breaklines=true,
  frame=L,
  xleftmargin=\parindent,
  language=Matlab,
  showstringspaces=false,
  basicstyle=\footnotesize\ttfamily,
  keywordstyle=\bfseries\color{black},
  commentstyle=\itshape\color{red!80!black},
  identifierstyle=\color{black},
  stringstyle=\color{orange},
}
\lstdefinestyle{overpass}{
  belowcaptionskip=1\baselineskip,
  breaklines=true,
  frame=L,
  xleftmargin=\parindent,
  language=C,
  showstringspaces=false,
  basicstyle=\footnotesize\ttfamily,
  keywordstyle=\bfseries\color{black},
  commentstyle=\itshape\color{red!80!black},
  identifierstyle=\color{black},
  stringstyle=\color{orange},
}

\title{Modelling Fire Incidents Response Times in {\AA}lesund}
\author{J. Christmas\thanks{University of Exeter, UK (Email: J.T.Christmas@exeter.ac.uk)}, R. Bergmann\thanks{Norwegian University of Science and Technology, Norway (Email: ronny.bergmann@ntnu.no)}, A. Zhakatayev\thanks{University of Agder, Norway (Email: altay.zhakatayev@uia.no)}, J. Rebenda\,\thanks{University of Agder, Norway, and Brno University of Technology, Czech Republic (Email: josef.rebenda@uia.no, rebenda@vutbr.cz)}, S. Singh \thanks{Norwegian University of Science and Technology, Norway (Email: shiprasingh384@gmail.com)} }

\date{June 2022}

\begin{document}

\maketitle

\begin{abstract}
In the ESGI-156 project together with \AA lesund Brannvesen we develop a model for response times to fire incidents on publicly available data for \AA lesund. We investigate different scenarios and a first step towards an interactive software for illustrating the response times.
\end{abstract}

\tableofcontents

\newpage
\section{Problem definition}
\label{sec:problem-definition}

Requirements on the response time in case of fire incidents for fire departments like the \AA lesund Brannvesen\footnote{\url{https://aabv.no}} in Norway are stated in the “Forskrift om organisering, bemanning og utrustning av brann- og redningsvesen og nødmeldesentralene”\footnote{\url{https://lovdata.no/dokument/SF/forskrift/2021-09-15-2755}} about regulations on the organization, staffing and equipment of the fire and rescue service and the emergency call centers. For our project the \emph{response time}, that is (cf. §2f) the time from the start of the emergency call until the firemen are at the scene, is the main focus. For this time §22 states that for densely populated areas with a particular danger of rapid and extensive fire spreading, hospitals, nursing homes and similar institutions that require assisted escape, and further important business operations, the time response shall not exceed 10 minutes. Under certain circumstances (§7 and 9) this might be exceeded. Furthermore within residential areas like cities or villages, the response time shall not exceed 20 minutes. Otherwise it shall not exceed 30 minutes.

All regulation requirements are actually met by the fire department. They have a list of all important buildings and areas in the department's district. They also know by experience how for example traffic density affects the response time.

The goal of this project is to model the response time for \AA lesund Brannvesen and their fire stations. The results should then be compared to actual data of response times. Then the model shall be used to investigate scenarios.
The scenarios might include changes of the following types:
\begin{itemize}
    \item 
Changes in the street infrastructure, that is, tolerance of the response time with respect to roads or bridges being closed down or affected by heavy traffic, for example due to construction.
\item 
Changes in operating fire stations. While \AA lesund has two main fire stations equipped full time, there are several part time fire stations, especially on different islands. Using the model, the effect of changing the mode of operation or moving fire stations can be investigated.
\end{itemize}
Finally, a “heat map” of response times, i.e. where the district of \AA lesund Brannvesen is colored in 10-minute, 20-minute, and 30-minute response time areas, together with a possibly interactive map to move or modify fire stations would allow for modelling several different scenarios that might indicate response time improvements.

This project aims to provide such a heat map of response times based on publicly available data as well as a comparison against real data to evaluate how accurate the model is.

\newpage
\section{Source data}
\label{sec:source-data}

An important requirement of the project was to work as much as possible on publicly available data. We decided to use OpenStreetMap\footnote{\url{https://www.openstreetmap.org/}} (OSM) as main source of geographical data, since it is publicly available and can be accessed by several APIs.

In this section we briefly explain which sources we were provided and how we unified the data to work on OSM data solely.

All provided data was preprocessed as explained in the following sections and stored in comma-separated-values (\lstinline!.csv!) files, such that the preprocessing (done in Julia using \lstinline!HTTP.jl!\footnote{\url{https://juliaweb.github.io/HTTP.jl/stable/}} and \lstinline!CSV.jl!\footnote{\url{https://csv.juliadata.org/stable/}}) can be easily implemented in other programming languages, where all headers of the \lstinline!.csv! files should be self-explanatory (e.g. \lstinline!city!, \lstinline!lon!, \lstinline!lat!, \lstinline!street! or \lstinline!!osm\_id!). All used files and their content are described in the \lstinline!Readme.md! and can be obtained from the authors upon request. 

\subsection{OpenStreetMap Data}

\begin{figure}[tb]
   \centering
   \subcaptionbox{map of the \AA lesund area\label{subfig:osm-example-area}}
      {\includegraphics[width=0.5\linewidth]{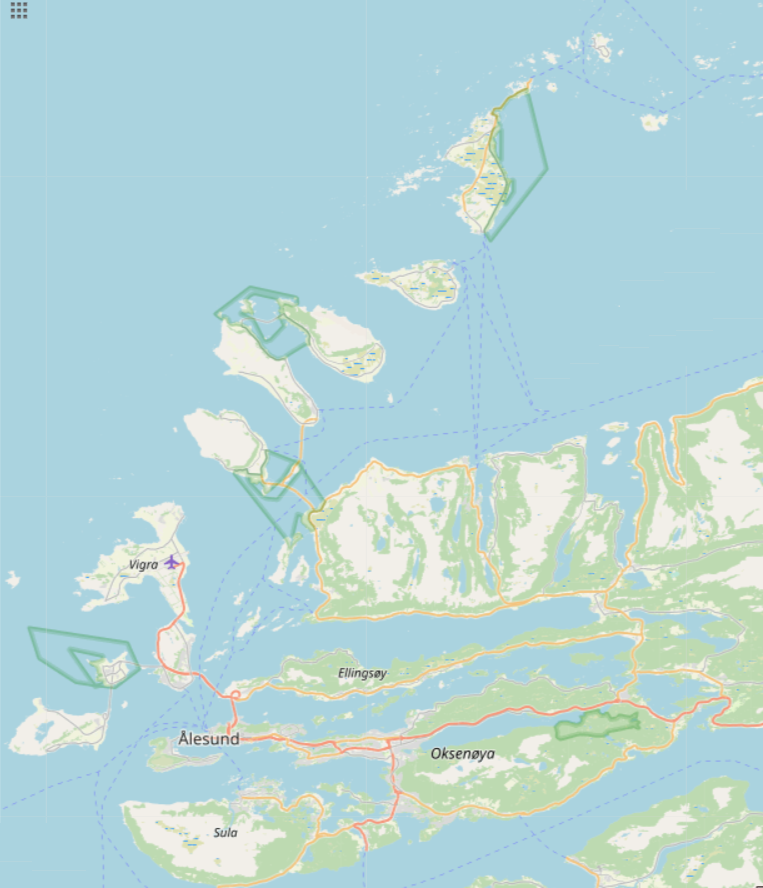}} \\[3mm]
   \subcaptionbox{example of the city street map\label{subfig:osm-example-streets}}
      {\includegraphics[width=\linewidth]{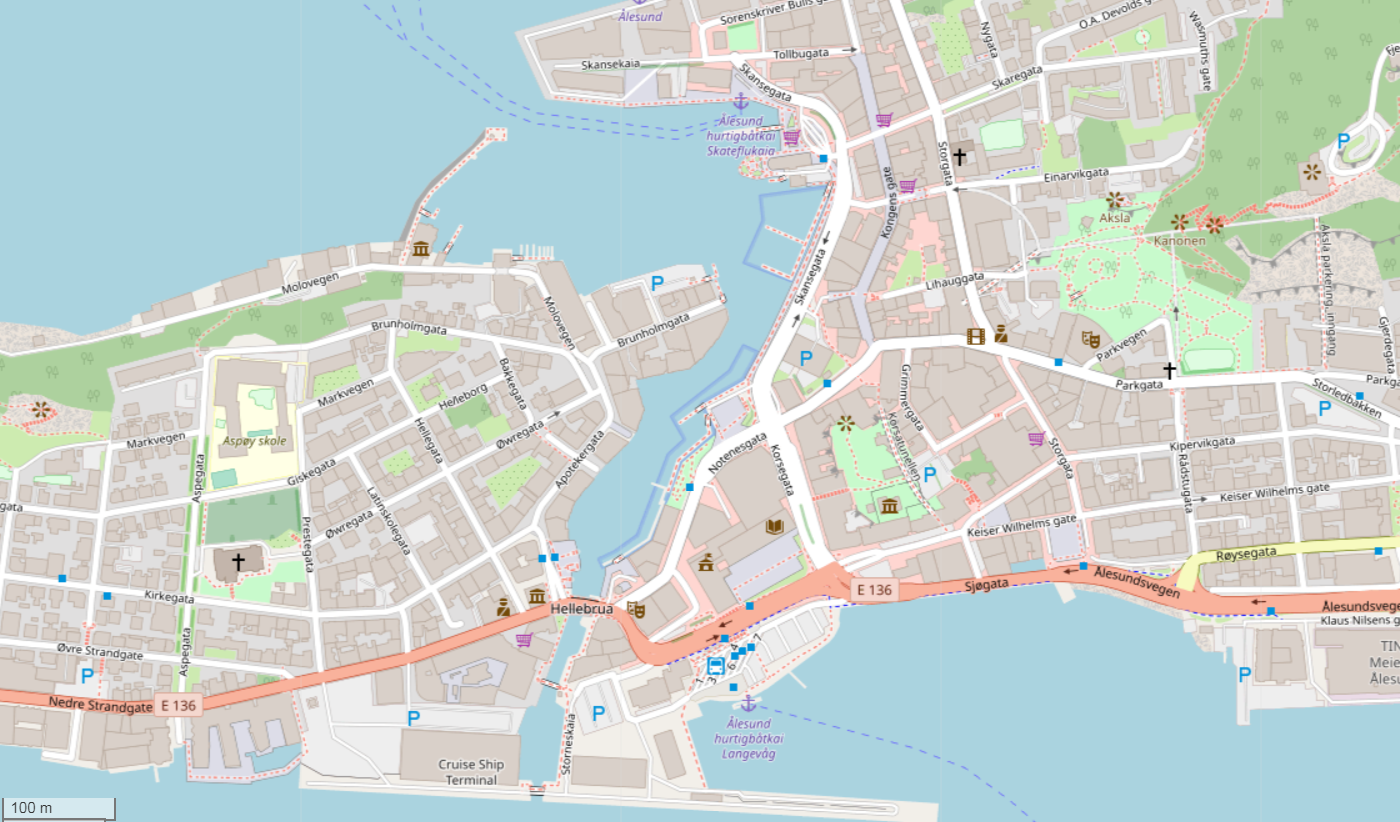}}
   \caption{OpenStreetMap Example, showing \subref{subfig:osm-example-area} the whole
      area, and \subref{subfig:osm-example-streets} a section of the streets.}
   \label{fig:osm-example}
\end{figure}

We were provided with a set of bounding boxes that together cover the whole area of all \AA lesund Brannvesen fire stations and their area of responsibility. In some exports one has to be careful, when using several bounding boxes such that tunnels and bridges that are included in at least one of the neighbouring areas. Figure \ref{fig:osm-example} provides an overview of the area and a detail of the inner city of \AA lesund.

While the overall data is quite large (5 \lstinline!.xml!-files of 216 MB data), we filtered these to only include streets. Addresses or points of interest (POI) were mapped to the closest street node.

\subsection{\AA lesund fire stations}

There is a map for the \AA lesund fire stations at \url{https://s.ntnu.no/alesund-brannstasjoner}, which we preprocessed by using the Names and addresses of the fire stations to map them\footnote{using the \url{https://nominatim.openstreetmap.org} API} to an OpenStreetMap node (point on the map) and its id, longitude, and latitude.

\subsection{Important locations}

We were provided an Excel sheet of objects that have to be within a radius of 10 minute response time.
We slightly modified this list of 55 locations, since some had two addresses (separated by “ and ”), so we again used the OSM API to match these 58 different addresses to longitude, latitude and OSM node id.

We used these points to check in certain scenarios of response times how many of the important locations can be reached in time and how much this time constraint is violated both in number of nodes not reached in time as well as the extent of time violation.

\subsection{Incident Statistics}

We were also provided a statistics of incidents for the years since 2016 until mid 2022, 7918 in total, with a very large amount of data.

We reduced this to the 732 cases reported as fires (“Brann”). For these we extracted the actual response time, longitude and latitude of the incident.

\afterpage{\clearpage}

\newpage
\section{Response time heatmaps}
\label{sec:heatmaps}

All the maps and road information used in this project were taken from \href{https://www.openstreetmap.org/relation/10148280}{OpenStreetMap (OSM)}\footnote{\url{https://www.openstreetmap.org/relation/10148280}}.  However, it proved easier to export the data from a different website called \href{http://overpass-turbo.eu/}{Overpass Turbo}\footnote{\url{http://overpass-turbo.eu/}} which is a tool for calling the Overpass API.

The main steps for creating the heatmaps are as follows:
\begin{itemize}[itemsep=1pt,label=--,leftmargin=*]
\item Define the geographical area of the {\AA}lesund Brannvesen's region of responsibility.
\item Extract the land map of this region.
\item Extract all the roads within this region, defined as a network of nodes joined by edges.
      Each node is specified by its longitude and latitude.
\item For each fire station,
       calculate the time required to drive from the fire station to every node in the network.
      
\item For each node in the network,
       find the fire station which is closest in drive time.
      
\item Draw the heatmap:
      \begin{itemize}
      \item fill the land map of this region in grey
      \item plot each node in a colour related to the drive time required to reach the closest fire station determined above.
      \item plot lines to show the locations of bridges and tunnels (to visualise connections to and between islands)
      \end{itemize}
\end{itemize}

\subsection{Geographical coordinate systems}

There are a number of different ways of uniquely specifying a location on the Earth's surface.  The most common, and the one used by OSM, are the angular longitude and latitude, which are the number of degrees west of the Greenwich Meridian and north of the Equator.  The disadvantage of this method is that \SI{1}{\degree} of longitude represents different distances (in, for example, metres) at different latitudes.  A second disadvantage is that the Earth is not spherical, so that converting from longitude/latitude to metres globally is not straightforward.  However, local conversion may be achieved using the Haversine formula, which uses a local longitude/latitude origin.  The Matlab code for achieving this is listed in appendix \ref{appendix:haversine}.

\setcounter{footnote}{0}

The {\AA}lesund Brannvesen region is defined at 
\href{https://kartserver.esunnmore.no/geoinnsyn/?project=Interessepunkter&layers=Flyfoto,Brannstasjon110&zoom=8&application=geoinnsyn&lat=6944482.35&lon=374169.89}{this website}\footnote{\url{https://kartserver.esunnmore.no/geoinnsyn/?project=Interessepunkter&layers=Flyfoto,Brannstasjon110&zoom=8&application=geoinnsyn&lat=6944482.35&lon=374169.89}}
using a different coordinate system, called Universal Transverse Mercator (UTM).  UTM divides the Earth into 60 zones and projects each zone to a plane as a basis for its coordinates. Specifying a location means specifying the zone and the $x$, $y$ coordinate in that plane.

The {\AA}lesund Brannvesen region coordinates are specified in UTM format, with M{\o}re og Romsdal falling within the 32N zone.  This is converted to standard longitude and latitude using a Matlab function \texttt{utmdeg.m}, provided by Mathworks. The calculation is not quite right because Norway is a \href{http://www.asprs.org/a/resources/grids/10-99-norway.pdf}{special case}\footnote{\url{http://www.asprs.org/a/resources/grids/10-99-norway.pdf}}; see the European map shown on \href{https://en.wikipedia.org/wiki/Universal_Transverse_Mercator_coordinate_system}{Wikipedia}\footnote{\url{https://en.wikipedia.org/wiki/Universal_Transverse_Mercator_coordinate_system}}.  The provided Matlab script does not appear to take this special case into account, but the resulting longitude/latitude values look appropriate when plotted over a map from OSM; see figure \ref{fig:region-polygon-map}.

\subsection{Define the fire service's region of responsibility}
\label{sec:region-of-responsibility}

The {\AA}lesund Brannvesen's region of responsibility used in this work has been extracted from \href{https://kartserver.esunnmore.no/geoinnsyn/?project=Interessepunkter&layers=Flyfoto,Brannstasjon110&zoom=8&application=geoinnsyn&lat=6944482.35&lon=374169.89}{this map}\footnote{\url{https://kartserver.esunnmore.no/geoinnsyn/?project=Interessepunkter&layers=Flyfoto,Brannstasjon110&zoom=8&application=geoinnsyn&lat=6944482.35&lon=374169.89}}, on which  all the fire stations are marked.

\begin{figure}[tb]
   \centering
   \fbox{\includegraphics[width=0.6\textwidth]{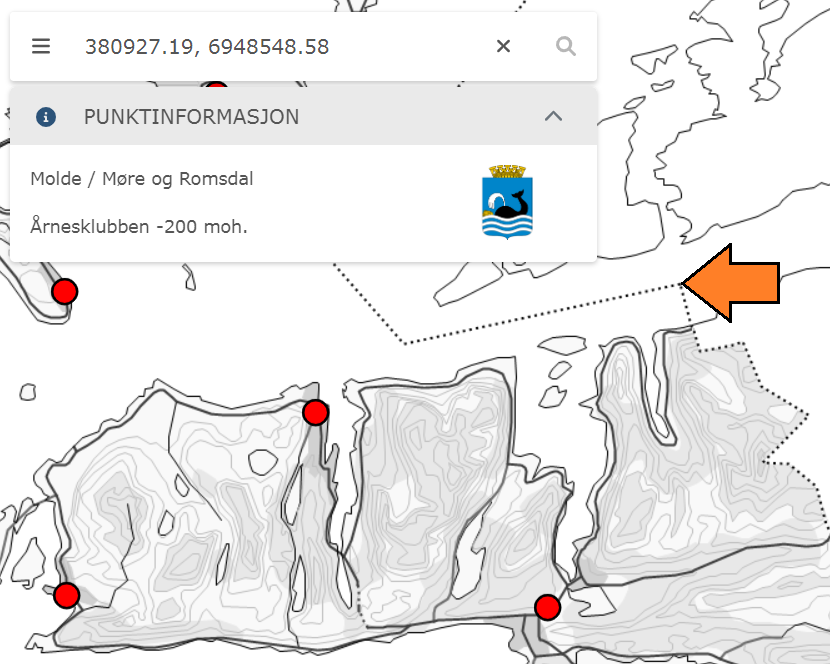}}
   \caption{Clicked on the map at the point indicated by the orange arrow.  The coordinates are displayed in the top left-hand corner:
            380927.19, 6948548.58.}
   \label{fig:define-region}
\end{figure}

\begin{figure}[p]
   \centering
   \includegraphics[width=0.6\textwidth]{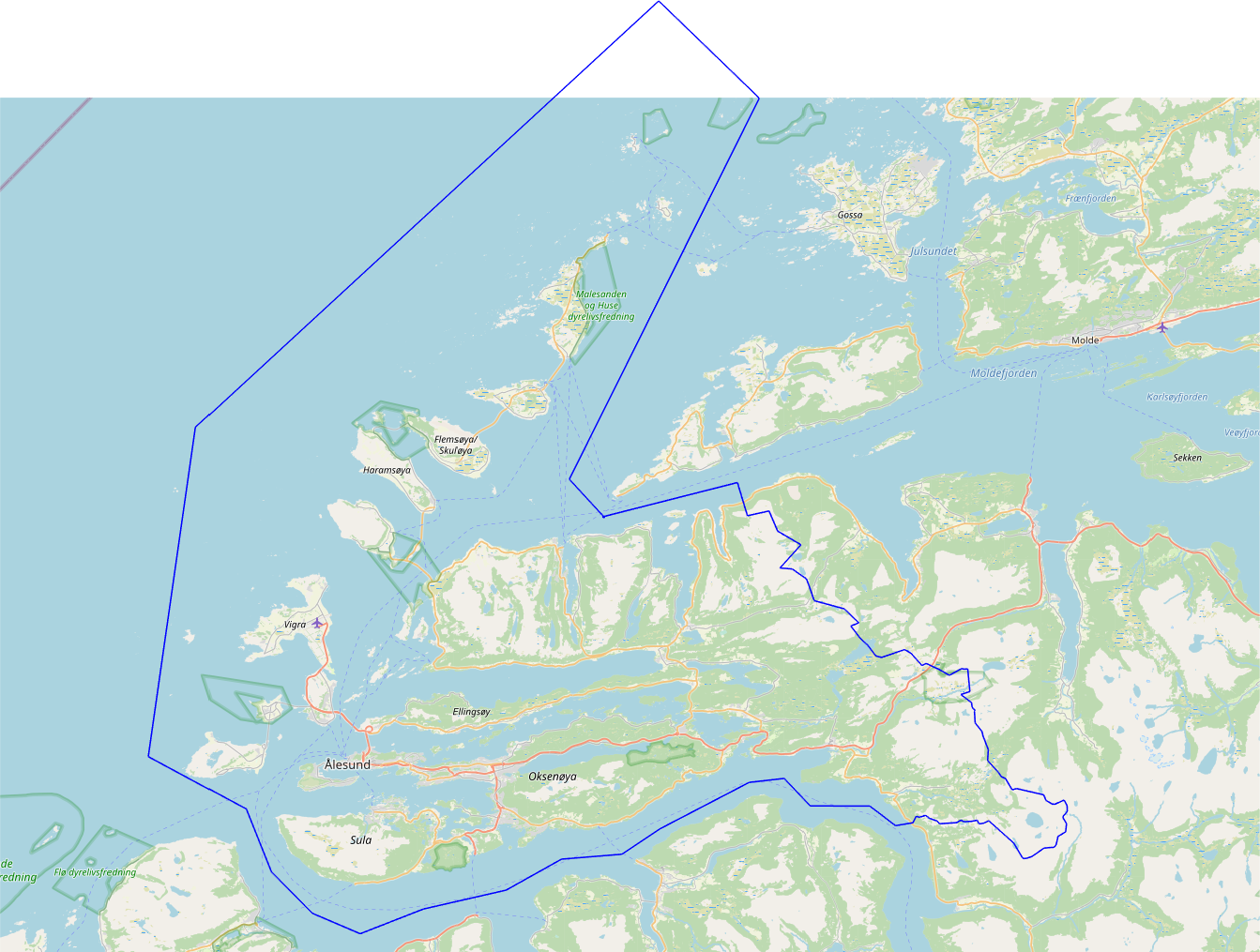}
   \caption{The fire service's region of responsibility superimposed on a map.}
   \label{fig:region-polygon-map}
\vspace{5mm}
   \centering
   \includegraphics[width=0.8\textwidth]{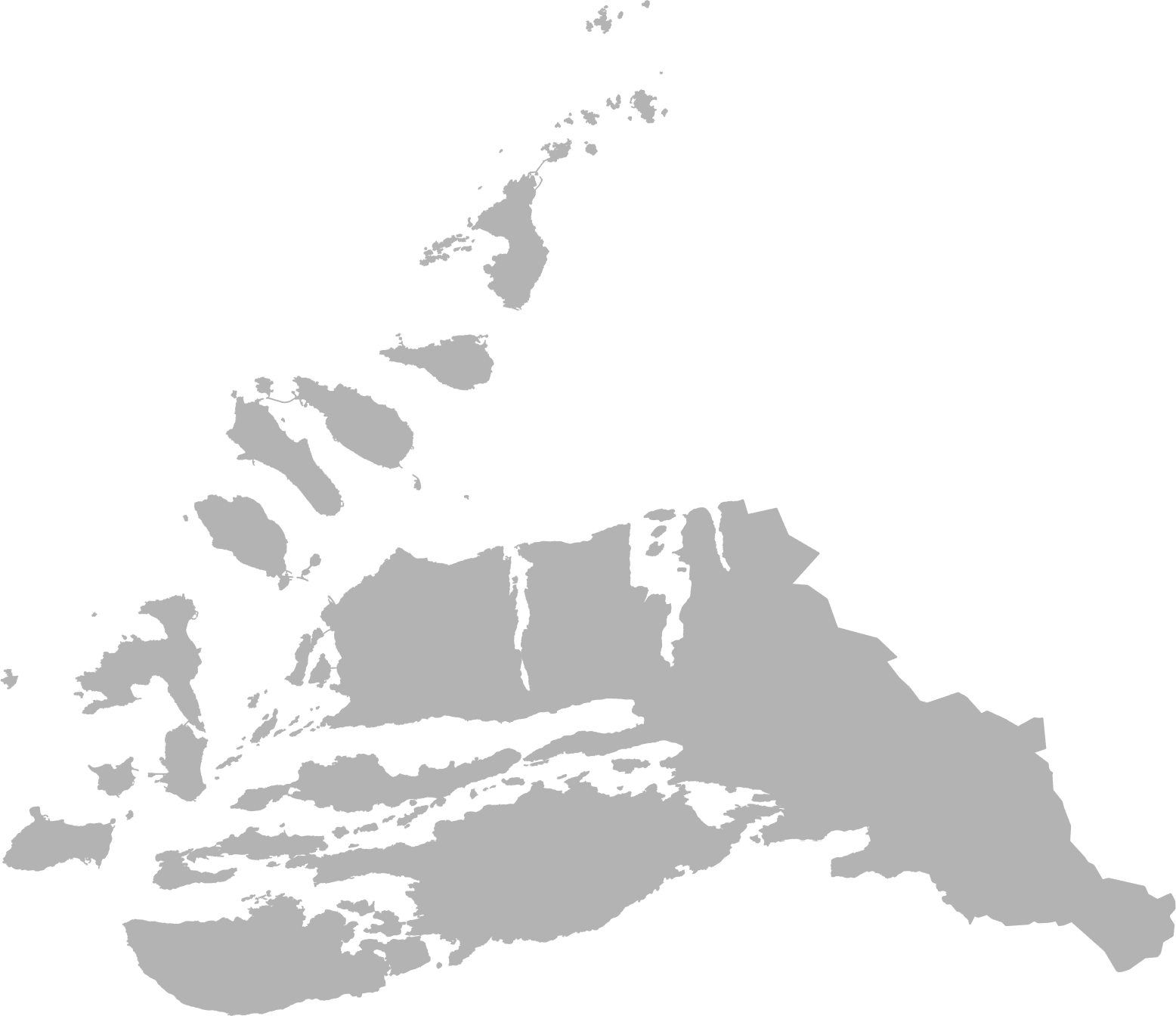}
   \caption{The fire service's geographical region of responsibility.}
   \label{fig:region}
\end{figure}

We clicked on each point around the boundary (in order) and recorded the coordinates of each point; an example is shown in figure \ref{fig:define-region}.  The coordinates are in UTM format which are then converted into global longitude and latitude.  The boundary is a single, complex polygon, as shown in figure \ref{fig:region-polygon-map}.

Running the Matlab script \texttt{aalesundPolygon.m} creates a Matlab file called \texttt{aalesundBrannvesen.mat} which defines the outline of the fire service's region of responsibility, as shown as a blue line in figure \ref{fig:region-polygon-map}.

\subsection{Extract the region boundaries from OSM}

Based on the polygon extracted in section \ref{sec:region-of-responsibility}, we want to extract a mask of the land covered by the fire service's region of responsibility.  This mask is generally coastline, but also includes the eastern land boundary with Vestnes.

In \href{http://overpass-turbo.eu/}{Overpass Turbo}, entering the following query extracts all coastline elements from within the rectangle defined by the bottom left corner with longitude 5.6838 and latitude 62.3536, and top right corner with longitude 7.3927 and latitude 62.8871.  This rectangle exceeds the fire service region in all directions.

\begin{lstlisting}[style=overpass]
[out:json][timeout:25];
(
  way["natural"="coastline"](62.3536,5.6838,62.8871,7.3927);
);
out body;
>;
out skel qt;
\end{lstlisting}
Running this query may result in a message like this:
\begin{lstlisting}[style=overpass]
Large amount of data
This query returned quite a lot of data (approx. 20 MB).
Your browser may have a hard time trying to render this.
Do you really want to continue?
\end{lstlisting}
If so, one needs to click on ``continue anyway''.  The map on the right-hand side of the window will become covered in markers.

To save the data generated by this query, it is necessary to click on ``Export'', and then by ``raw OSM data'' click on ``download'' option.  The downloaded file should then be saved as ``coastlineExport.json''.

The \texttt{makeBetterCoastline.m} script calls three functions:
\begin{description}
\item[\texttt{extractCoastlineFromJSON}] extracts all the coastline nodes and edges from the JSON file.
\item[\texttt{cropCoastline}] extracts only those nodes and edges inside the fire service's region, and adds edges to complete the perimeter
    according to the polygon.
\item[\texttt{convertNEtoPolygons}] converts the nodes+edges format into a faces+vertices format so that Matlab's \texttt{patch} function can
    draw it efficiently.  It also removes small islands.
\end{description}
Finally, calling the \texttt{plotRegion.m} script will cause the fire service's region to be plotted, as shown in figure \ref{fig:region}.

\subsection{Extract the highways from OSM}

Based on the polygon extracted in section \ref{sec:region-of-responsibility}, we want to extract all the roads contained within the fire service's region of responsibility.  Each road is defined as a series of nodes joined by edges.

In \href{http://overpass-turbo.eu/}{Overpass Turbo}, entering the following query extracts all road (``highway'') elements from within the rectangle defined by the bottom left corner with longitude 5.6838 and latitude 62.3536, and top right corner with longitude 7.3927 and latitude 62.8871.  This rectangle exceeds the fire service region in all directions.

\begin{lstlisting}[style=overpass]
[out:json][timeout:25];
(
  node["highway"](62.3536,5.6838,62.8871,7.3927);
  way["highway"](62.3536,5.6838,62.8871,7.3927);
  relation["highway"](62.3536,5.6838,62.8871,7.3927);
);
out body;
>;
out skel qt;
\end{lstlisting}
Running this query may result in a message like this:
\begin{lstlisting}[style=overpass]
Large amount of data
This query returned quite a lot of data (approx. 30 MB).
Your browser may have a hard time trying to render this.
Do you really want to continue?
\end{lstlisting}
If so, then one needs to click on ``continue anyway''.  The map on the right-hand side of the window will become covered in markers.

To save the data generated by this query, it is necessary to click on ``Export'', and then by ``raw OSM data'' click on ``download'' option.  The downloaded file should then be saved as ``highwaysExport.json''.

The \texttt{makeBetterHighways.m} script calls two functions:
\begin{description}
\item[\texttt{extractHighwaysFromJSON}] extracts all the nodes and edges from the JSON file that represent roads.
\item[\texttt{cropHighways}] extracts only those nodes and edges inside the fire service's region.
\end{description}

Within OSM, each road is classified as a ``highway'' with a specific type.  Not all types represent the sort of roads a vehicle could reasonably drive along.  Table \ref{tab:highway-types} lists all the different highway types currently found in the map of {\AA}lesund and whether or not they are included in our extract.

\begin{table}
   \centering
   \begin{tabular}{lcc}\toprule
   \textbf{highway type} & \textbf{included?} & \textbf{default max speed} \\
   \midrule
   construction    & \cross & \\
   cycleway        & \cross & \\
   footway         & \cross & \\
   living\_street  & \tick  & 50 \\
   path            & \cross & \\
   pedestrian      & \cross & \\
   platform        & \cross & \\
   primary         & \tick  & 50 \\
   primary\_link   & \tick  & 50 \\
   proposed        & \cross & \\
   raceway         & \tick  & 50 \\
   residential     & \tick  & 20 \\
   secondary       & \tick  & 50 \\
   secondary\_link & \tick  & 50 \\
   service         & \tick (see caption) & 10 \\
   steps           & \cross & \\
   trunk           & \tick  & 50 \\
   trunk\_link     & \tick  & 50 \\
   track           & \tick  & 5 \\
   tertiary        & \tick  & 50 \\
   unclassified    & \tick  & 50 \\
   \bottomrule
   \end{tabular}
   \caption{A list of the different highway types currently found in the OSM map of {\AA}lesund and whether they are included in our extract.
            Service roads with a service type of ``parking\_aisle'' are not classified as highways.}
   \label{tab:highway-types}
\end{table}

Four useful tags associated with highways are: \emph{tunnel}, \emph{bridge}, \emph{oneway}, and \emph{maxspeed}.  The first three are set to the value ``yes'' if a section of highway is a tunnel, a bridge or a one-way road respectively.  If a highway is not marked as one-way, then each edge is replicated in the network to show that the road is two-way.  The \emph{maxspeed} tag, if present, is set to the speed limit for that section of highway.

In order to convert longitude and latitude to metres, we need a local point of origin; this is the mean longitude and latitude of all the nodes in the region.

\begin{figure}[p]
   \centering
   \includegraphics[width=1\textwidth]{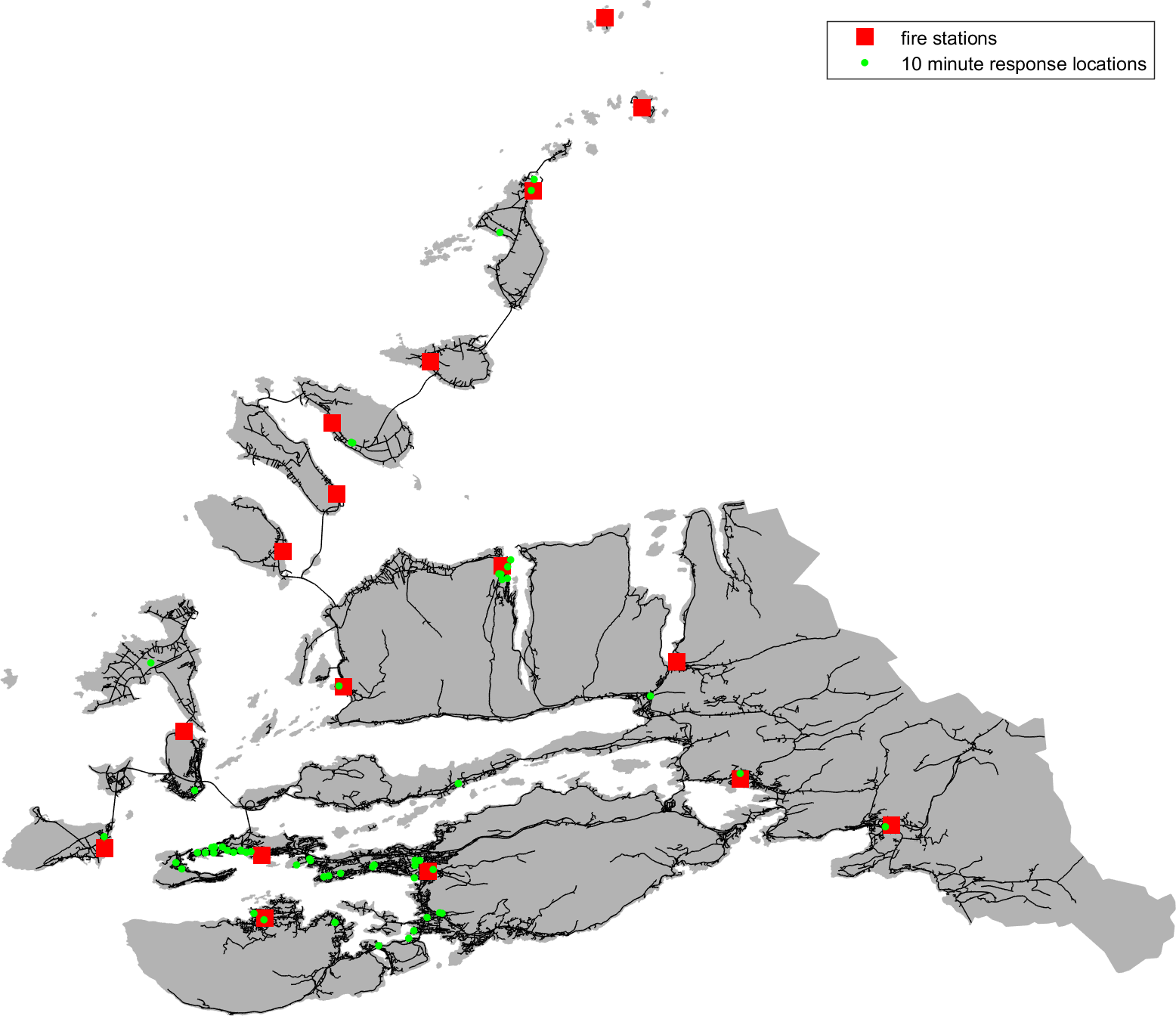}
   \caption{The roads within the fire service's geographical region of responsibility.}
   \label{fig:highways-fs-sl}
   \vspace{10mm}
   \includegraphics[width=0.8\textwidth]{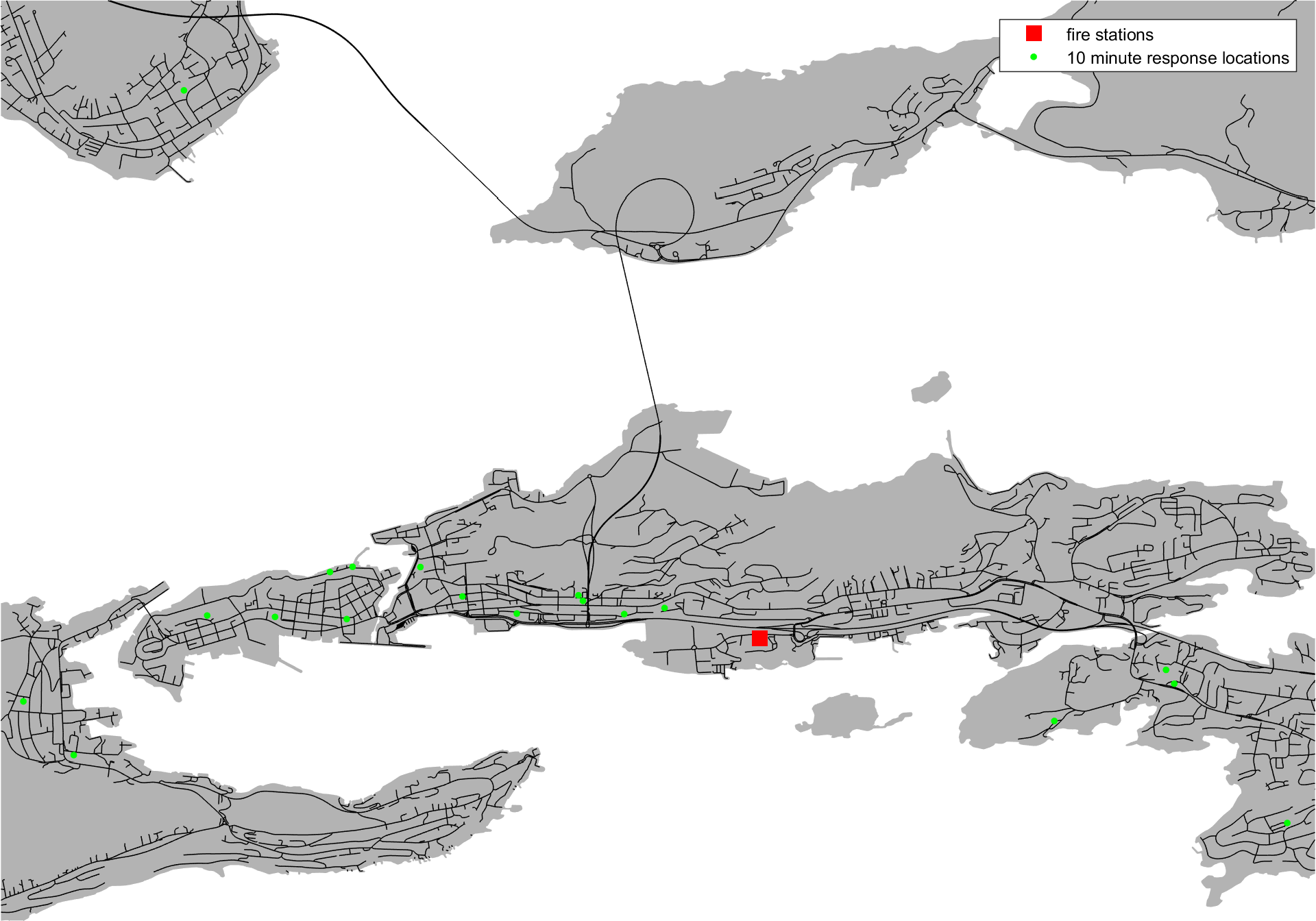}
   \caption{A close-up of the above.}
   \label{fig:highways-fs-sl-zoom}
\end{figure}

Calling the \texttt{plotHighways.m} script will cause the fire service's region to be plotted overlaid with all the roads, as shown in figures \ref{fig:highways-fs-sl}--\ref{fig:highways-fs-sl-zoom}.  Here the locations of the fire stations have been marked with red squares, and the high-priority locations (those which must be responded to within 10 minutes) with green points.


\begin{figure}[tb]
   \centering
   \includegraphics[width=\textwidth]{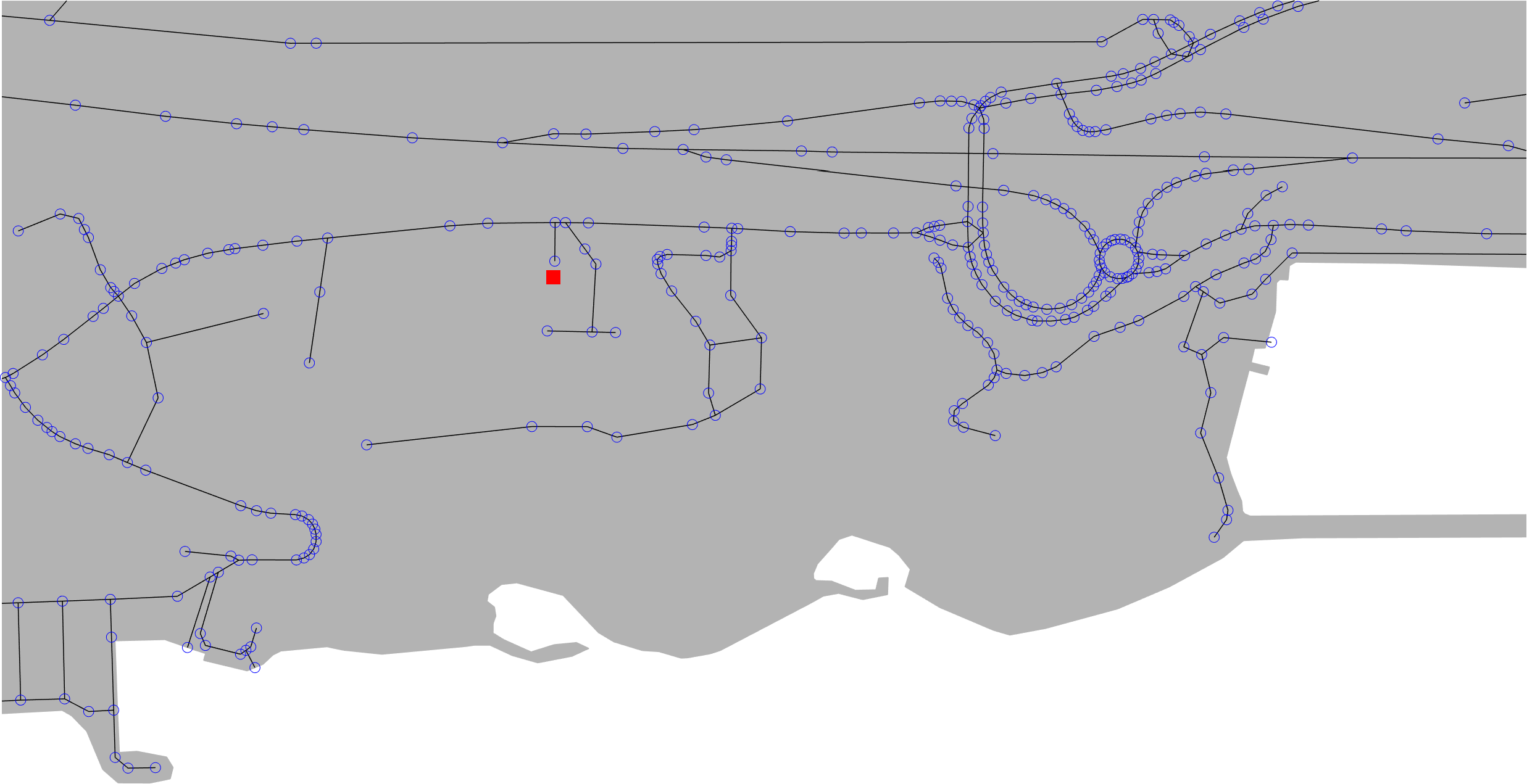}
   \caption{A close-up of the area around the main {\AA}lesund Brannstasjon on Sj{\o}mannsvegen (marked with a red square).
            Roads are shown as black lines and nodes as blue circles.  This shows the complexity of the graph generated from the OSM
            road network, as nodes do not just represent road junctions, but may also represent, for example, the curve of the road,
            a change in the speed limit, or a pedestrian crossing.  Each section of road between two adjacent nodes is represented as
            a single edge in the network.}
   \label{fig:highways-nodes-zoom}
\end{figure}

The complexity of the resulting network is striking; it contains \numprint{124943} nodes and \numprint{249176} edges.  The reasons for the high numbers are two-fold.  Firstly, each edge is unidirectional, so a standard two-way road segment must be represented by two edges.  Secondly, an edge is a section of road between two adjacent nodes, but the nodes represent much more than road junctions; they may also represent, for example, the shape of a curve in the road, a change in the speed limit, or a pedestrian crossing.  As can be seen from figure \ref{fig:highways-nodes-zoom}, where nodes are shown as blue circles and edges as black lines, reducing the size of the network, without changing its essential features, is not straightforward.  For example, the roundabout to the right of the fire station is not easily reduced to a single road junction consisting of one node.

\subsection{Calculating response times heatmaps}
\label{sec:resp_times}

The road map is represented as a network consisting of nodes and edges.  We can, therefore, use standard path-finding algorithms to calculate the distance from each fire station to every other node in the network.  There are many different algorithms, each of which operates most efficiently under different circumstances.  We have chosen to use Dijkstra's algorithm \citep{dijkstra1959} for two reasons: firstly it is guaranteed to find the shortest path between two nodes, and secondly it naturally finds the shortest path from one node to all others.  The well-known A* algorithm \citep{hart1968,dechter1985} is also guaranteed to find the shortest path between two nodes, but is less efficient if we are looking for paths to all nodes.  Dijkstra's algorithm is also easy to understand and implement.

\begin{figure}[p]
   \centering
   \includegraphics[width=0.8\textwidth]{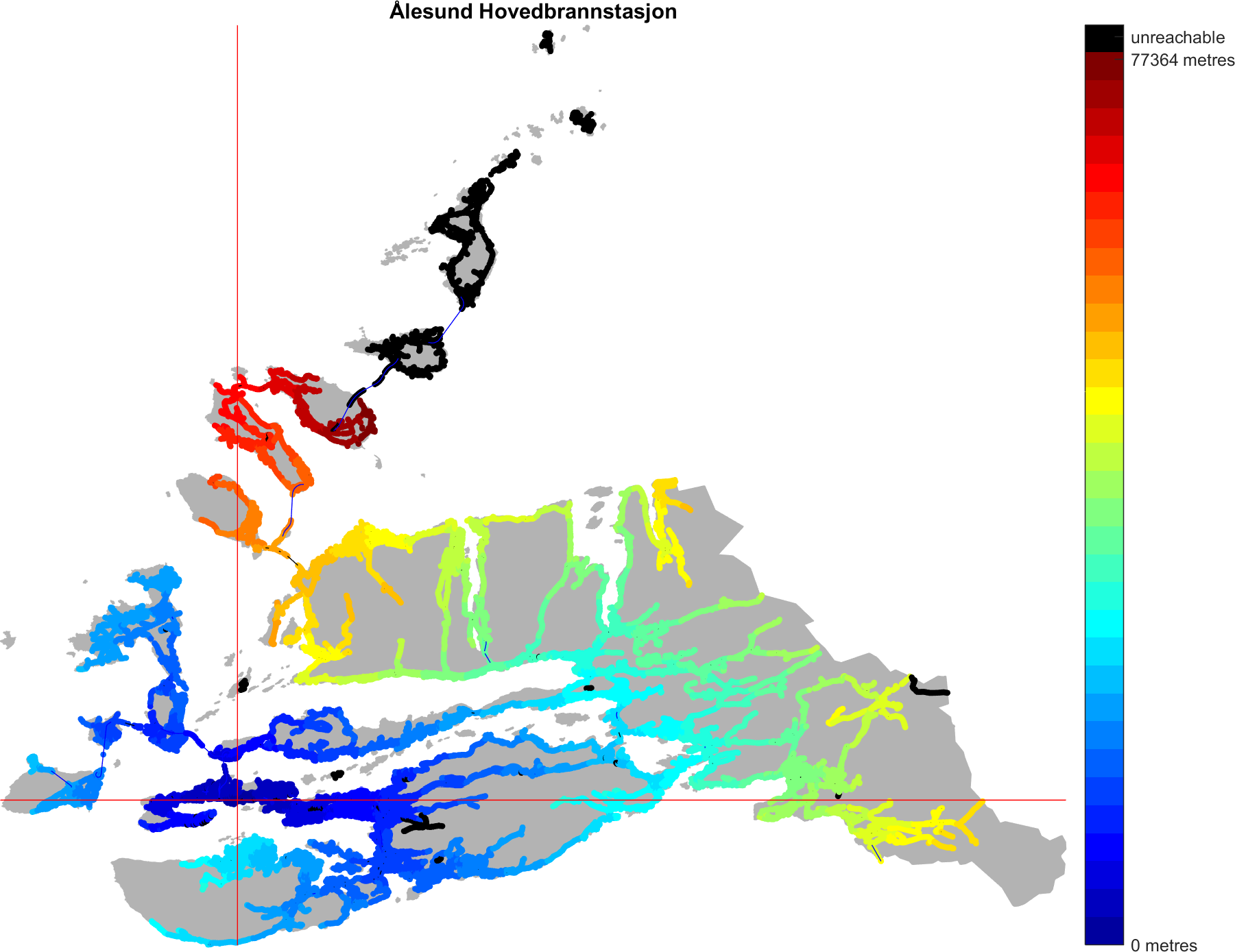}
   \caption{Each node in this map has been coloured according to its distance from the {\AA}lesund Hovedbrannstasjon (marked with the red
            cross-hairs).  The colour key is shown on the right-hand side.}
   \label{fig:main-station-metres}
\vspace{5mm}
   \centering
   \includegraphics[width=0.8\textwidth]{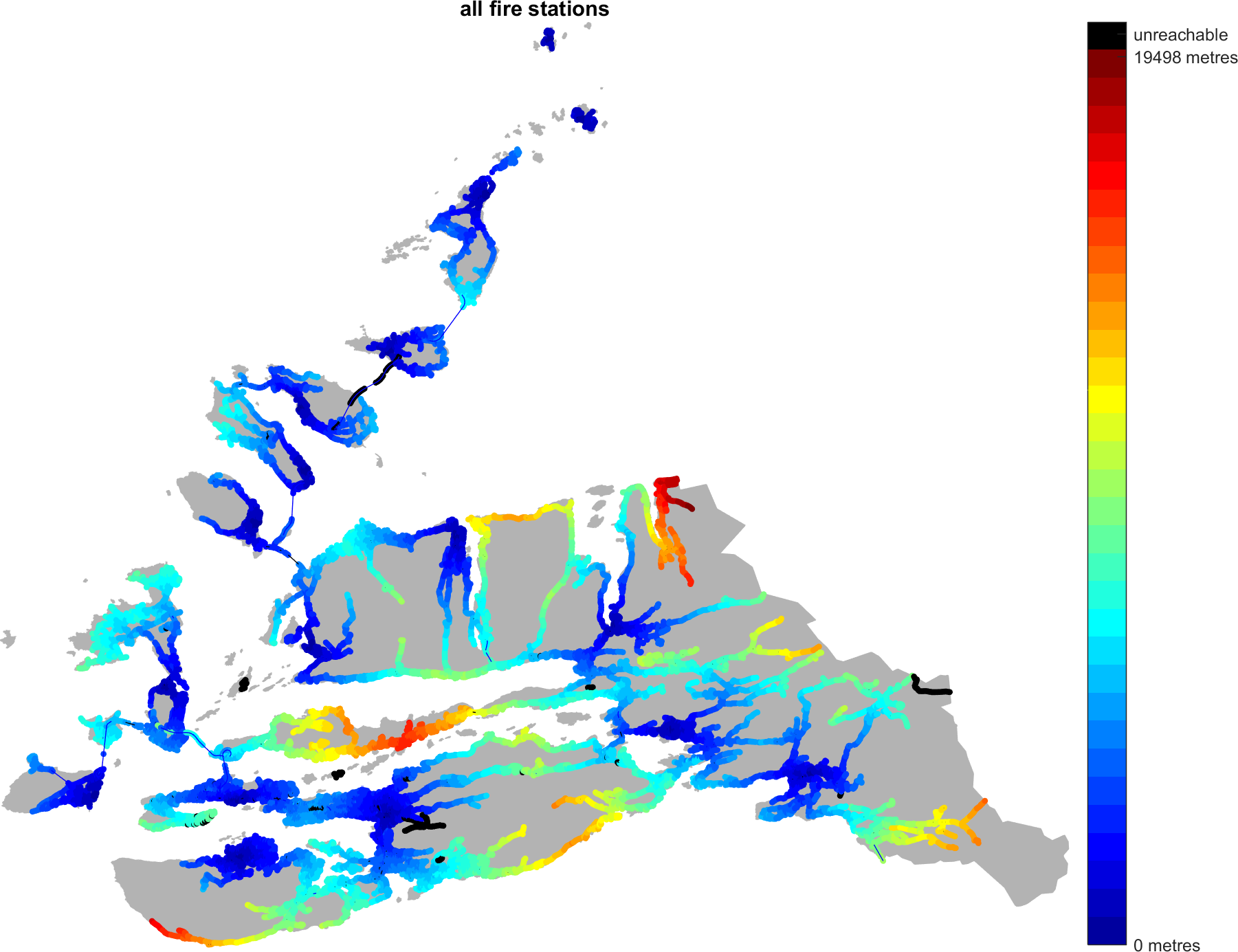}
   \caption{Each node in this map has been coloured according to its shortest distance from any fire station.
            The colour key is shown on the right-hand side.}
   \label{fig:all-stations-metres}
\end{figure}

\begin{figure}[p]
   \centering
   \includegraphics[width=0.8\textwidth]{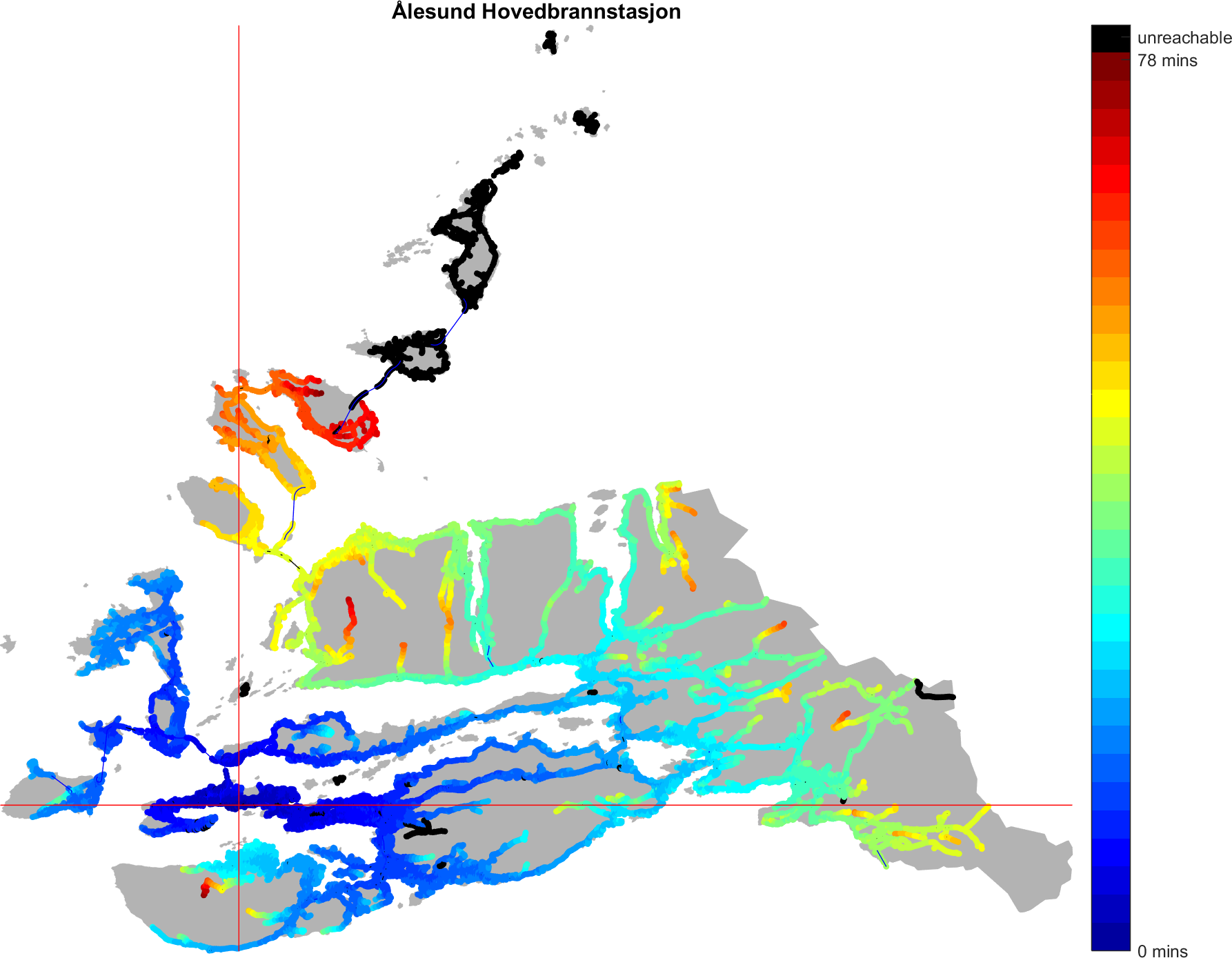}
   \caption{Each node in this map has been coloured according to its distance in time from the {\AA}lesund Hovedbrannstasjon (marked with the
            red cross-hairs).  The colour key is shown on the right-hand side.}
   \label{fig:main-station-time}
\vspace{5mm}
   \centering
   \includegraphics[width=0.8\textwidth]{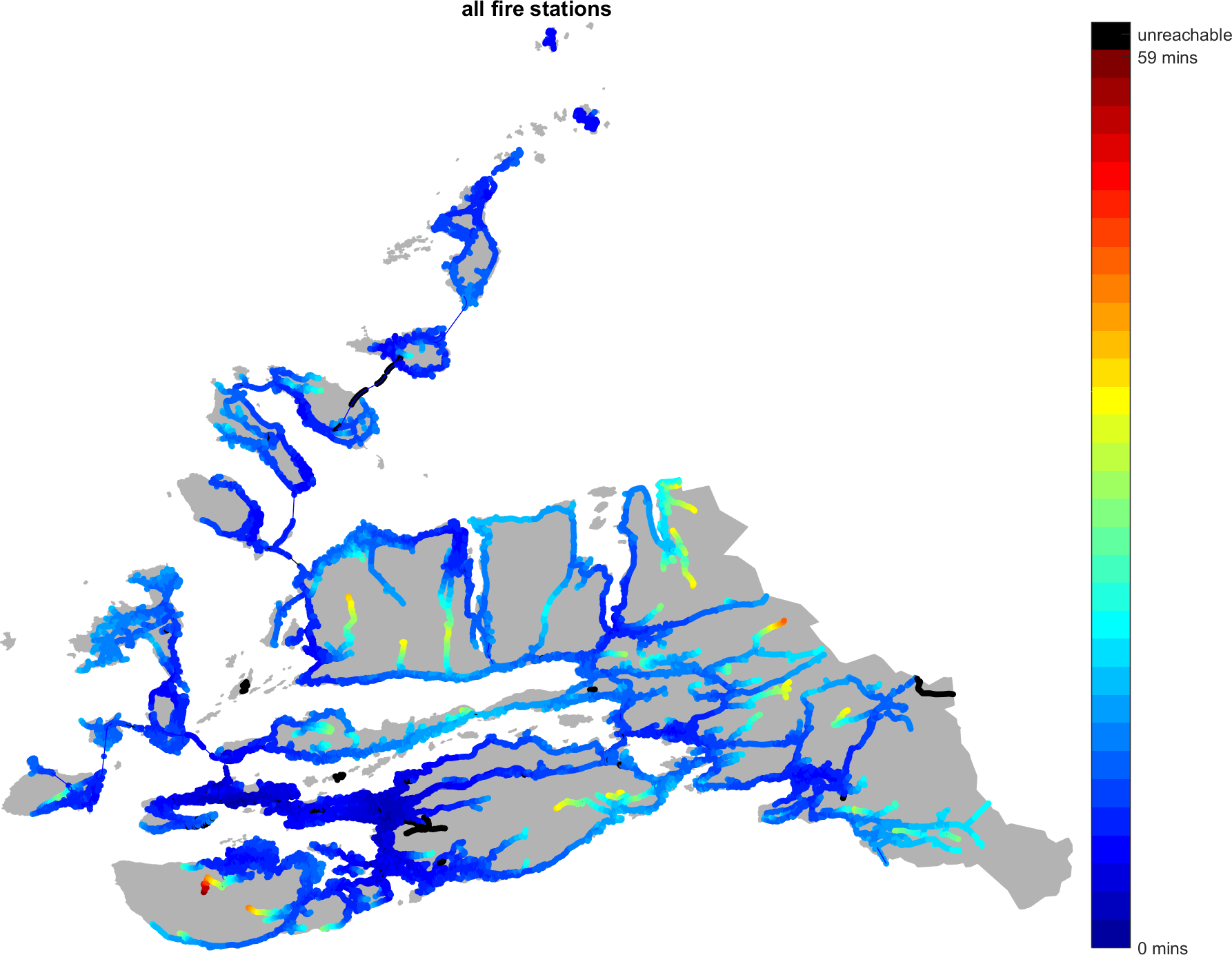}
   \caption{Each node in this map has been coloured according to its shortest distance in time from any fire station.
            The colour key is shown on the right-hand side.}
   \label{fig:all-stations-time}
\end{figure}

The first step in calculating response times is to calculate the shortest distance from each fire station to every other node in the network.  We can now create a map of the network nodes coloured by how far away each one is from a given fire station, as shown in figure \ref{fig:main-station-metres}.

From this map, showing all distances from the {\AA}lesund Hovedbrannstasjon, we can see that the furthest reachable distance is \SI{77.364}{\kilo\metre}, on the north-eastern edge of Flems{\o}ya.  This is easily validated by comparison with Google Maps, which gives a distance of \SI{78.1}{\kilo\metre}.  There are a number of black nodes on this map; these represent locations that are not reachable from that particular fire station because there is no route available.  For the northern-most islands this is clearly because there is no road link, either by bridge or tunnel.  There seems to be a connection between Flems{\o}ya and Fj{\o}toft, but this is showing as unreachable.  This is because although the tunnel had been completed at the time this map was constructed, the link roads had not.  There are other small sections of unreachable road: at the eastern edge this road has been cut off by the boundary, while the cross-shaped section east and south of the fire station is a track mislabelled as a road.

Combining the distance data from all fire stations, we can create a similar map that shows the minimum distance to any fire station; see figure \ref{fig:all-stations-metres}.  Now we can see that the maximum distance is about \SI{20}{\kilo\metre} to Rekdal.

\begin{figure}[tb]
   \centering
   \includegraphics[width=0.8\textwidth]{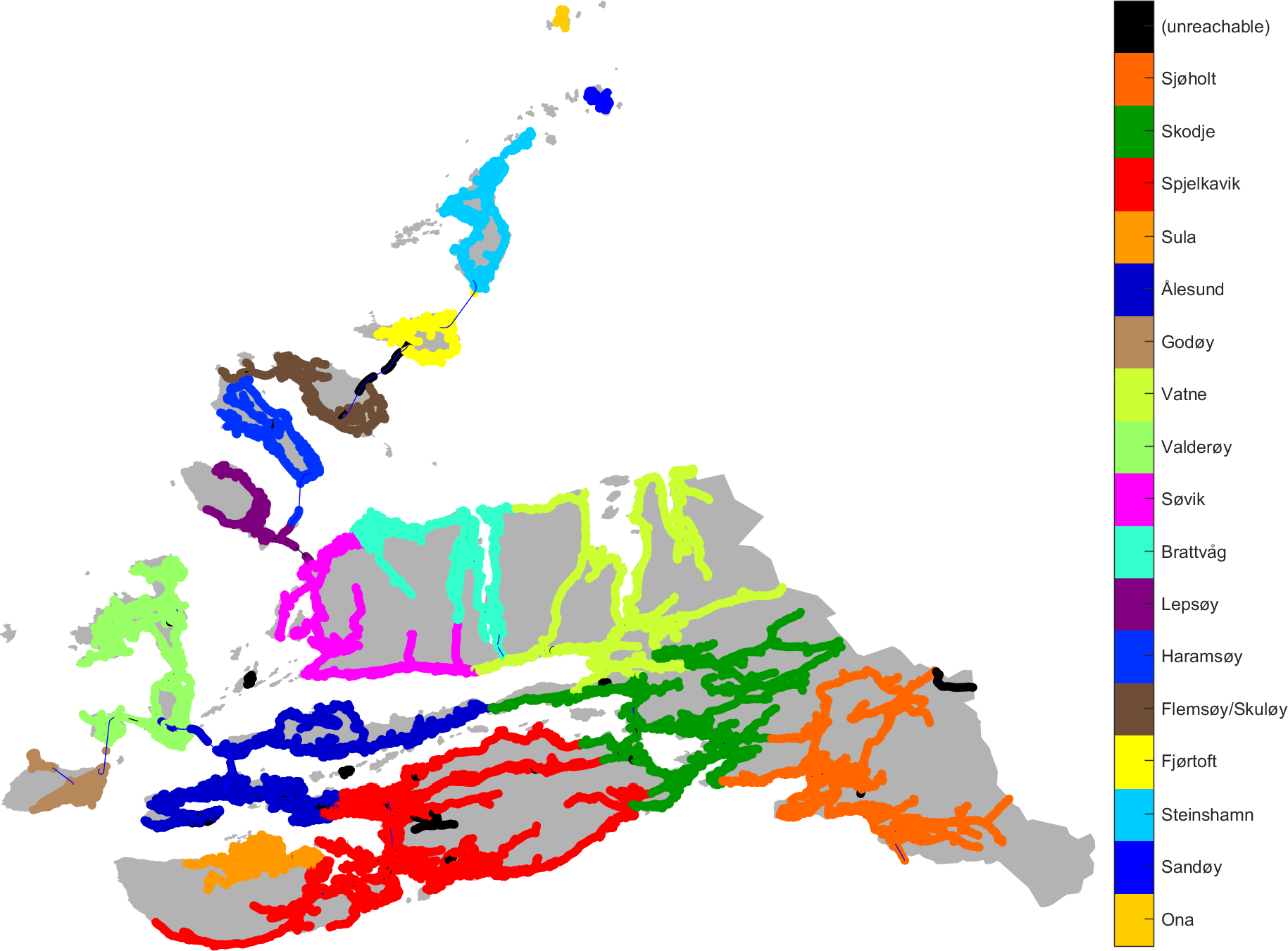}
   \caption{The areas closest (in terms of travel time) to each fire station.}
   \label{fig:areas}
\end{figure}

However, distance is not the key measure, while time is, so we need to convert these distances to time.  Nearly every highway in the OSM map has a maximum speed associated with it.  Instead of calculating distances in metres, we can use this speed combined with the length of an edge to calculate how long a vehicle would take to travel along the edge (the distance in time, still using Dijkstra's algorithm), assuming that vehicles were travelling at that maximum speed.  For roads not labelled with maximum speeds, we have applied a default value as shown in table \ref{tab:highway-types}.  Figures \ref{fig:main-station-time} and \ref{fig:all-stations-time} are the travel time equivalents of figures \ref{fig:main-station-metres} and \ref{fig:all-stations-metres}.

Each fire station has an area that they are closest to in terms of travel time.  These areas are shown in figure \ref{fig:areas}.

\begin{figure}[tb]
   \centering
   \includegraphics[width=0.8\textwidth]{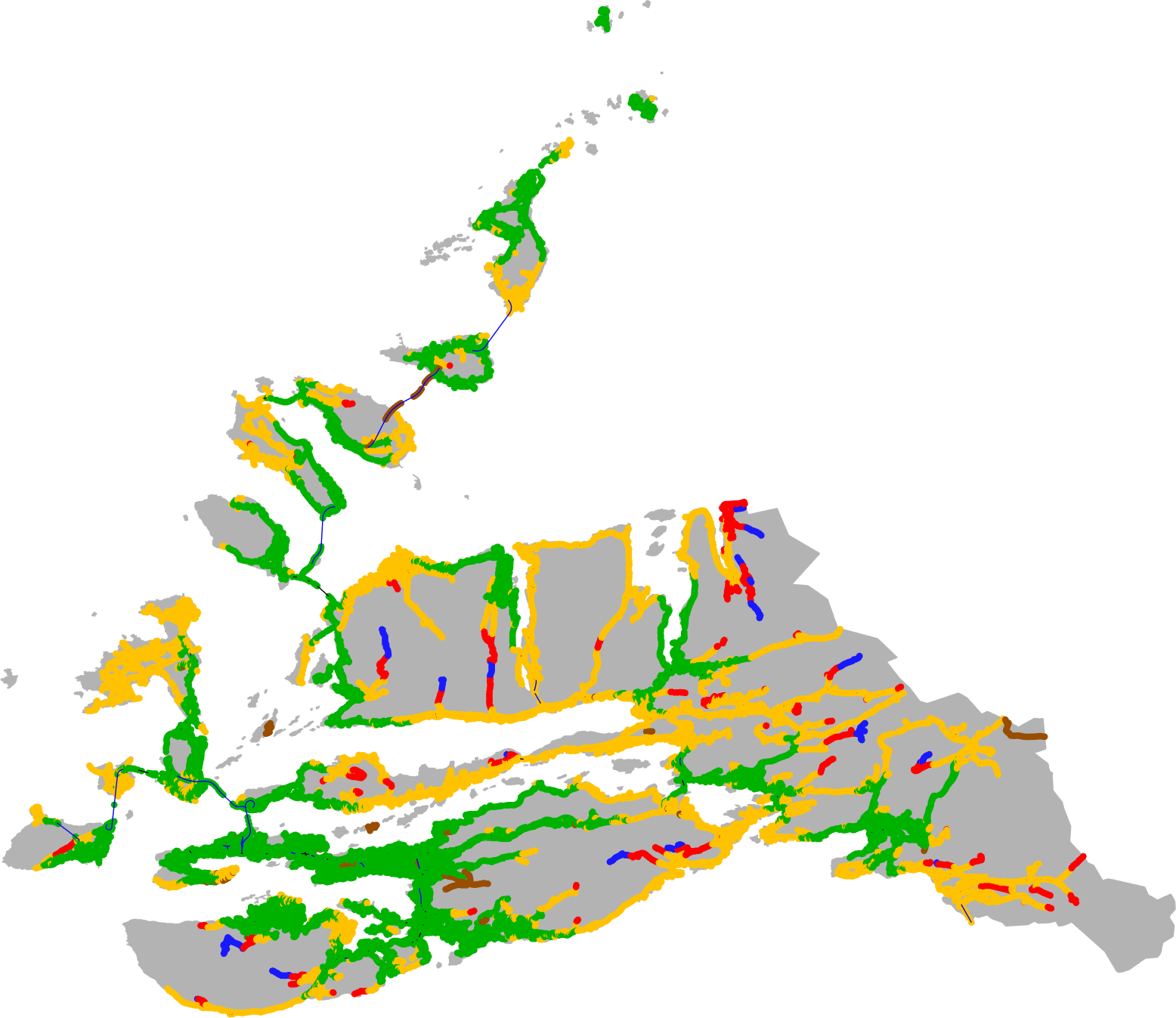}
   \caption{The time-banded equivalent of figure \ref{fig:all-stations-time} showing the time bands for response from the nearest
            fire station.  This represents the current situation, and is the baseline for later comparisons.}
   \label{fig:baseline}
\end{figure}

The time distances to the nodes are then discretised into the required time bands: less than 10 minutes (green), 10-20 minutes (amber), 20-30 minutes (red) and over 30 minutes (blue).  Two additional bands are included: nodes that are always unreachable (coloured black in figure \ref{fig:all-stations-time}) which we now colour brown, and nodes that are conditionally unreachable because of the nature of a test scenario (perhaps an island fire station has been closed, for example) which we colour black.  The banded equivalent of figure \ref{fig:all-stations-time} is shown in figure \ref{fig:baseline}.

\afterpage{\clearpage}

\newpage
\section{``What if'' scenarios, part 1}
\label{sec:scenarios-1}

With this number of nodes and edges, calculating the distances is a relatively slow procedure, taking about 3 or 4 minutes for each fire station.  If the road network changes then the distances from each fire station to every node must be recalculated, and if a new fire station opens or an existing one changes location, then the distance from that new location to every node must be recalculated.  These calculation times were recorded when running unoptimised Matlab code on a standard laptop.  On a larger server the process could be speeded up, certainly calculating the distances for each fire station in parallel rather than sequentially, but there are also versions of Dijkstra's algorithm that are designed for parallel execution (e.g. \citep{crauser1998}).

\begin{figure}[p]
   \centering
   Fj{\o}rtoft Brannstasjon (part-time) \\[-4mm]
   \includegraphics[width=0.38\textwidth]{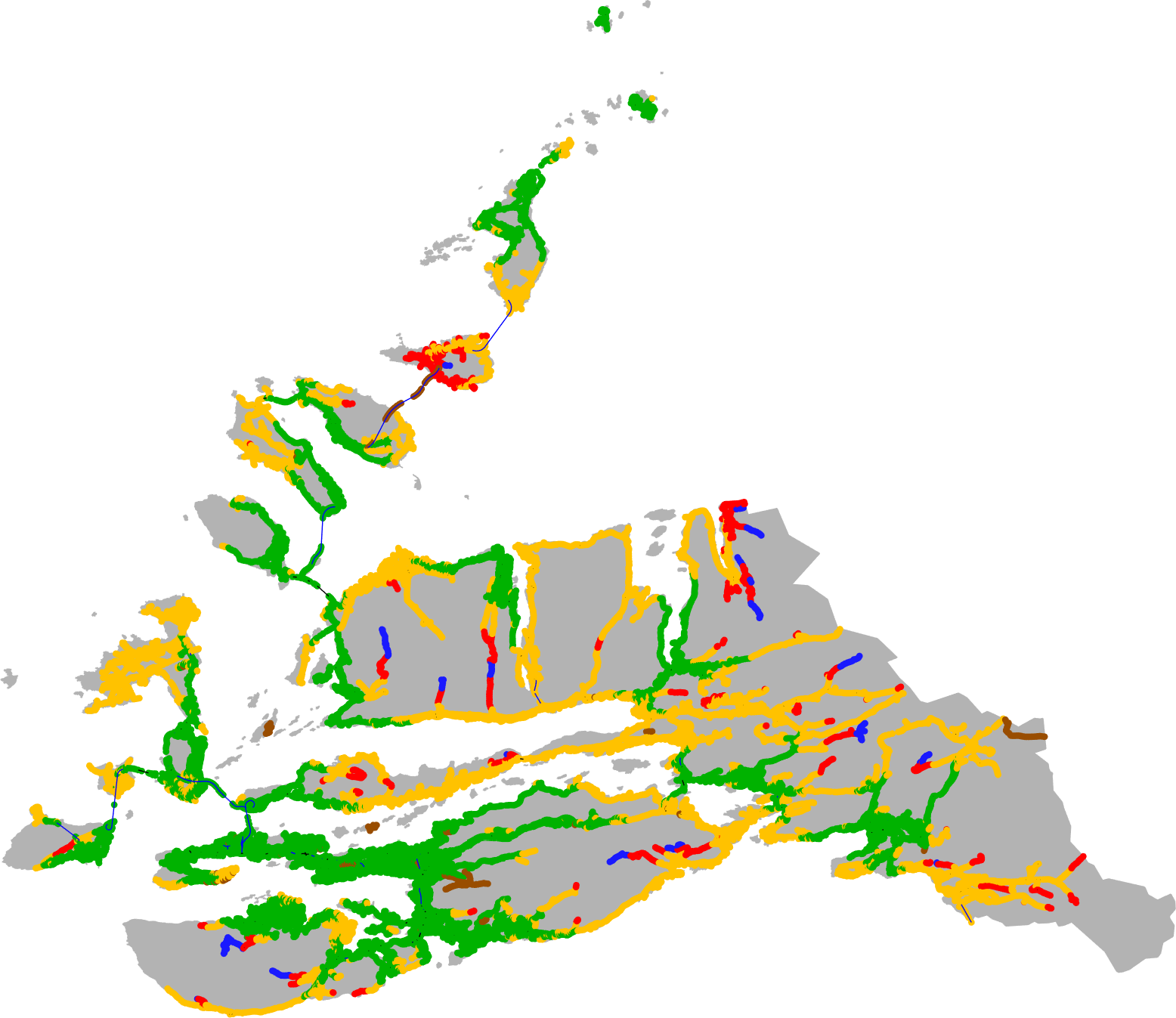}
   \hfill\includegraphics[width=0.38\textwidth]{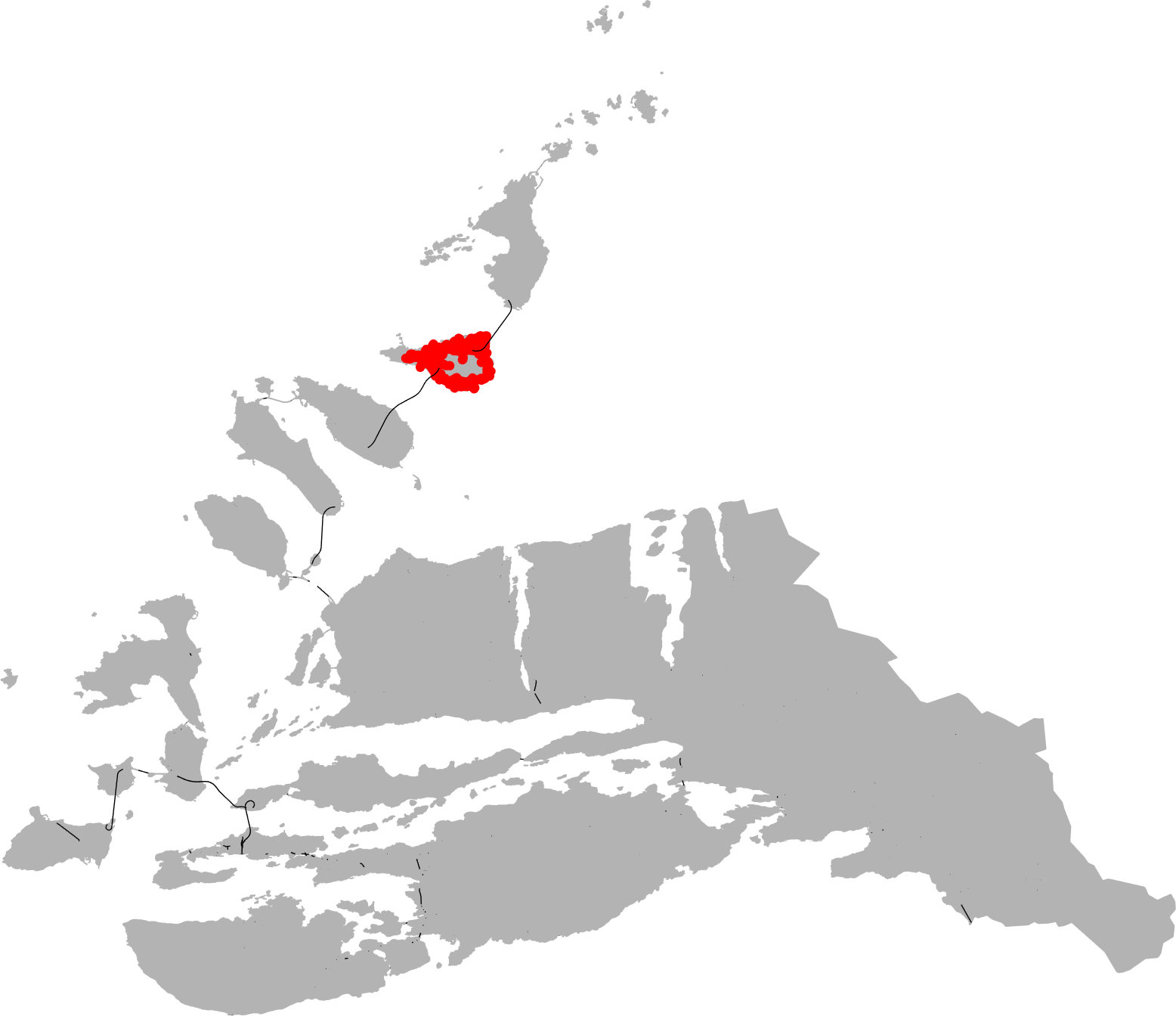} \\[5mm]
   Vatne Brannstasjon (part-time) \\[-4mm]
   \includegraphics[width=0.38\textwidth]{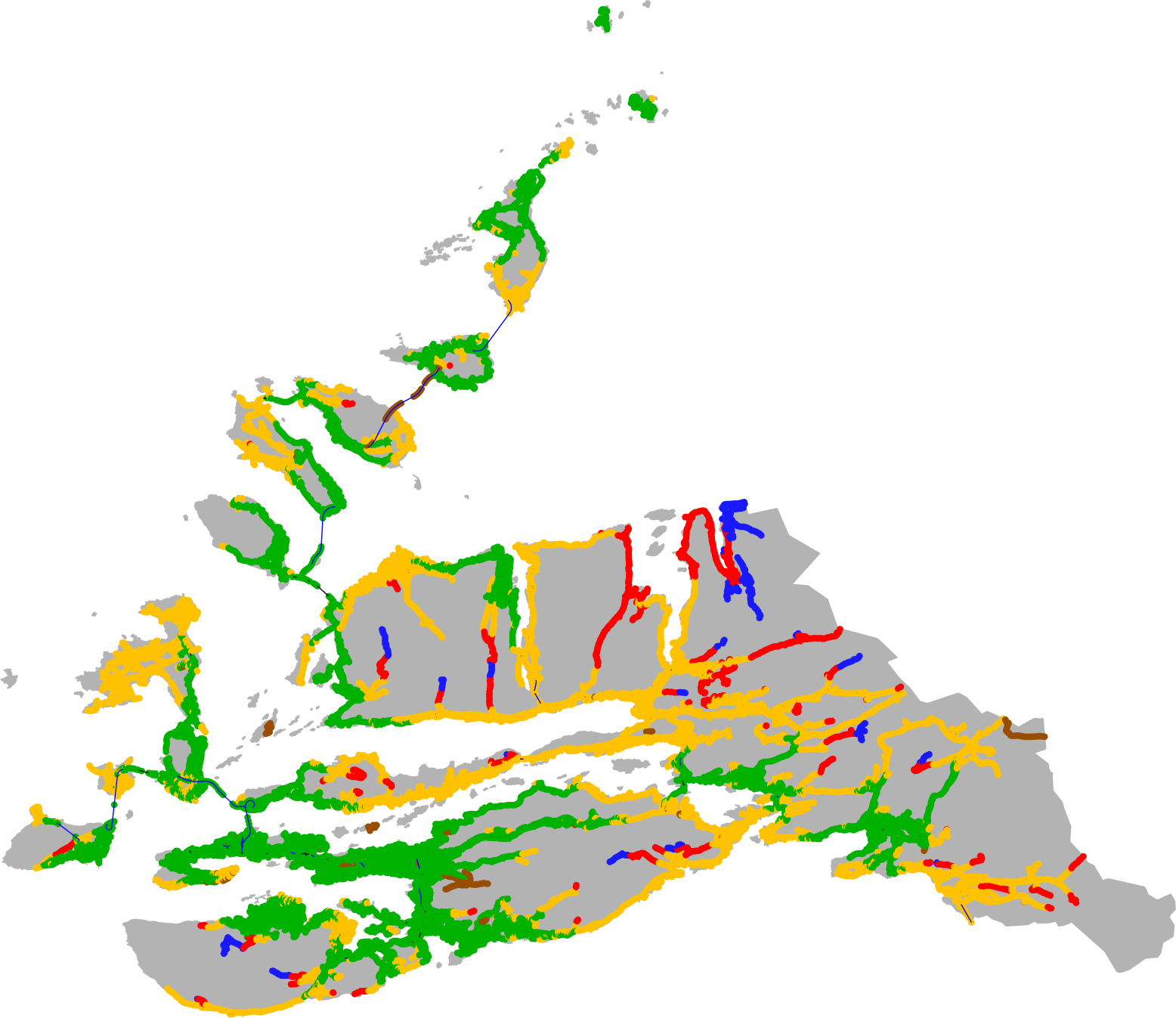}
   \hfill\includegraphics[width=0.38\textwidth]{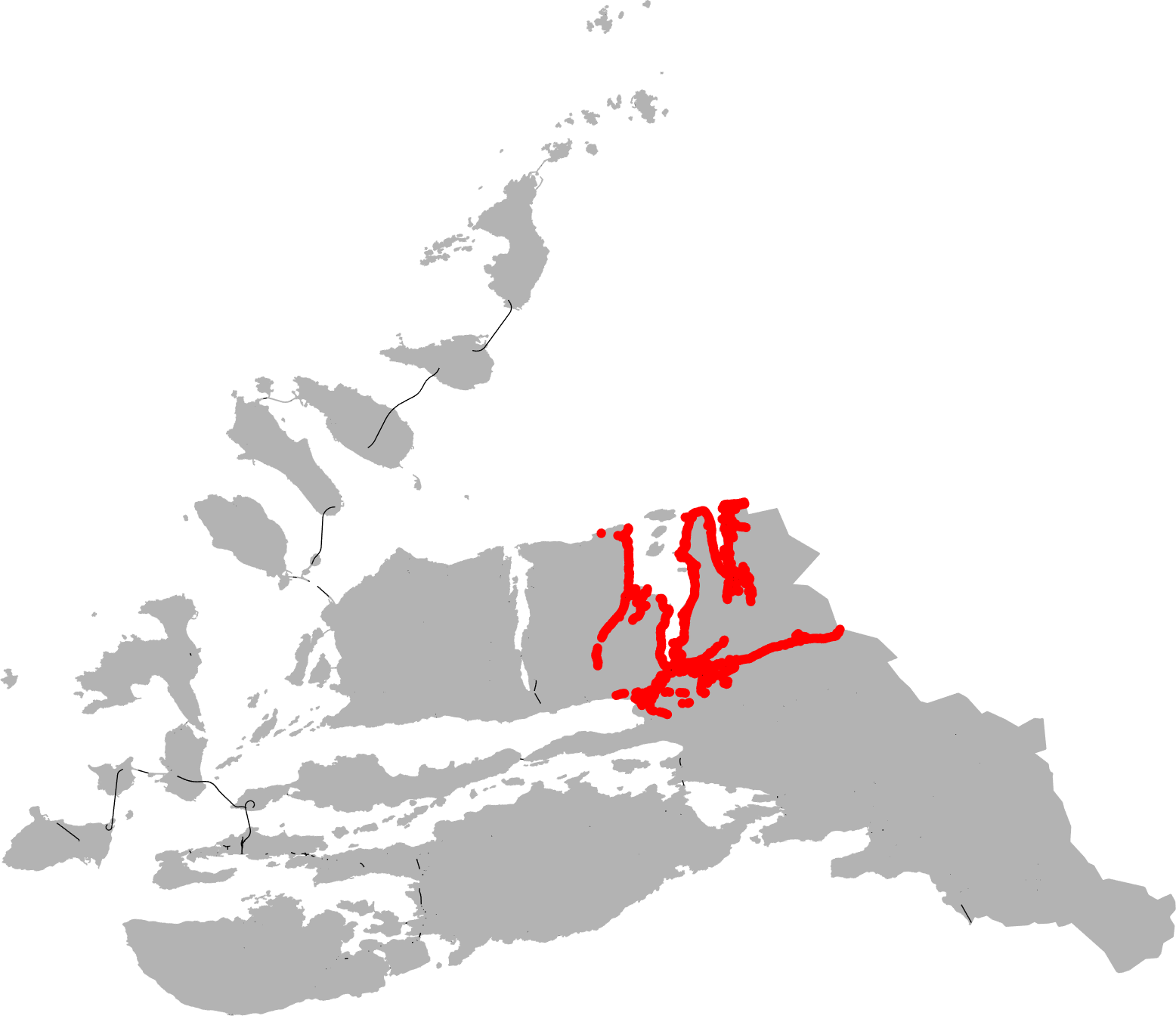} \\[5mm]
   {\AA}lesund Hovedbrannstasjon (full-time) \\[-4mm]
   \includegraphics[width=0.38\textwidth]{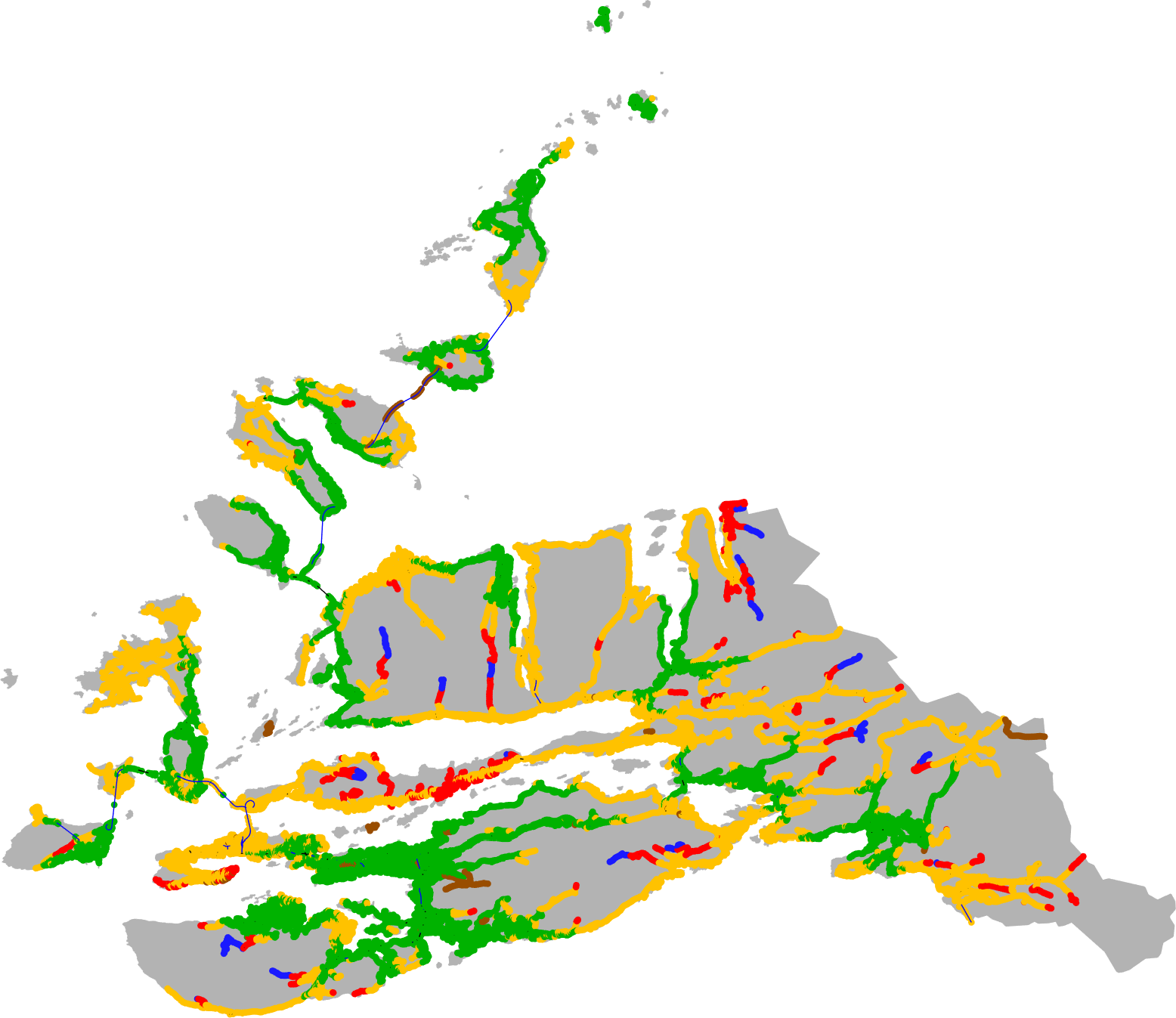}
   \hfill\includegraphics[width=0.38\textwidth]{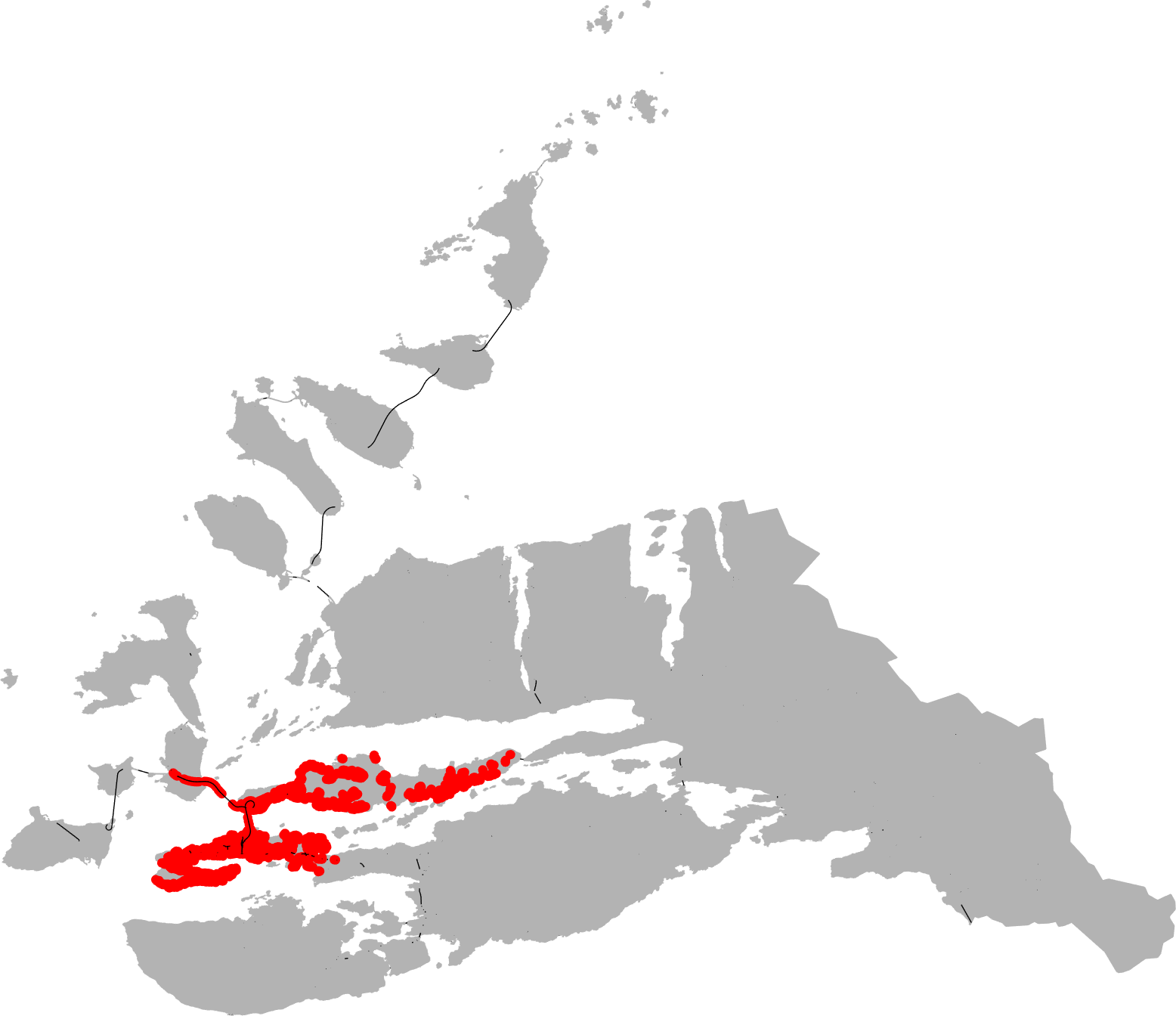} \\[5mm]
   Spjelkavik Brannstasjon (full-time) \\[-4mm]
   \includegraphics[width=0.38\textwidth]{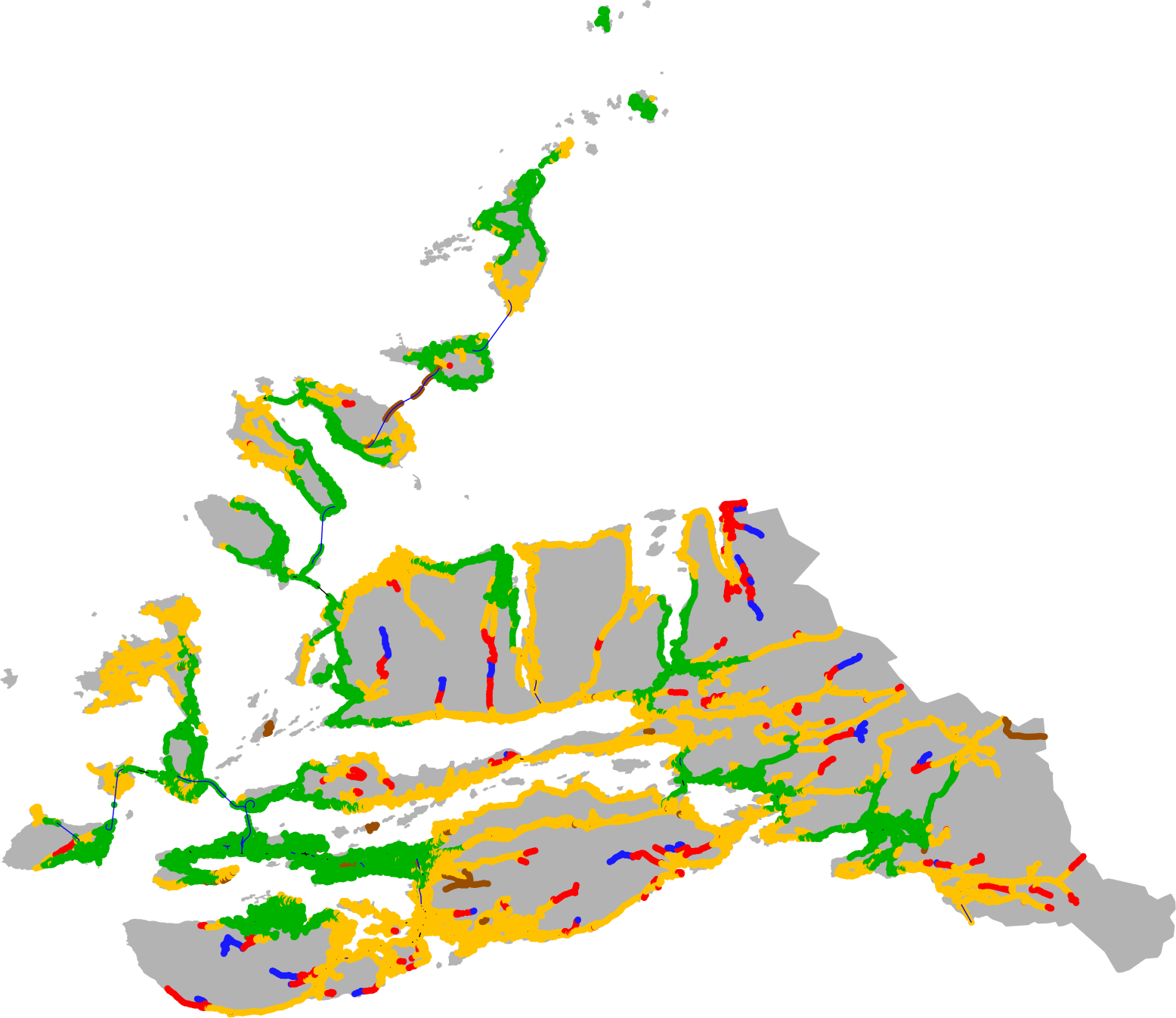}
   \hfill\includegraphics[width=0.38\textwidth]{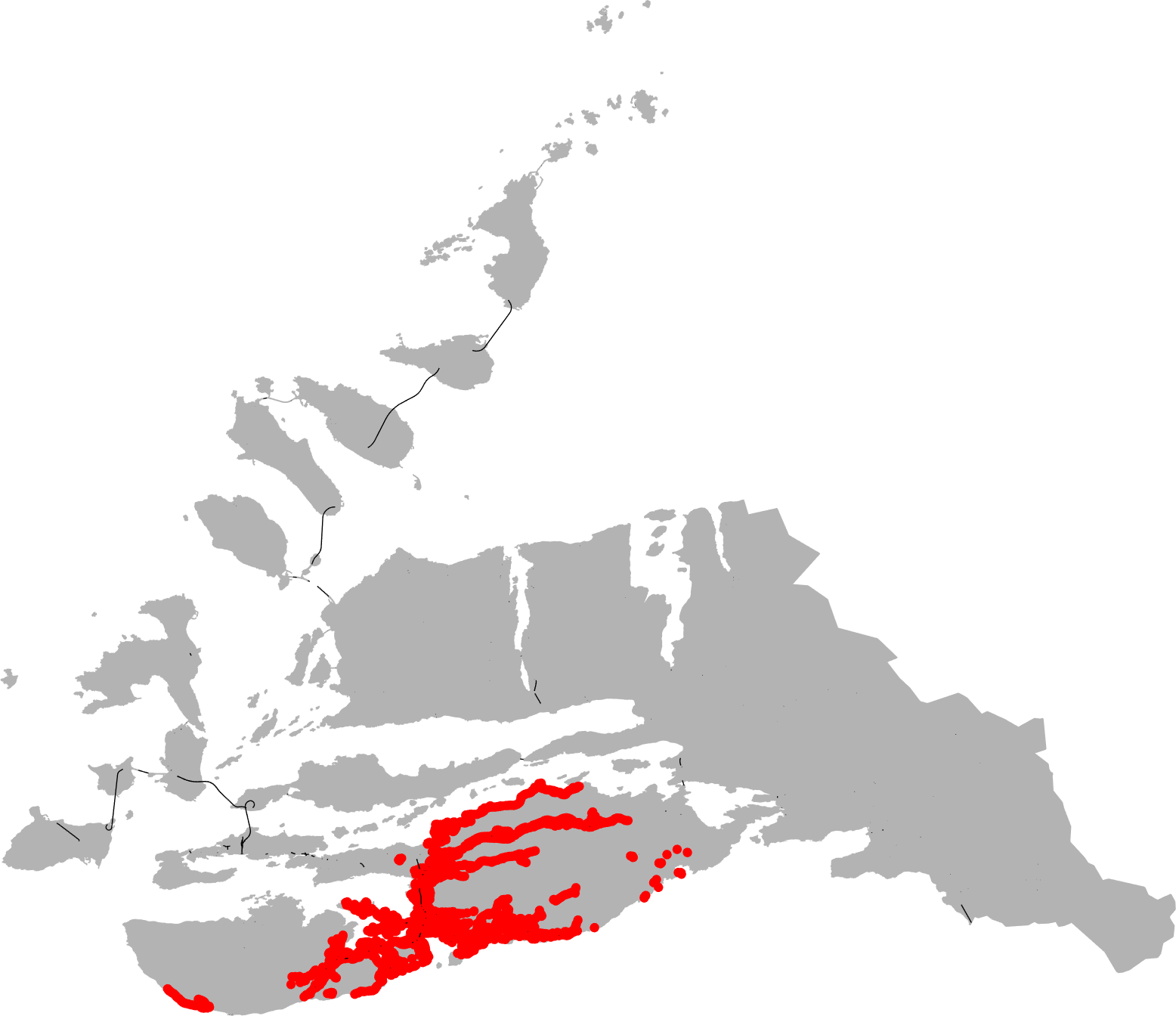}
   \caption{The effect on time of closing individual fire stations.
            The left-hand column shows the time heatmap. The right-hand column shows the associated difference map.}
   \label{fig:closing-one-station}
\end{figure}

\begin{figure}[p]
   \centering
   Fj{\o}rtoft Brannstasjon (part-time) \\[-4mm]
   \includegraphics[width=0.38\textwidth]{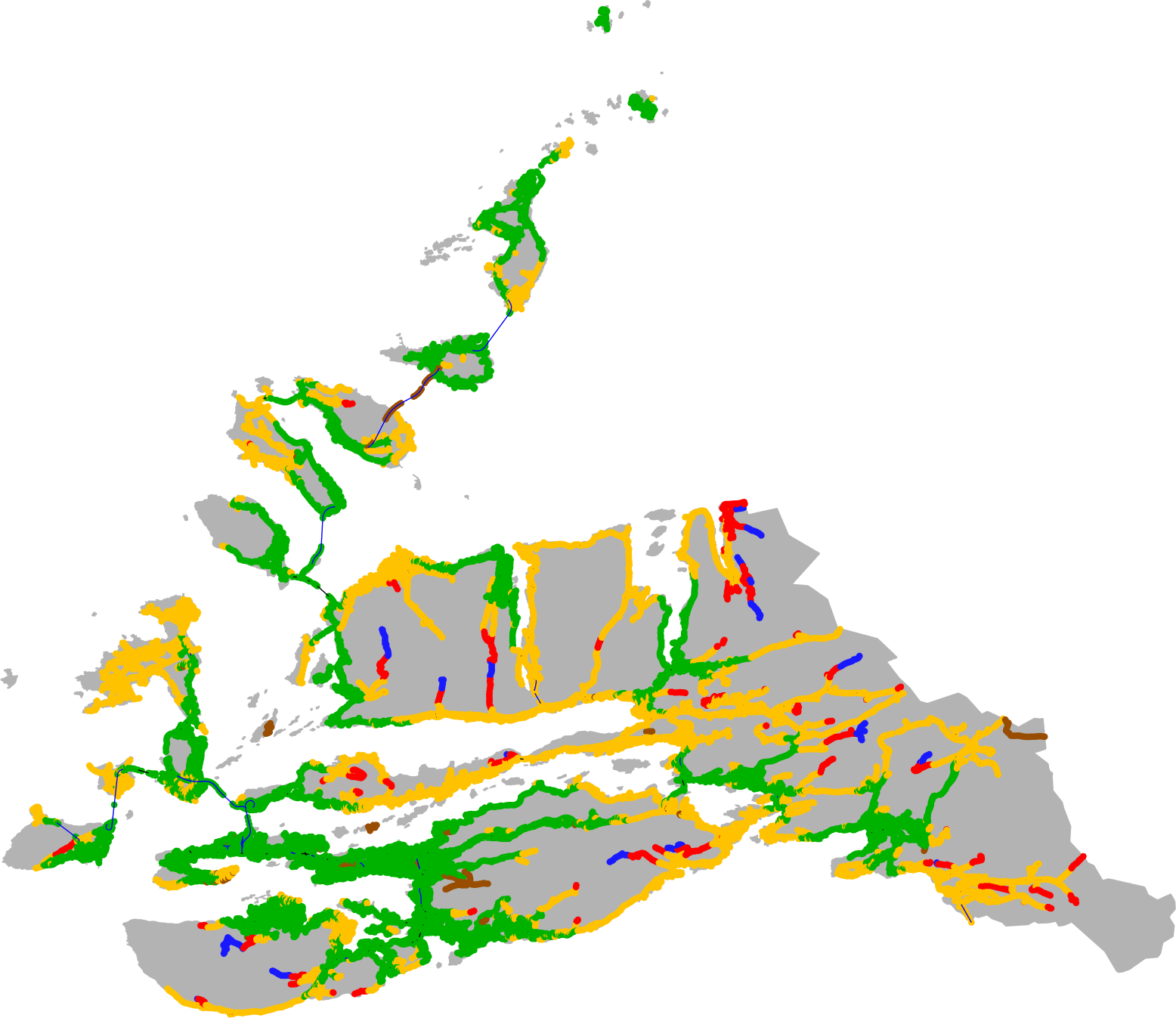}
   \hfill\includegraphics[width=0.38\textwidth]{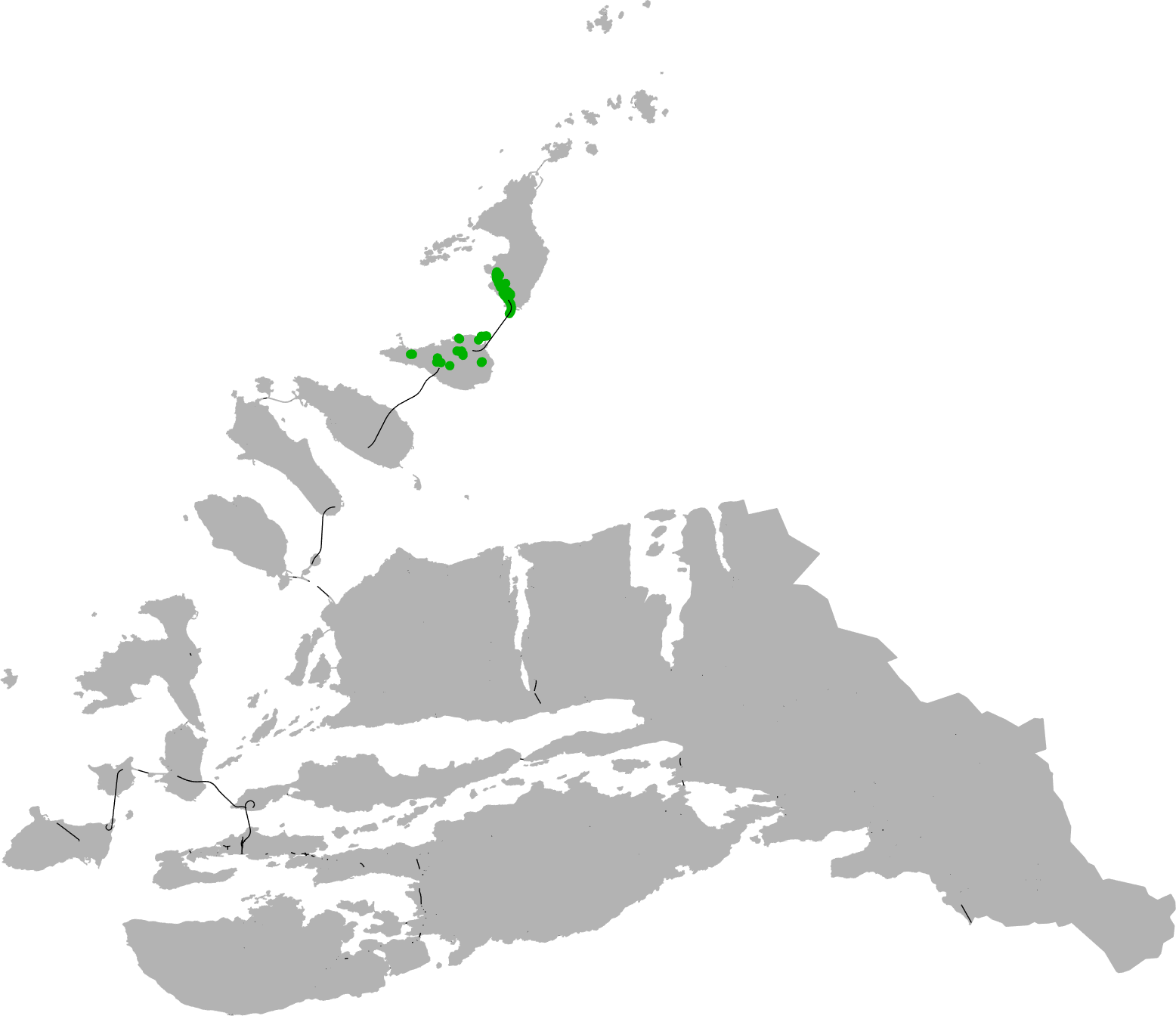} \\[5mm]
   Vatne Brannstasjon (part-time) \\[-4mm]
   \includegraphics[width=0.38\textwidth]{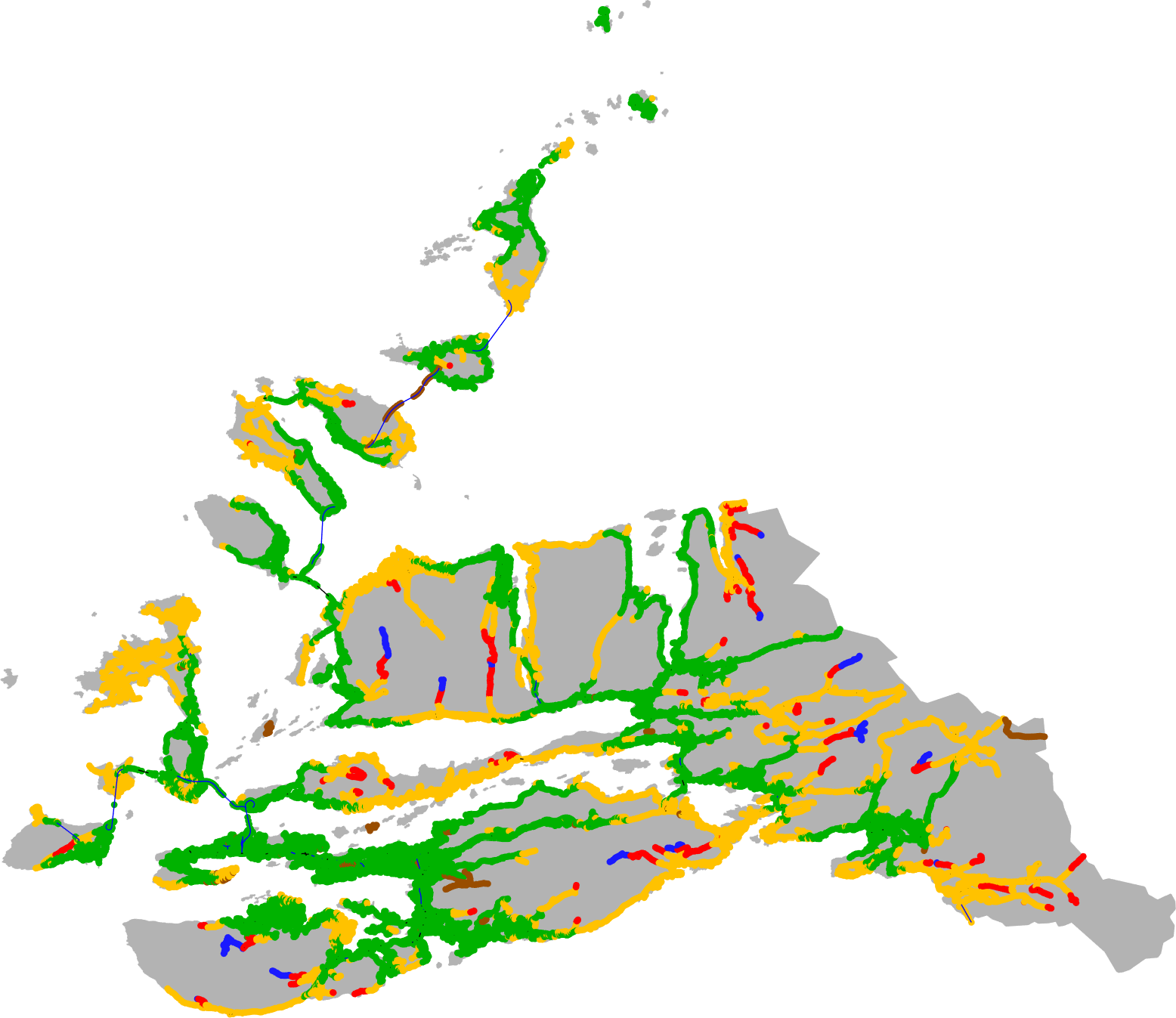}
   \hfill\includegraphics[width=0.38\textwidth]{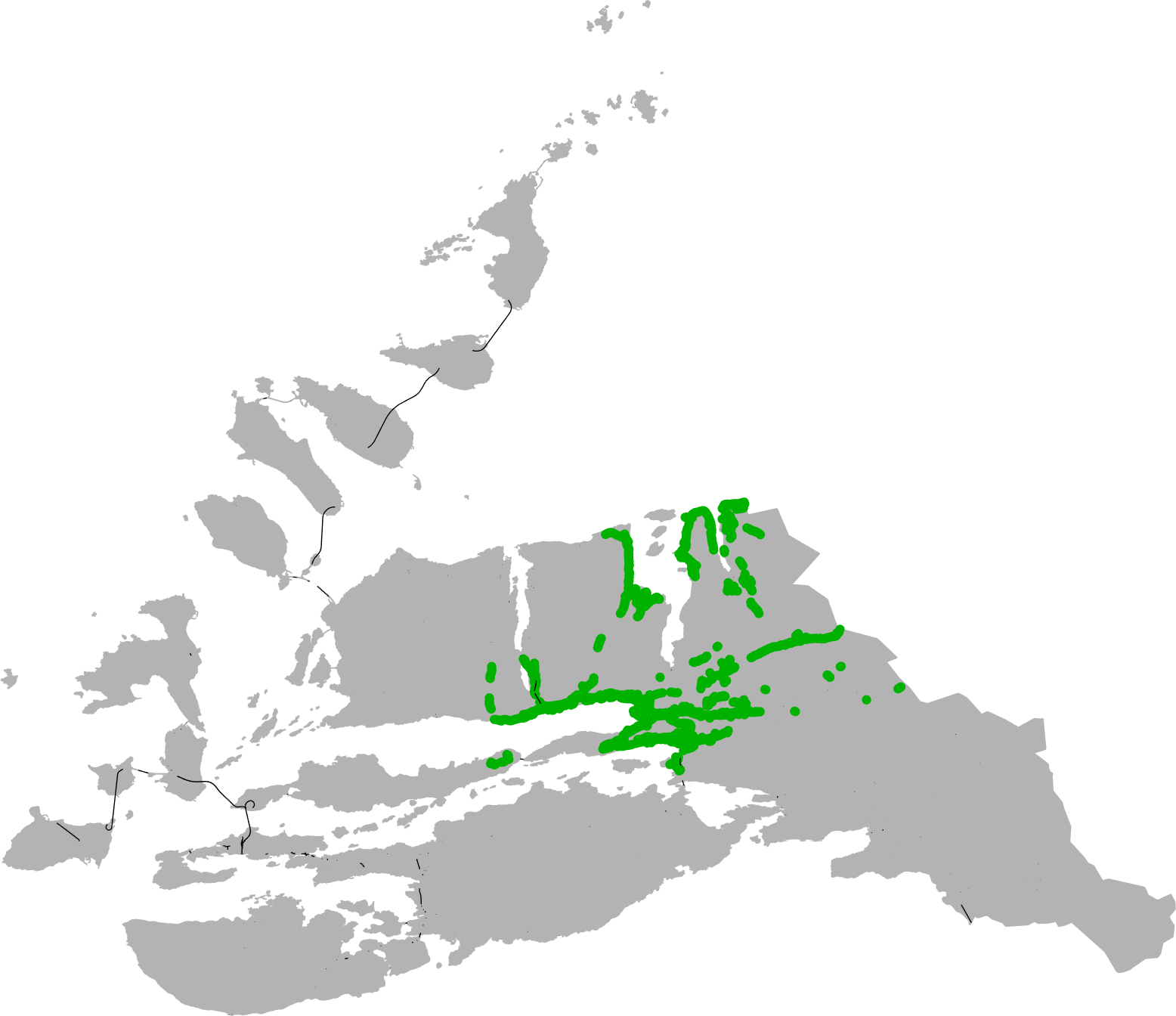} \\[5mm]
   {\AA}lesund Hovedbrannstasjon (full-time) \\[-4mm]
   \includegraphics[width=0.38\textwidth]{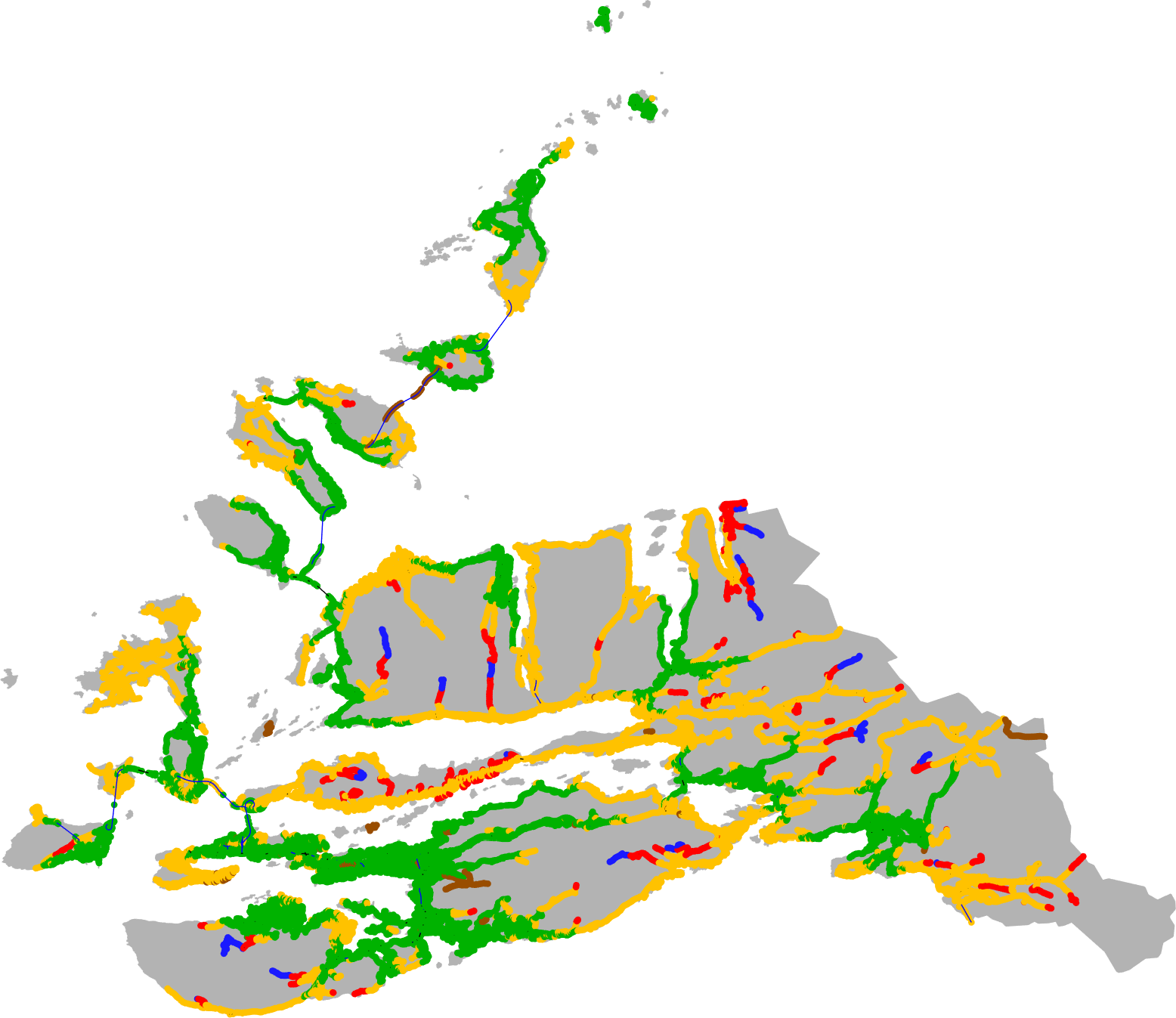}
   \hfill\includegraphics[width=0.38\textwidth]{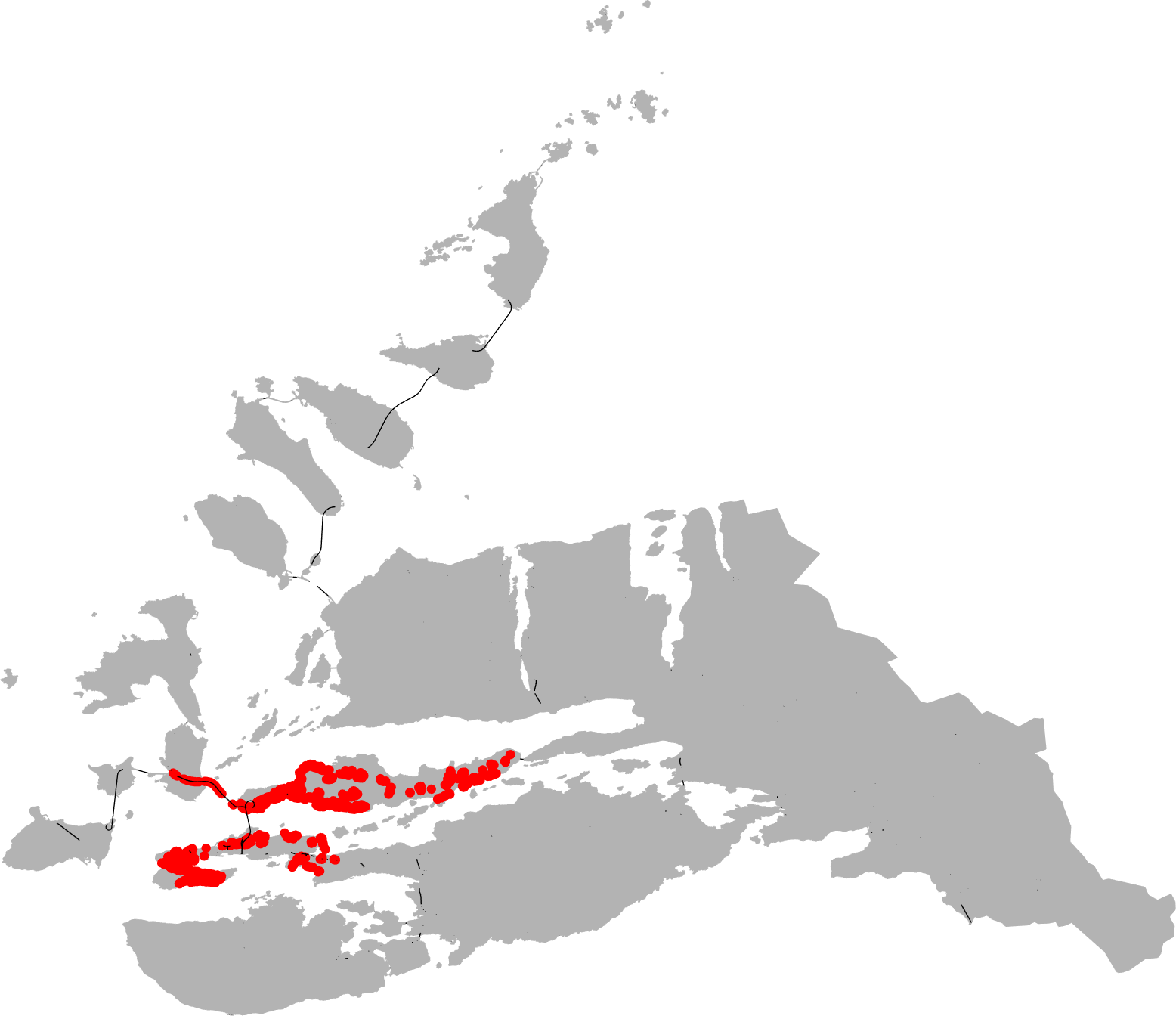} \\[5mm]
   Spjelkavik Brannstasjon (full-time) \\[-4mm]
   \includegraphics[width=0.38\textwidth]{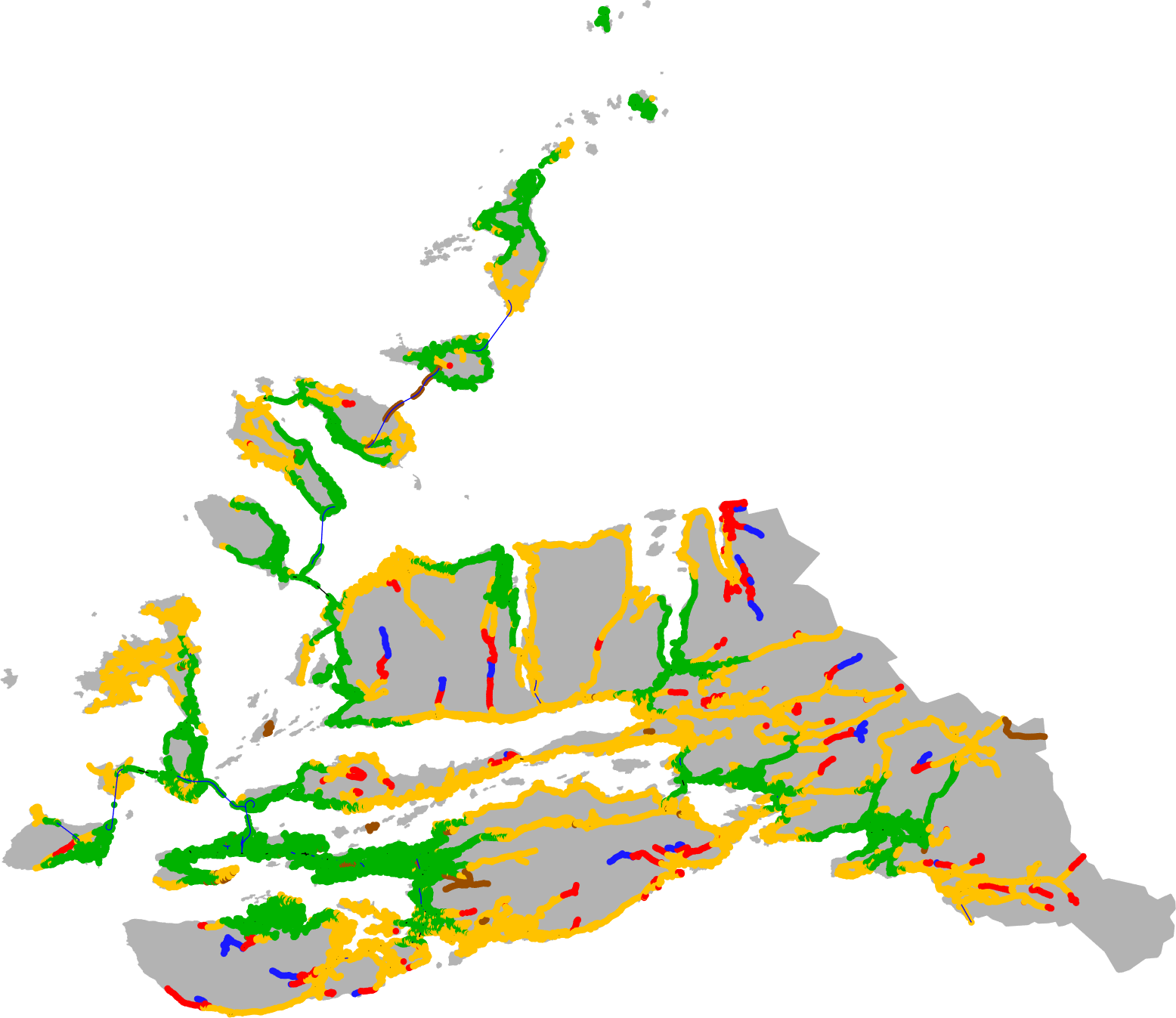}
   \hfill\includegraphics[width=0.38\textwidth]{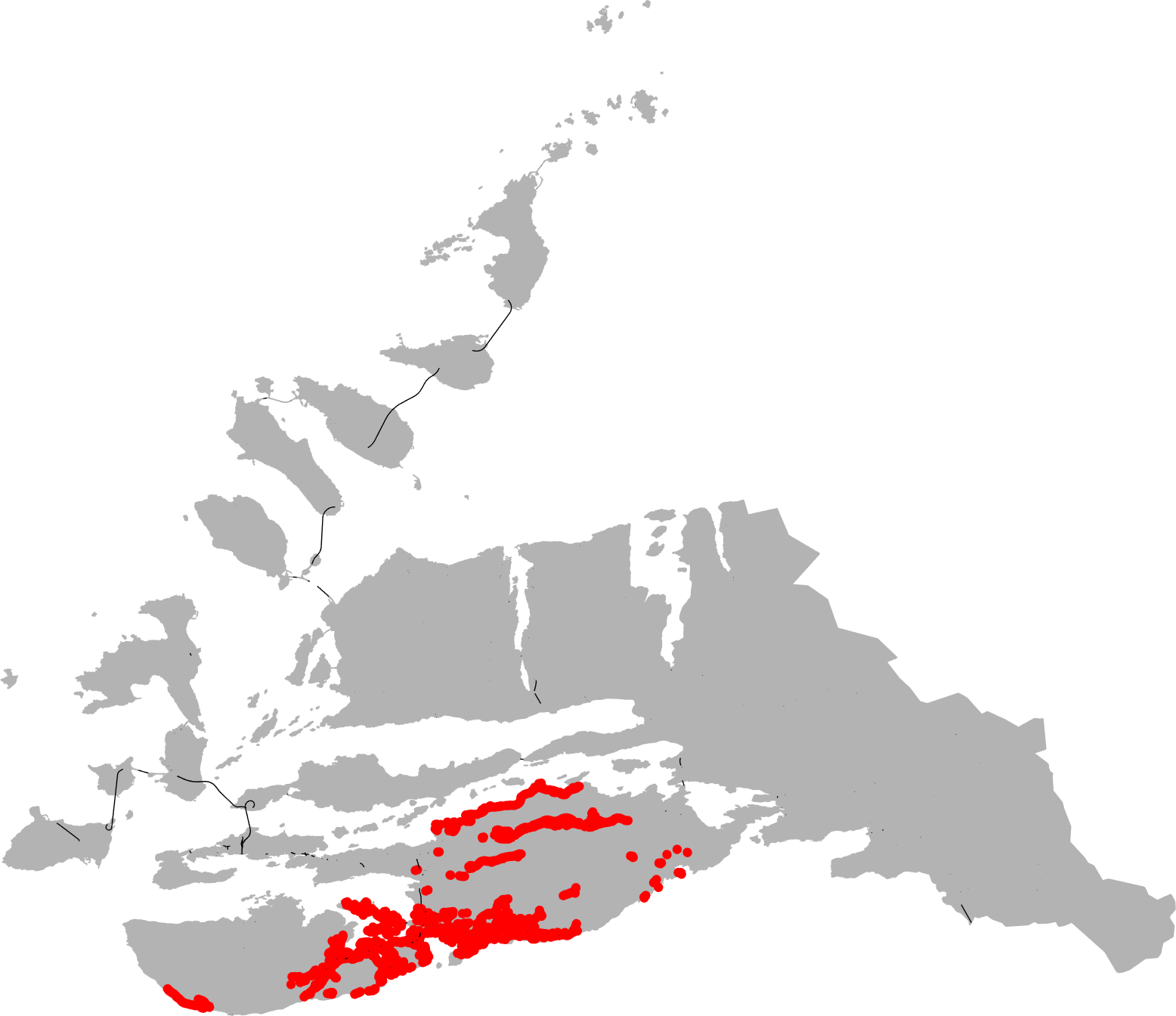}
   \caption{The effect on time of switching individual fire stations from part-time to full-time, or vice versa.
            The left-hand column shows the time heatmap. The right-hand column shows the associated difference map.}
   \label{fig:switching-one-station}
\end{figure}

\begin{figure}[p]
   \centering
   1\hspace{-1em}\includegraphics[width=0.38\textwidth]{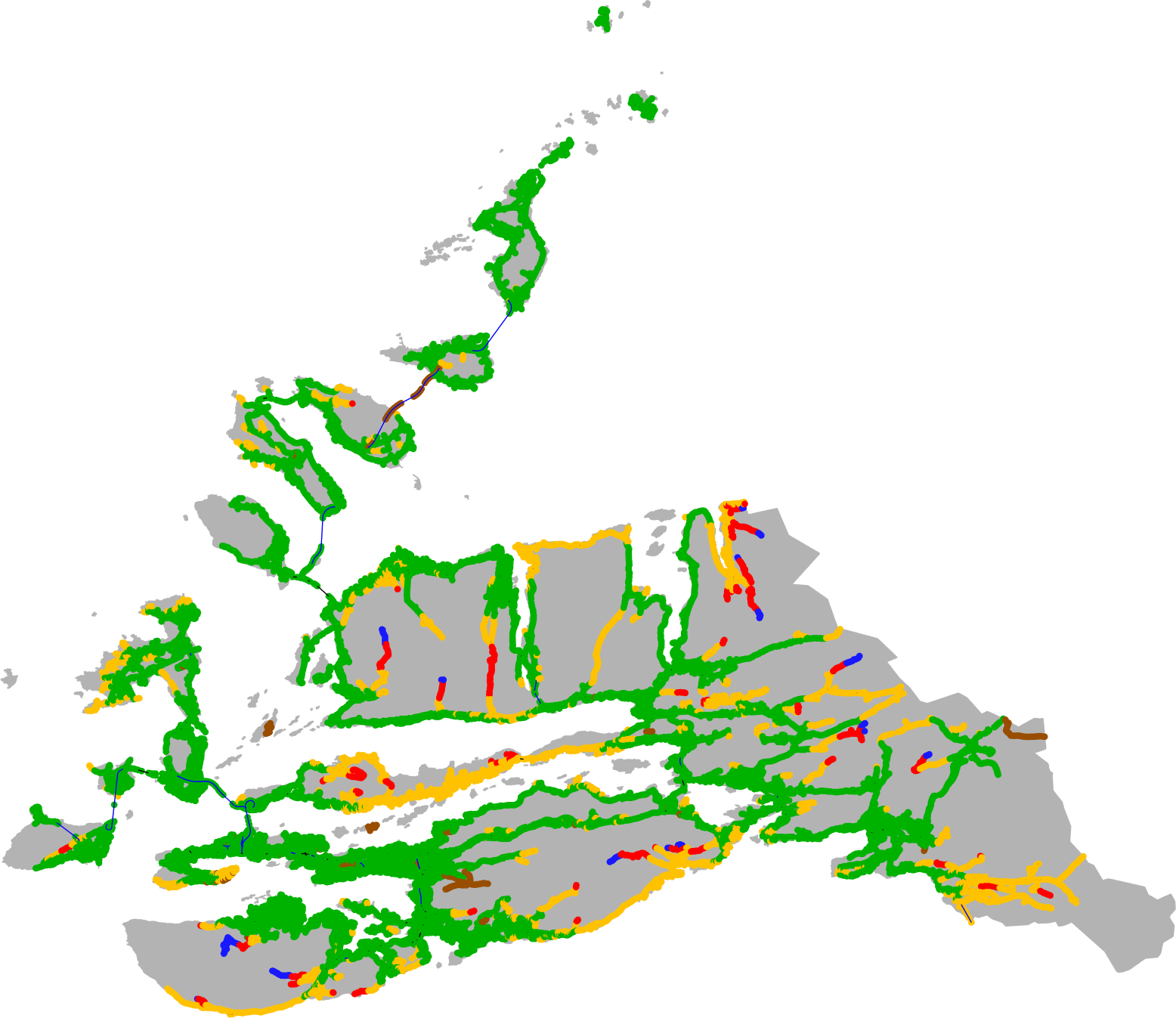}
   \hspace{10mm}
   2\hspace{-1em}\includegraphics[width=0.38\textwidth]{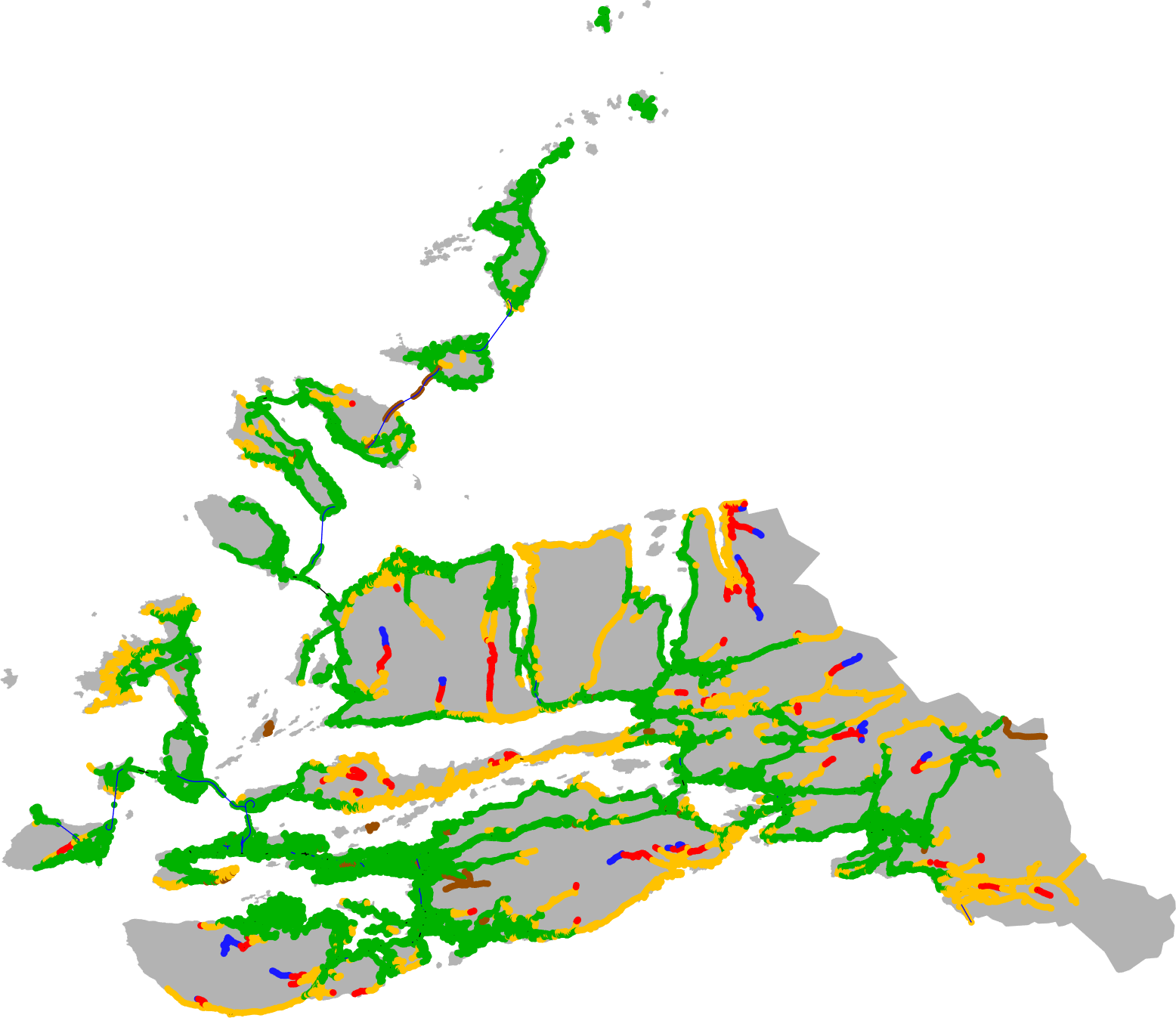} \\[-3mm]
   \hspace{15mm}3\hspace{-1em}\includegraphics[width=0.38\textwidth]{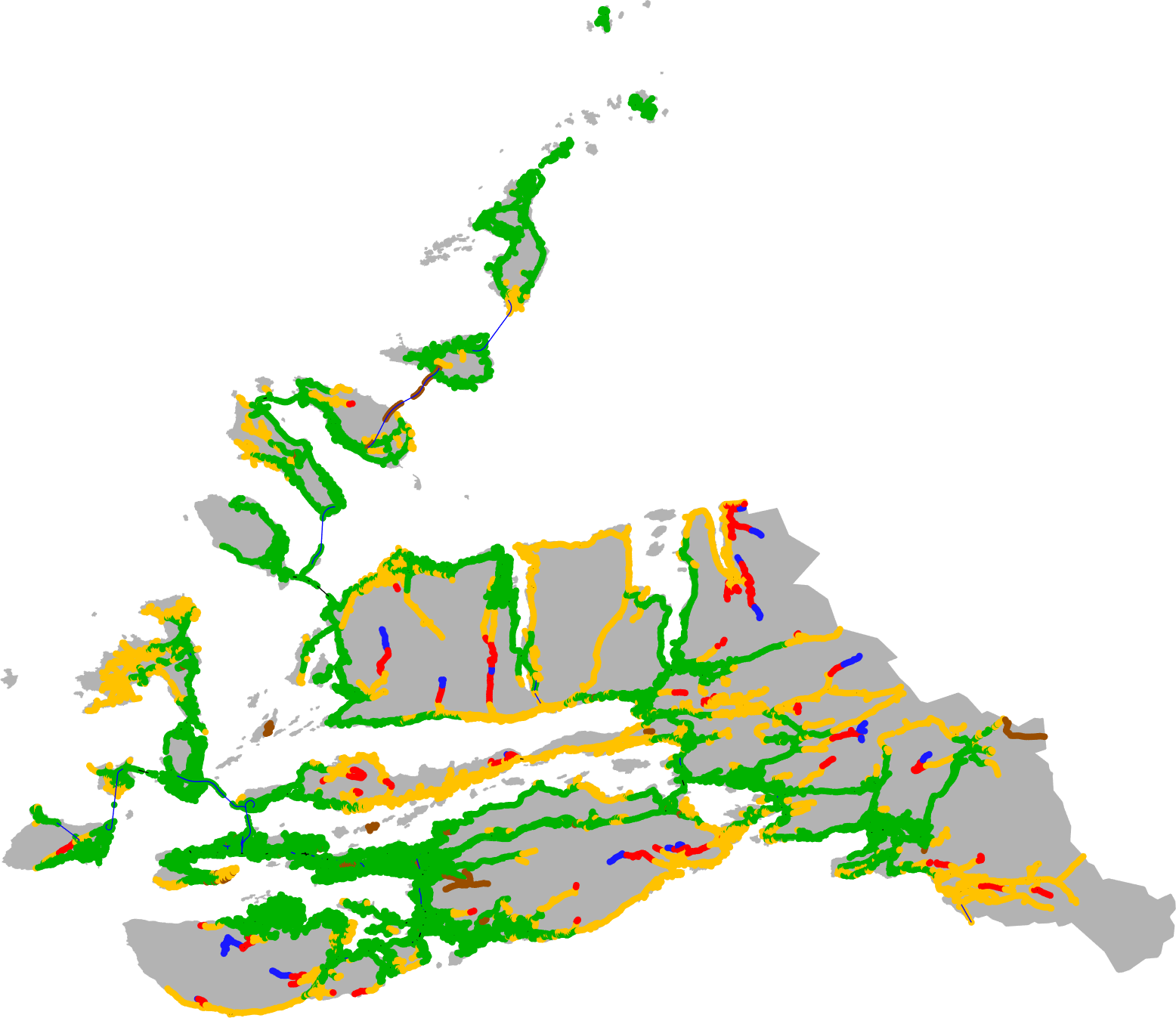}
   \hspace{10mm}
   4\hspace{-1em}\includegraphics[width=0.38\textwidth]{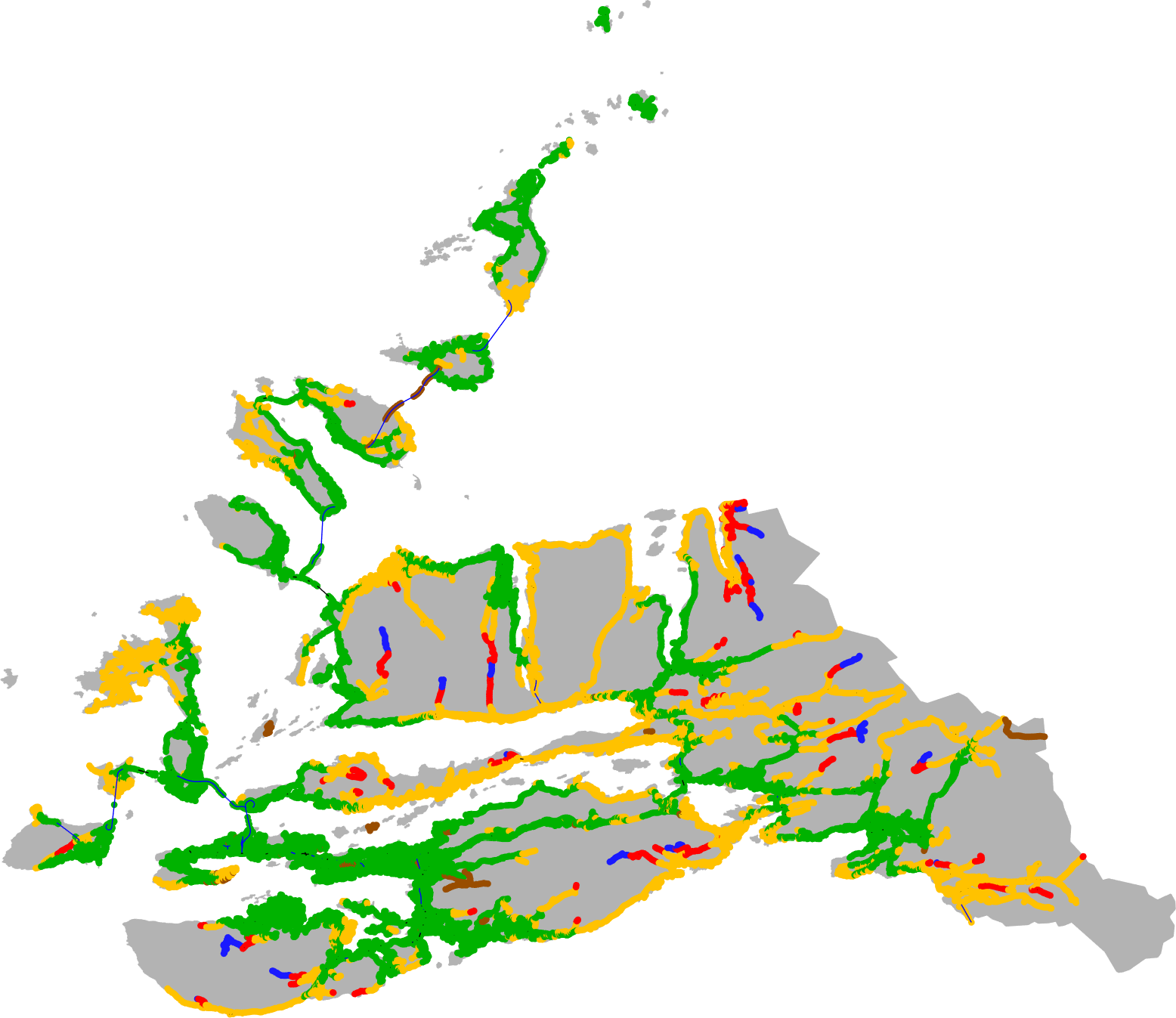} \\[-1mm]
   5\hspace{-1em}\includegraphics[width=0.38\textwidth]{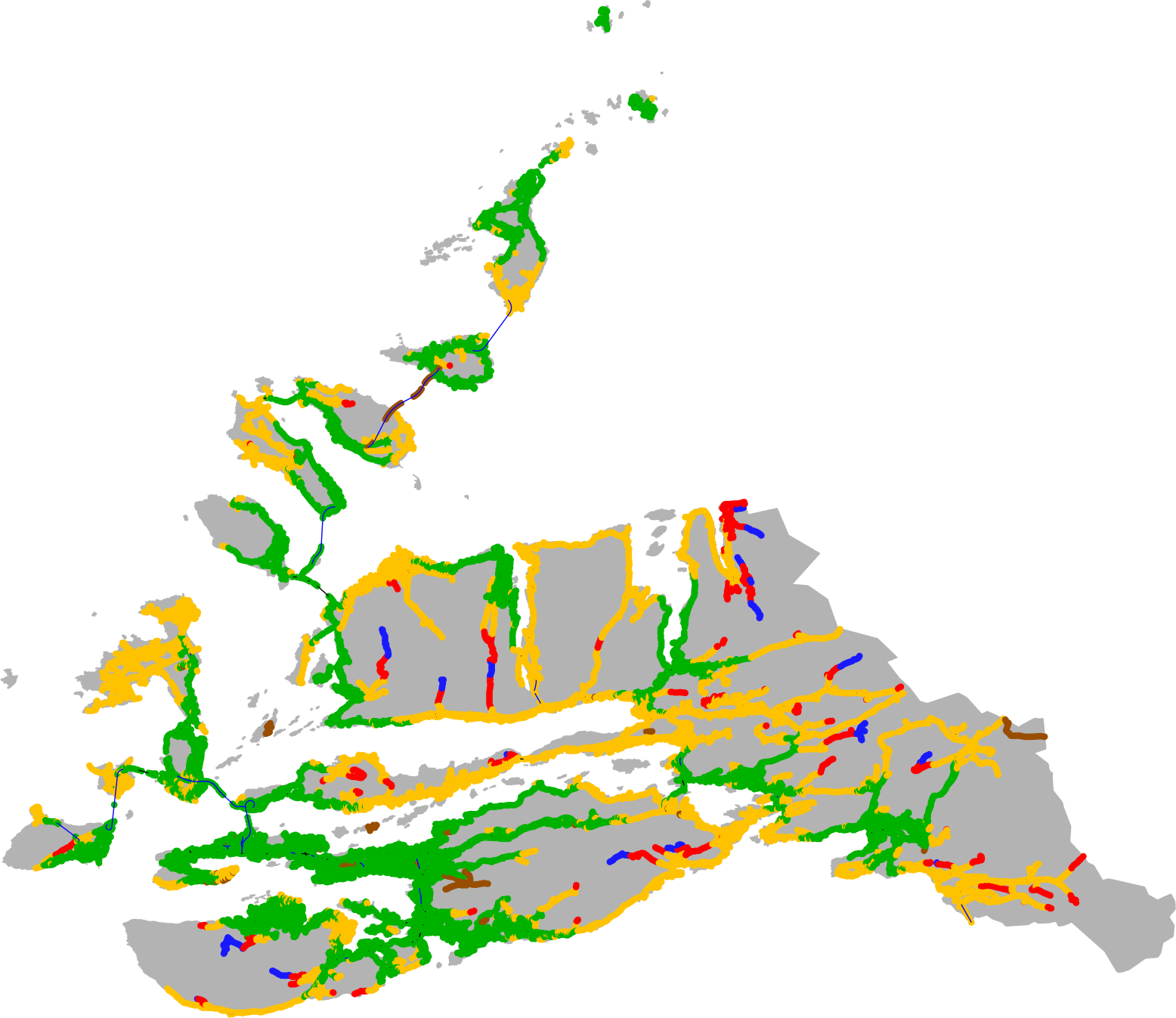} \\[-4mm]
   6\hspace{-1em}\includegraphics[width=0.38\textwidth]{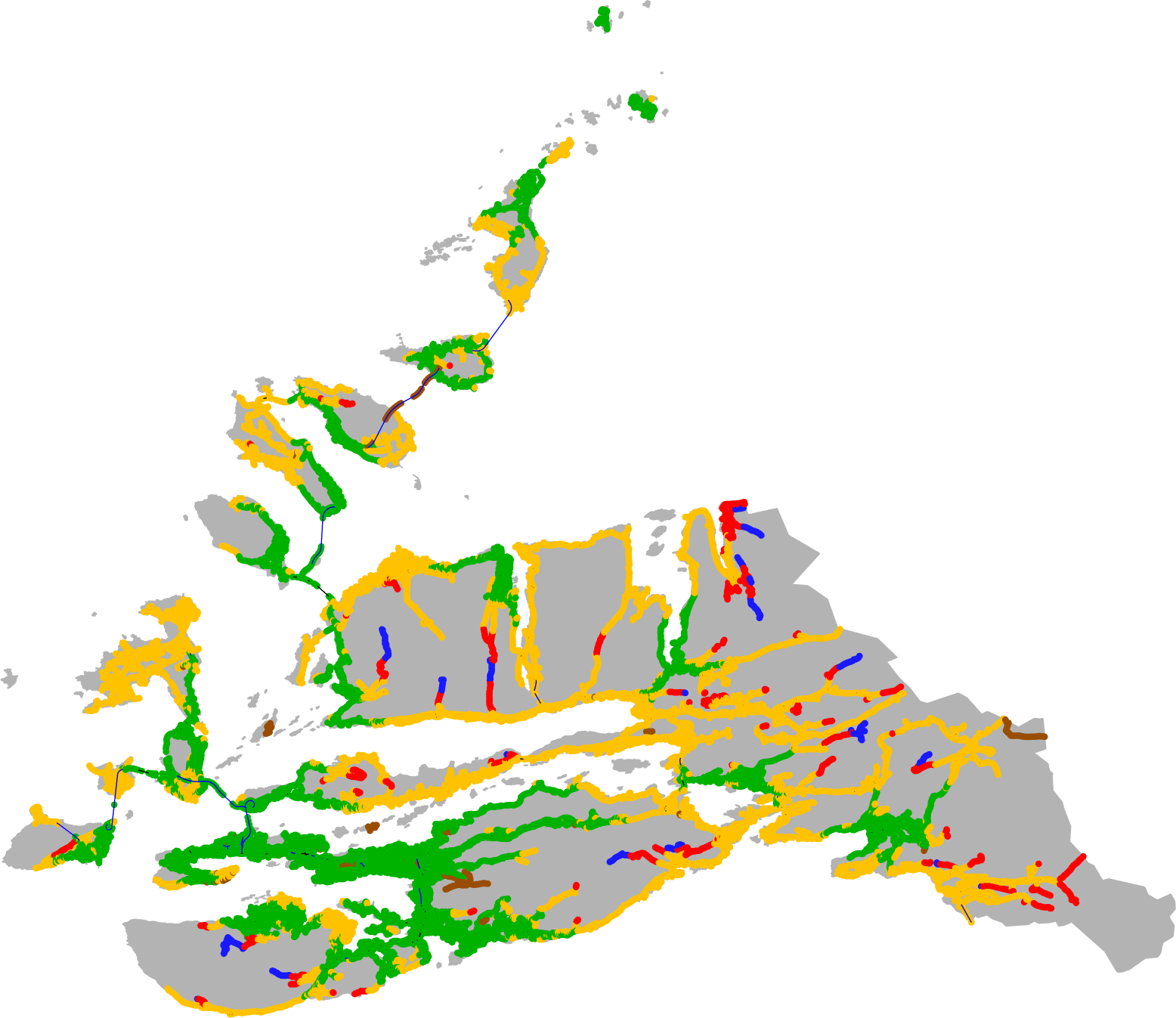}
   \hspace{10mm}
   7\hspace{-1em}\includegraphics[width=0.38\textwidth]{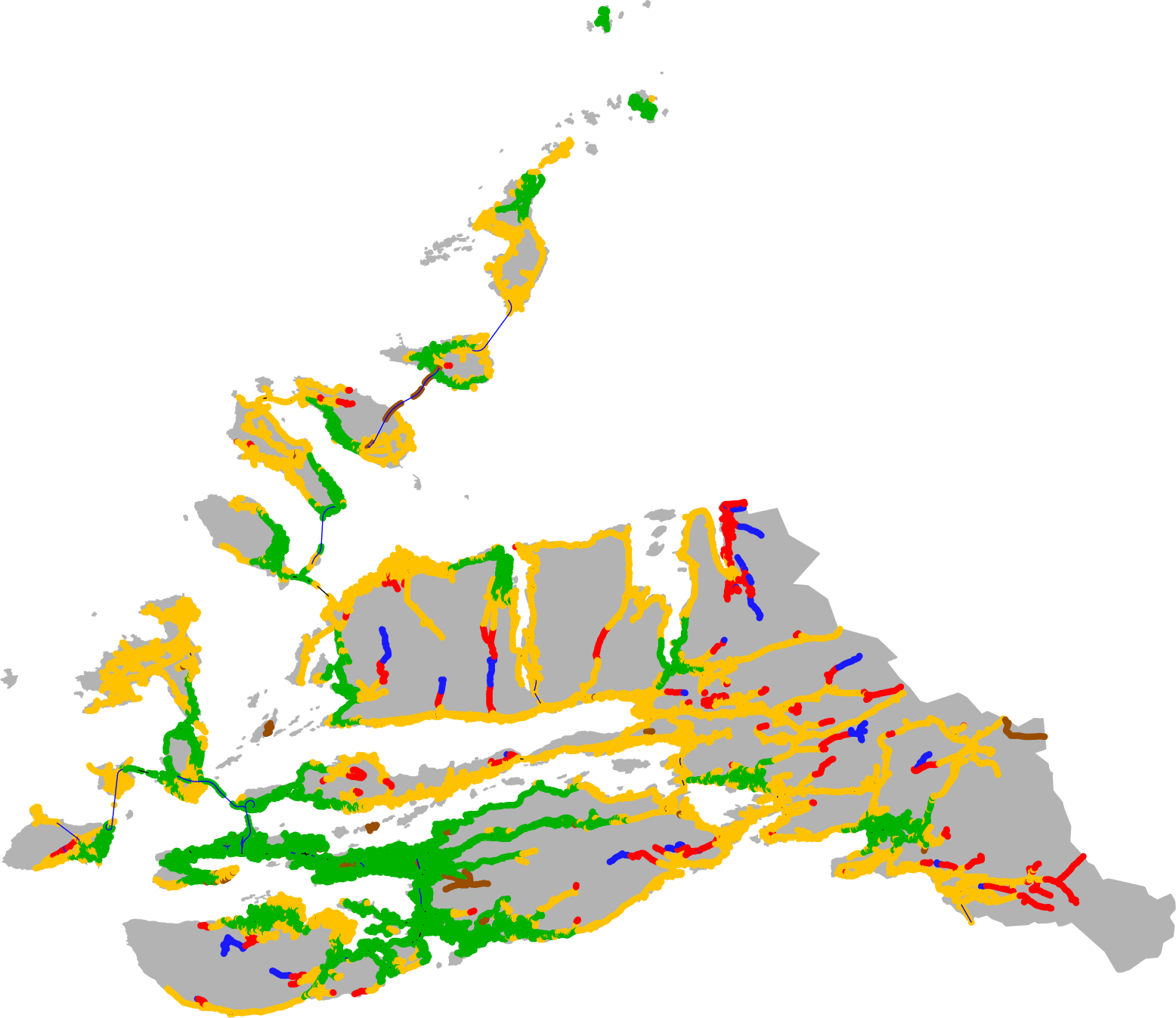} \\[-3mm]
   \hspace{15mm}8\hspace{-1em}\includegraphics[width=0.38\textwidth]{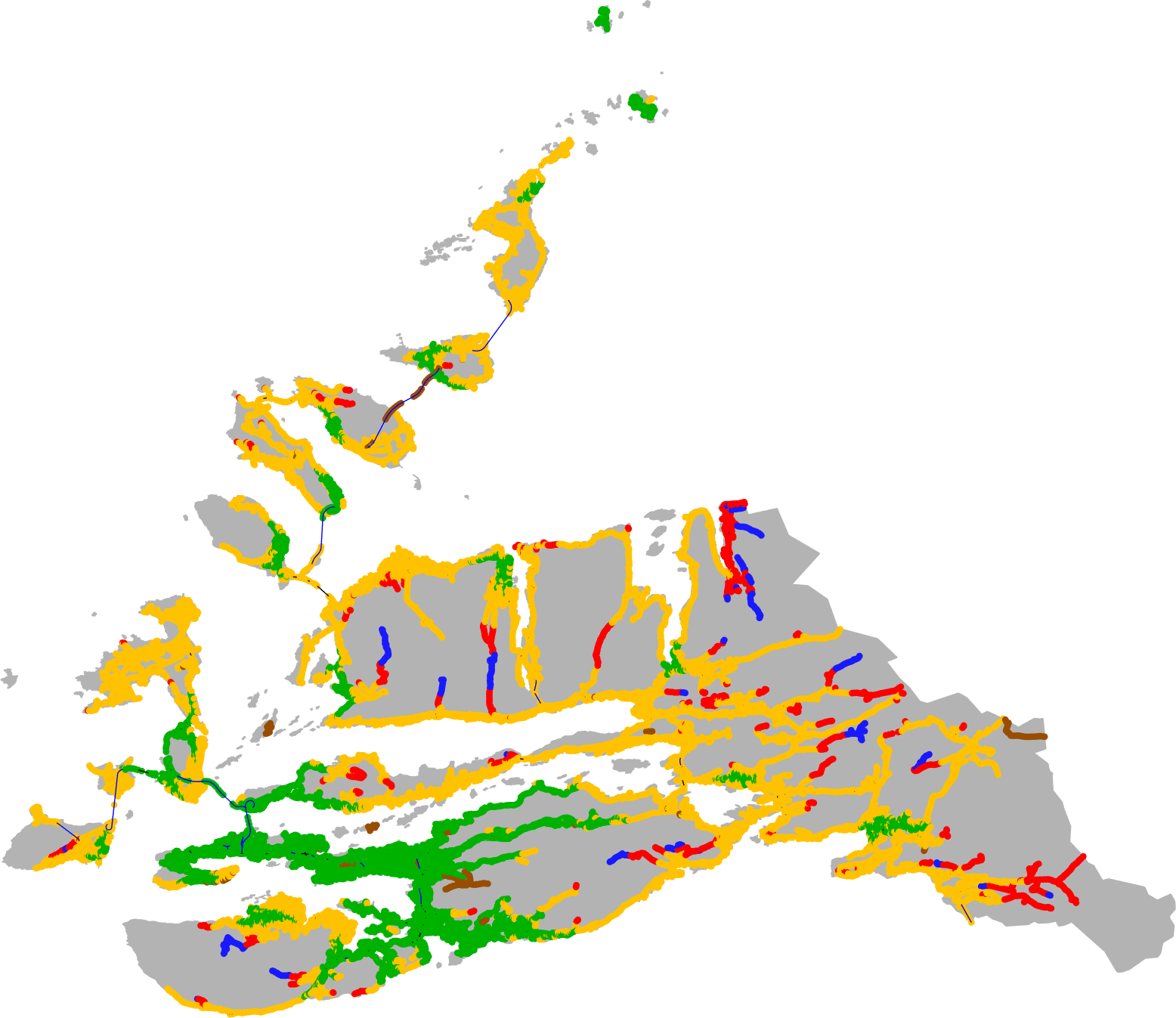}
   \hspace{10mm}
   9\hspace{-1em}\includegraphics[width=0.38\textwidth]{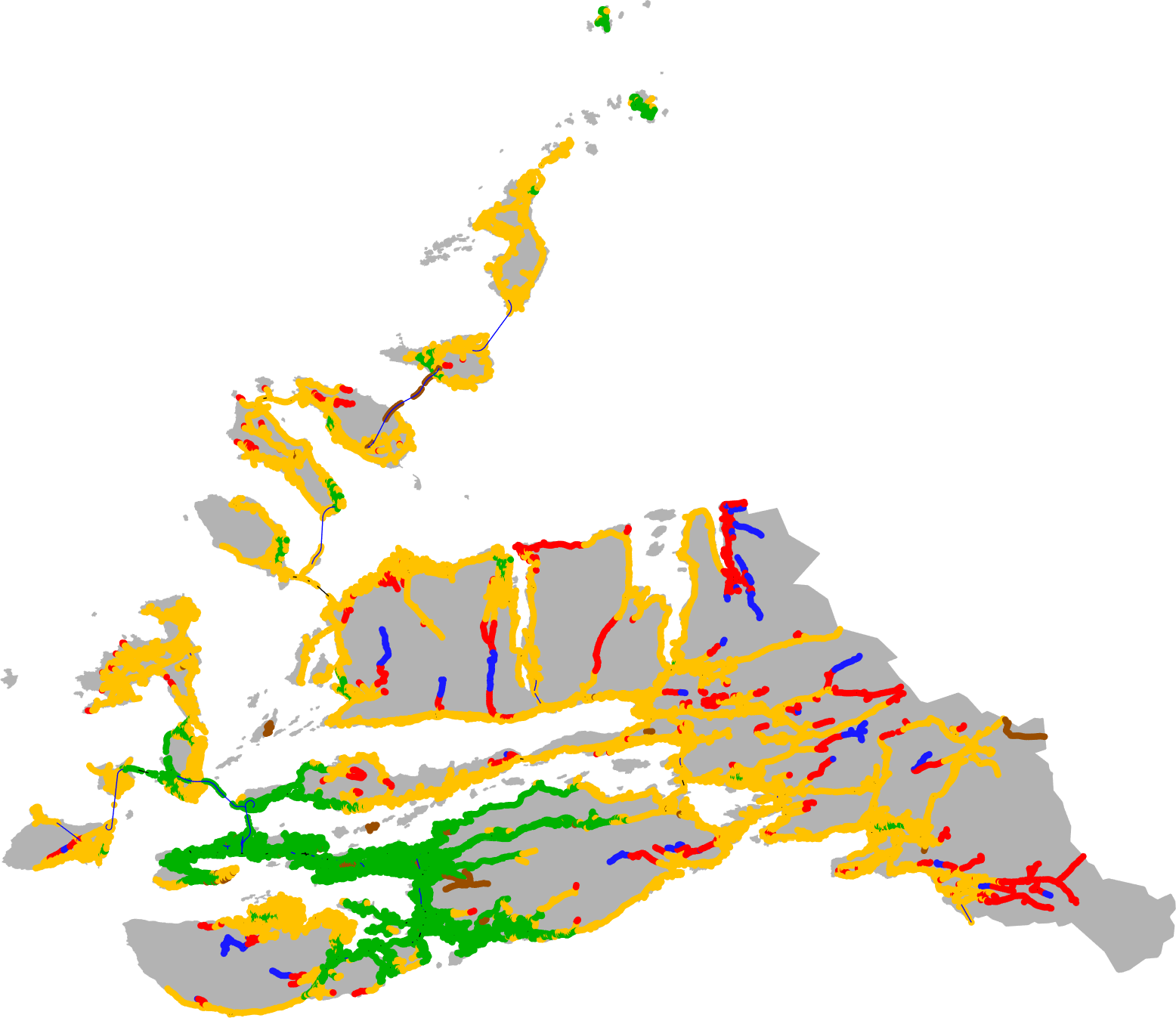}
   \caption{Heatmaps showing the effect on time of setting the part-time fire stations' call-out delay to 1 to 9 minutes, as labelled.
            The baseline model is 5 minutes.}
   \label{fig:part-time-callout-delays}
\end{figure}

\begin{figure}[p]
   \centering
   1\hspace{-1em}\includegraphics[width=0.38\textwidth]{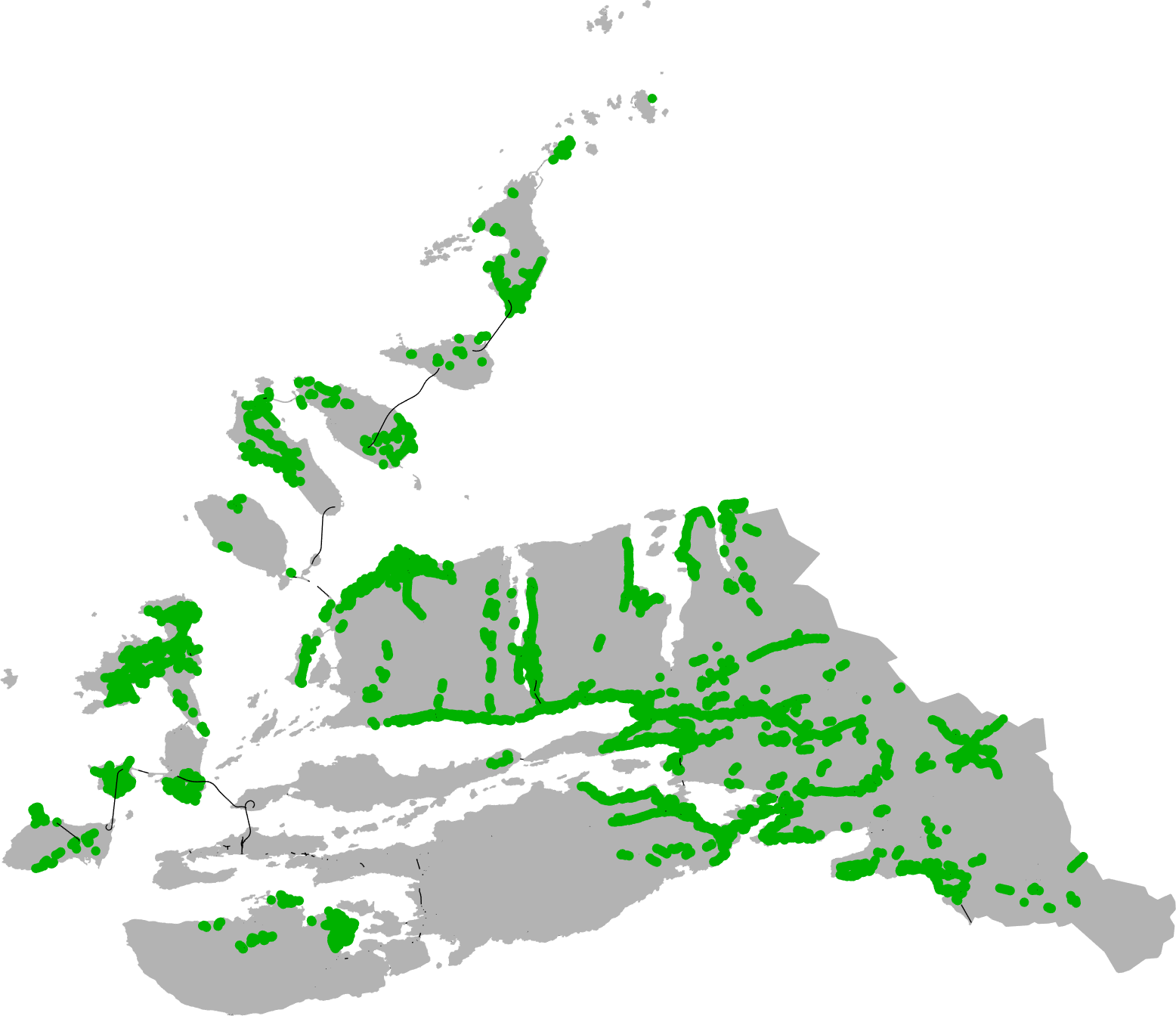}
   \hspace{10mm}
   2\hspace{-1em}\includegraphics[width=0.38\textwidth]{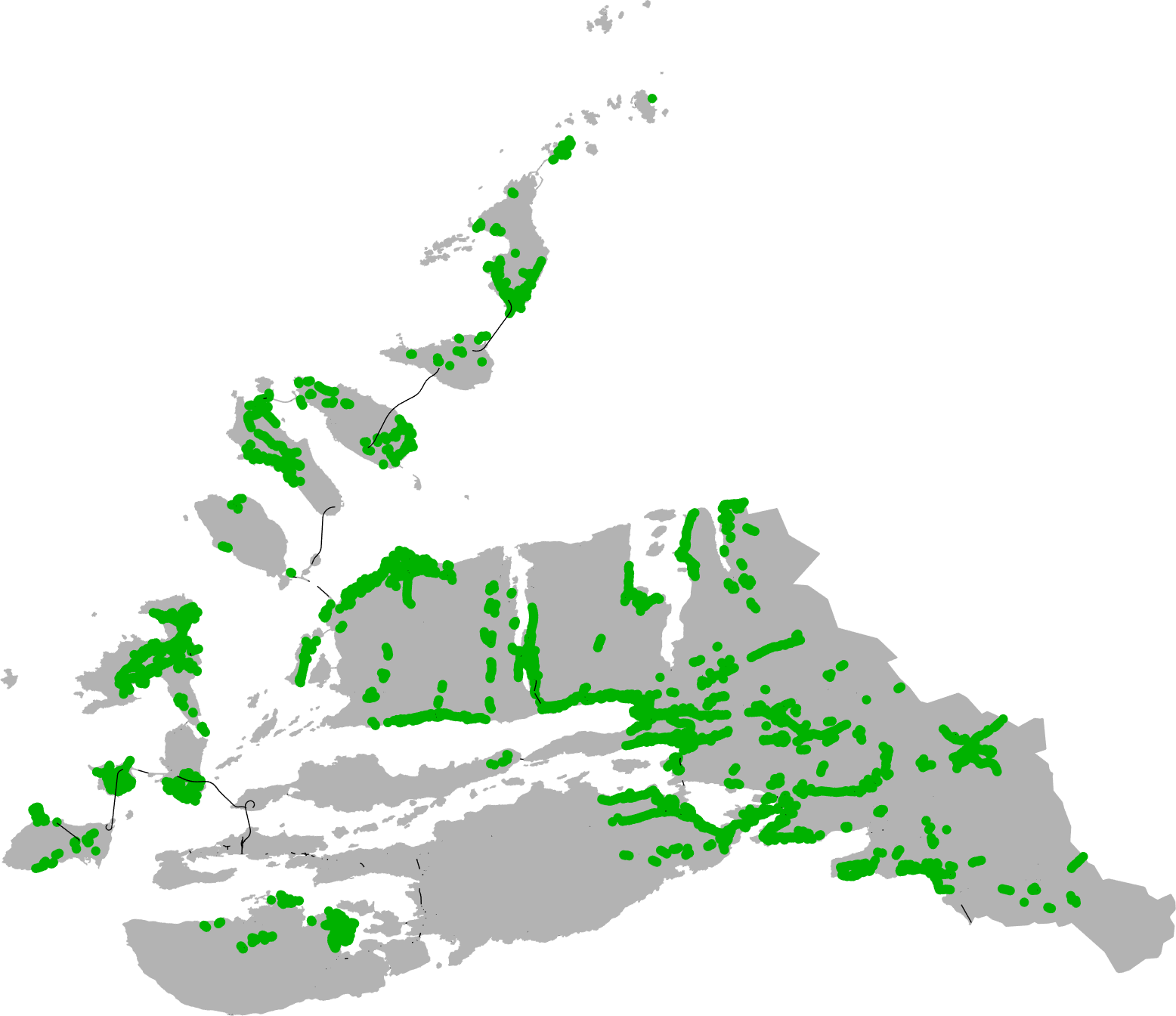} \\[-3mm]
   \hspace{15mm}3\hspace{-1em}\includegraphics[width=0.38\textwidth]{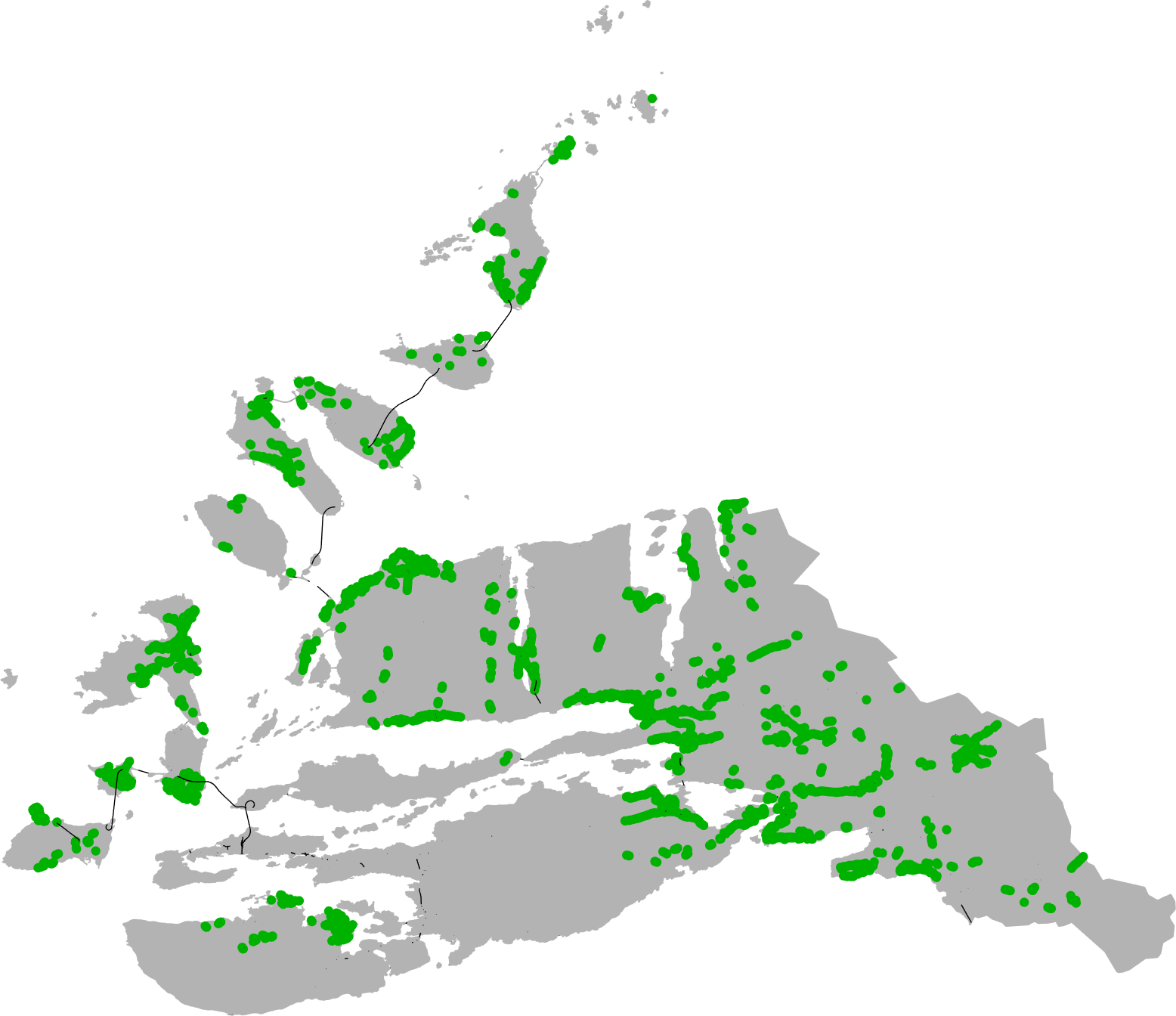}
   \hspace{10mm}
   4\hspace{-1em}\includegraphics[width=0.38\textwidth]{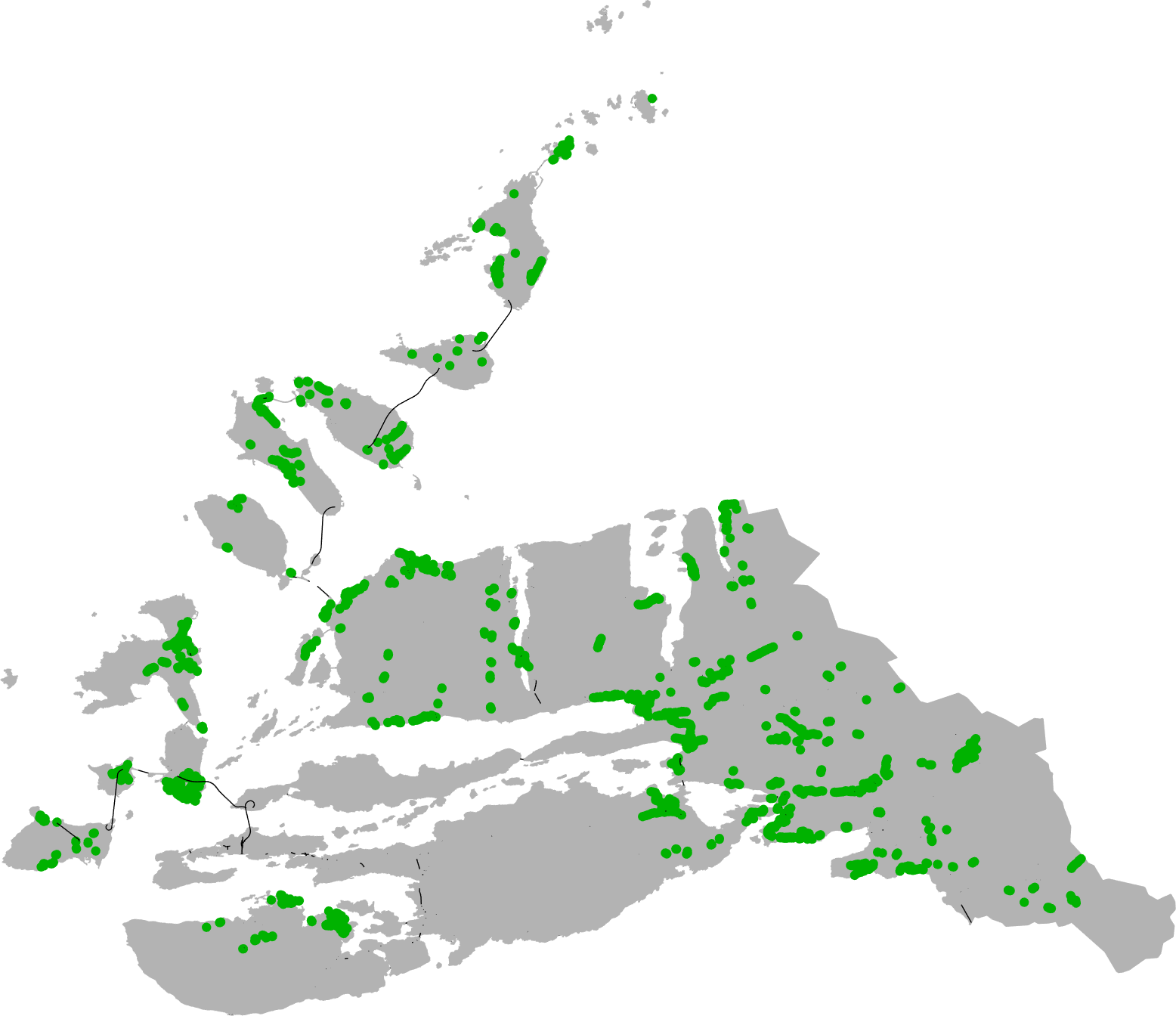} \\[-1mm]
   5\hspace{-1em}\includegraphics[width=0.38\textwidth]{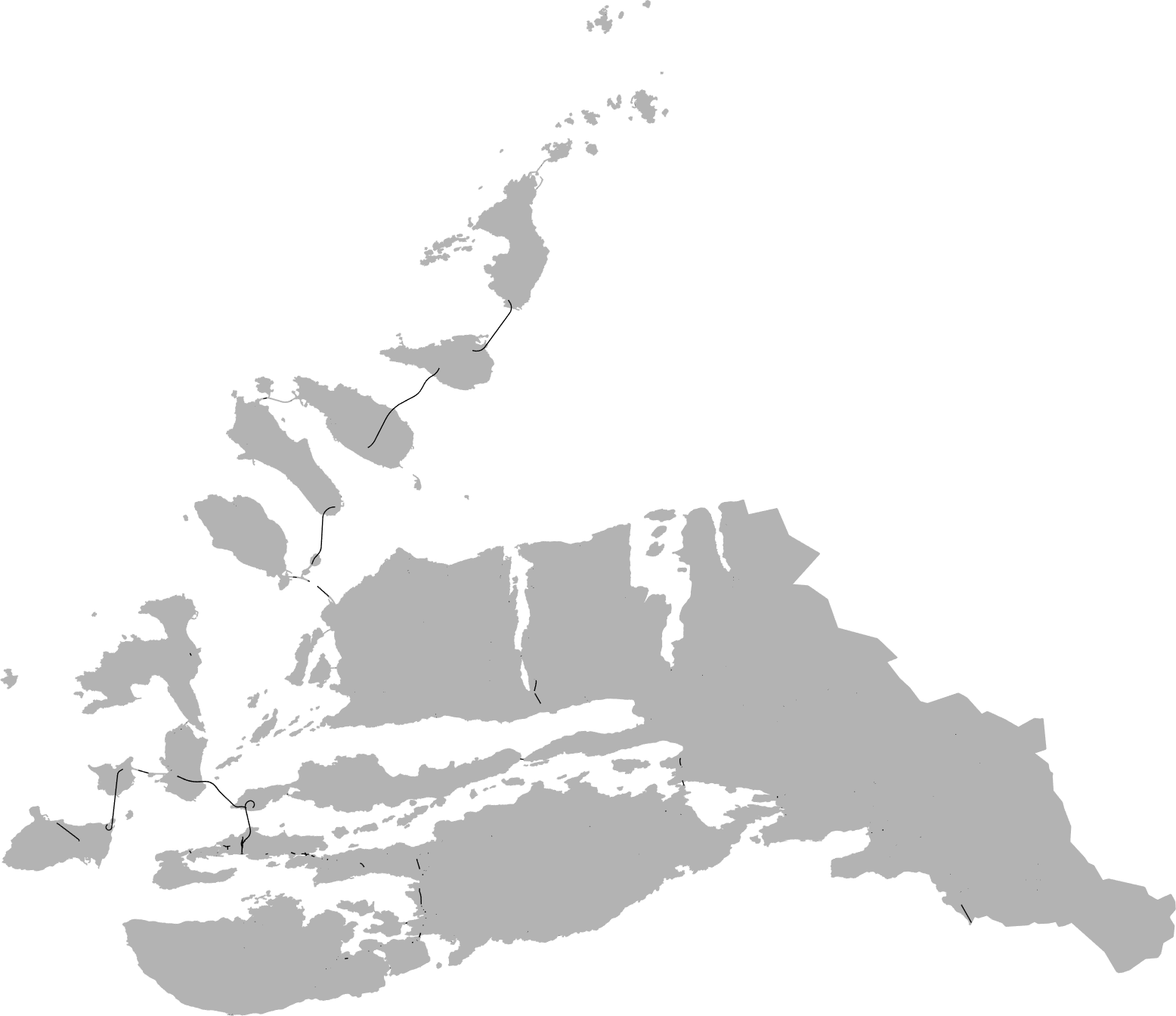} \\[-4mm]
   6\hspace{-1em}\includegraphics[width=0.38\textwidth]{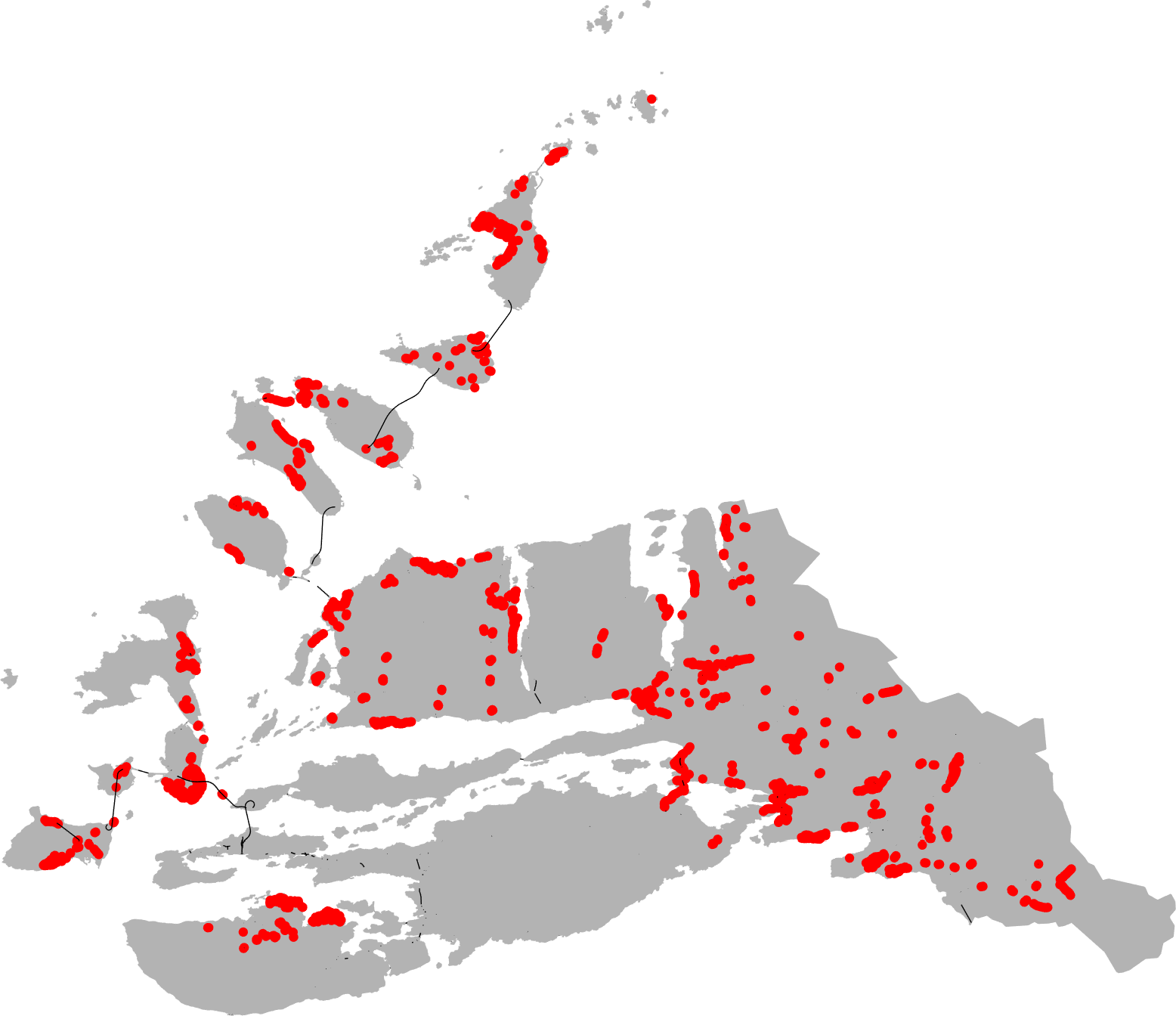}
   \hspace{10mm}
   7\hspace{-1em}\includegraphics[width=0.38\textwidth]{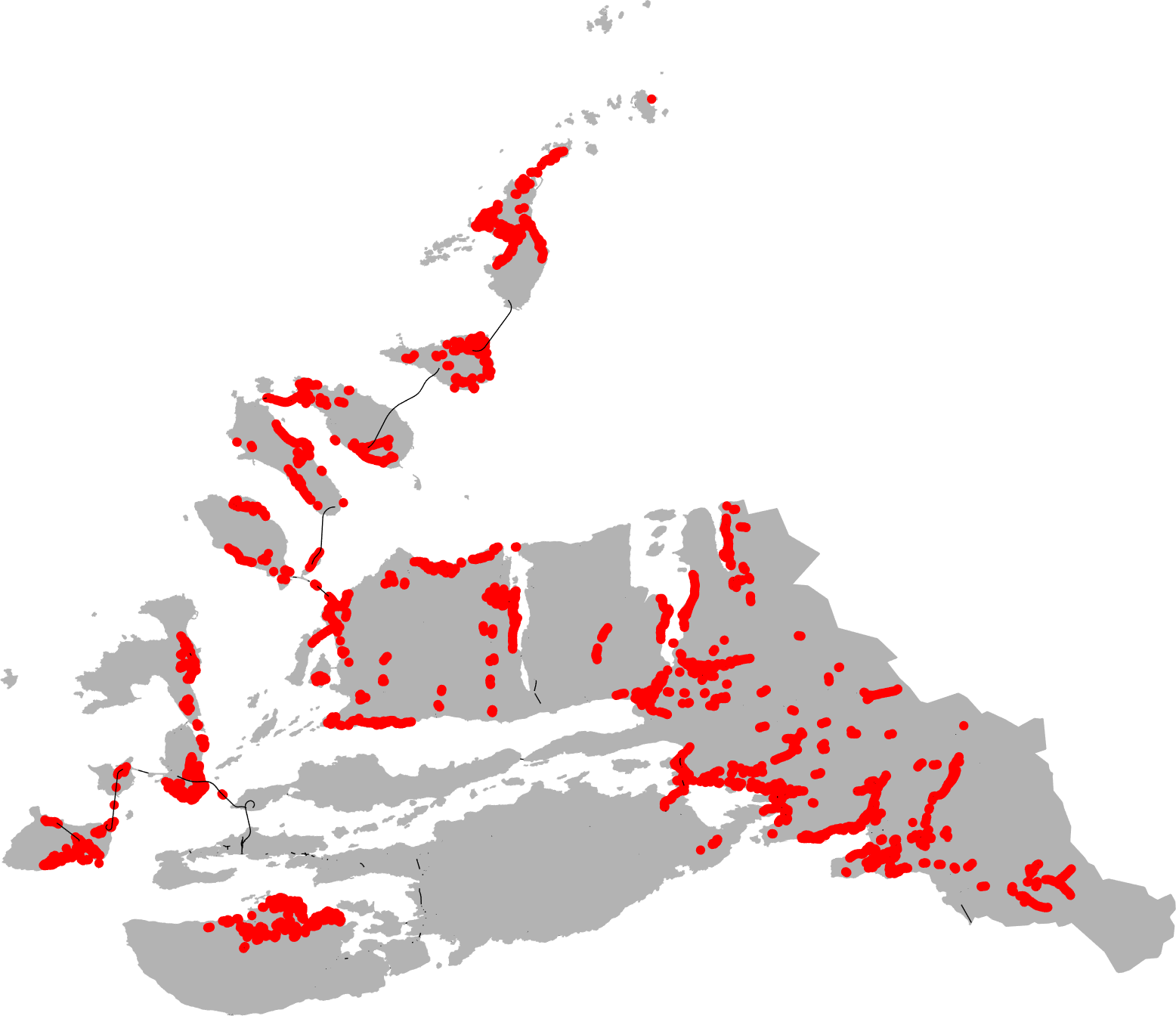} \\[-3mm]
   \hspace{15mm}8\hspace{-1em}\includegraphics[width=0.38\textwidth]{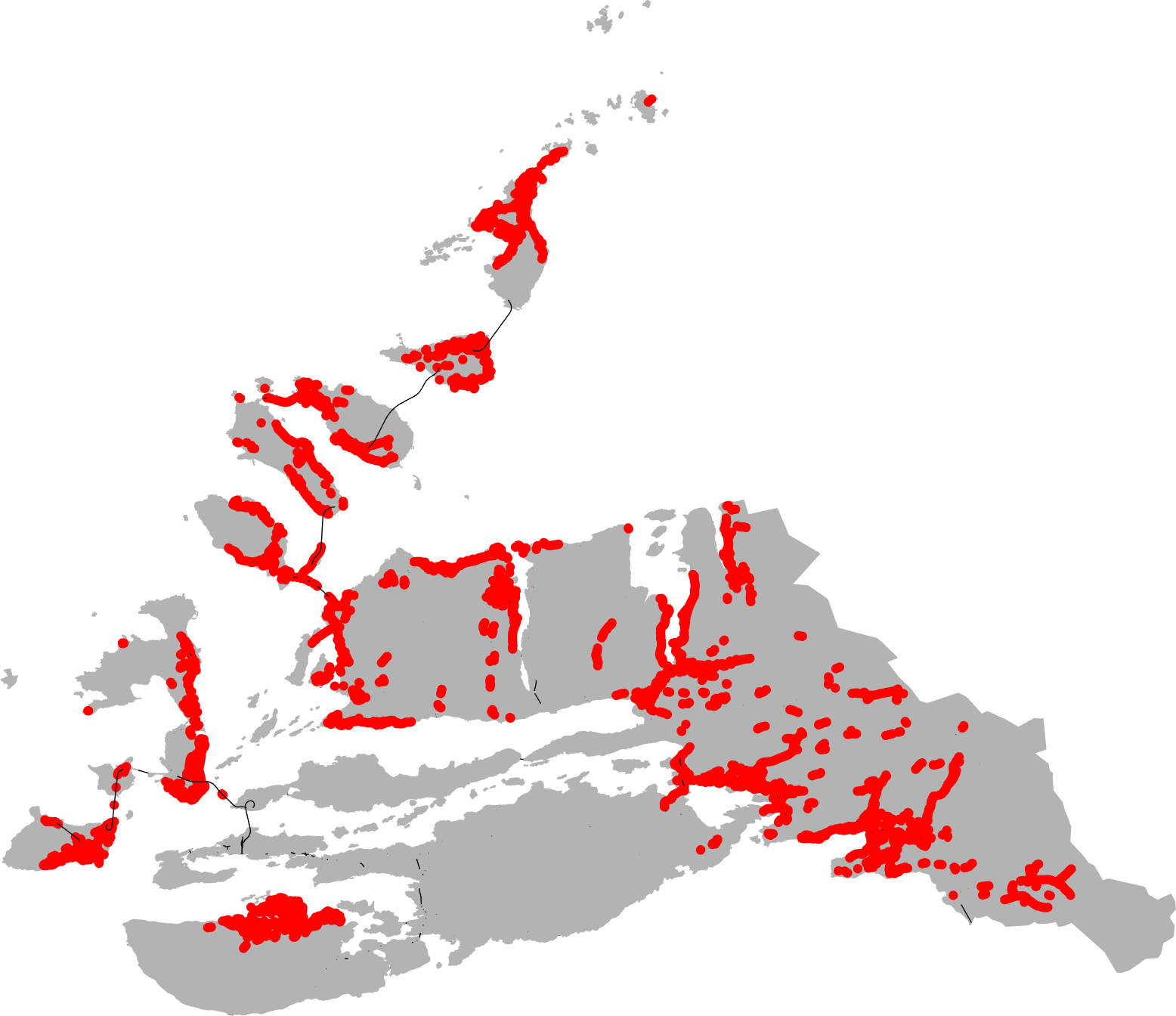}
   \hspace{10mm}
   9\hspace{-1em}\includegraphics[width=0.38\textwidth]{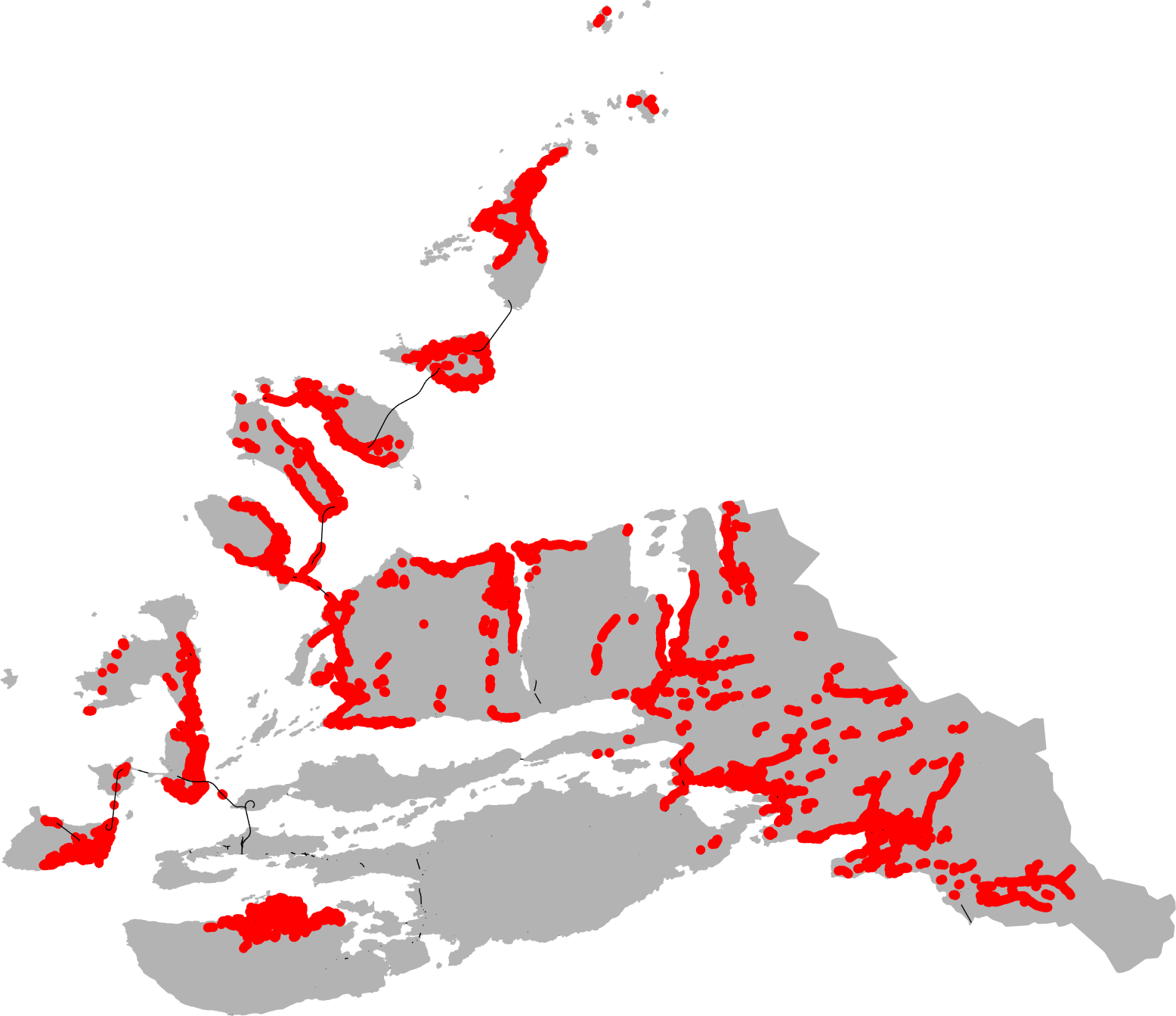}
   \caption{The difference maps associated with the heatmaps shown in figure \ref{fig:part-time-callout-delays}.}
   \label{fig:part-time-callout-diffs}
\end{figure}

\begin{figure}[p]
   \centering
   0\hspace{-1em}\includegraphics[width=0.38\textwidth]{fig_station_speedTime_scenario0}
   \hfill
   1\hspace{-1em}\includegraphics[width=0.38\textwidth]{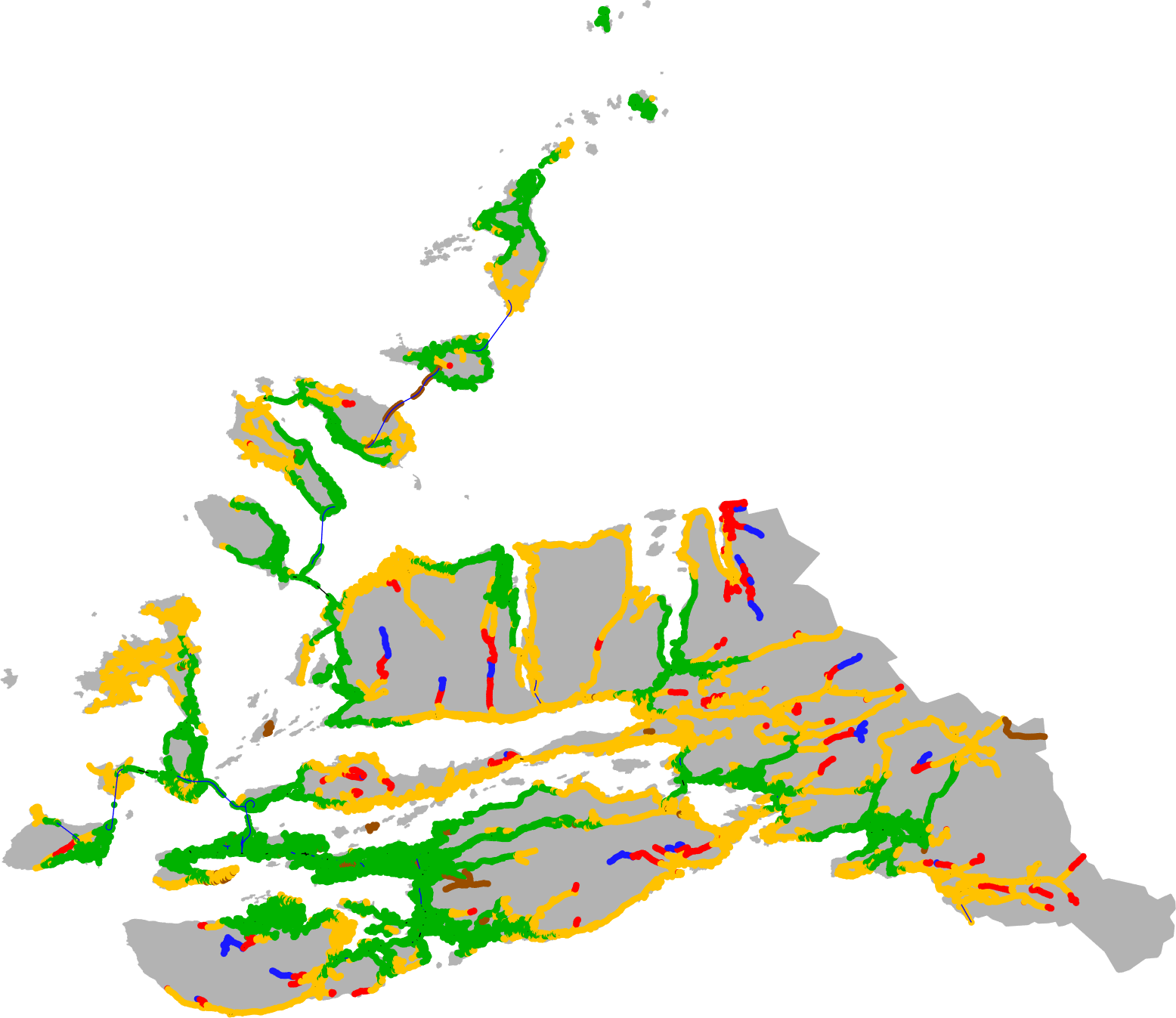} \\
   2\hspace{-1em}\includegraphics[width=0.38\textwidth]{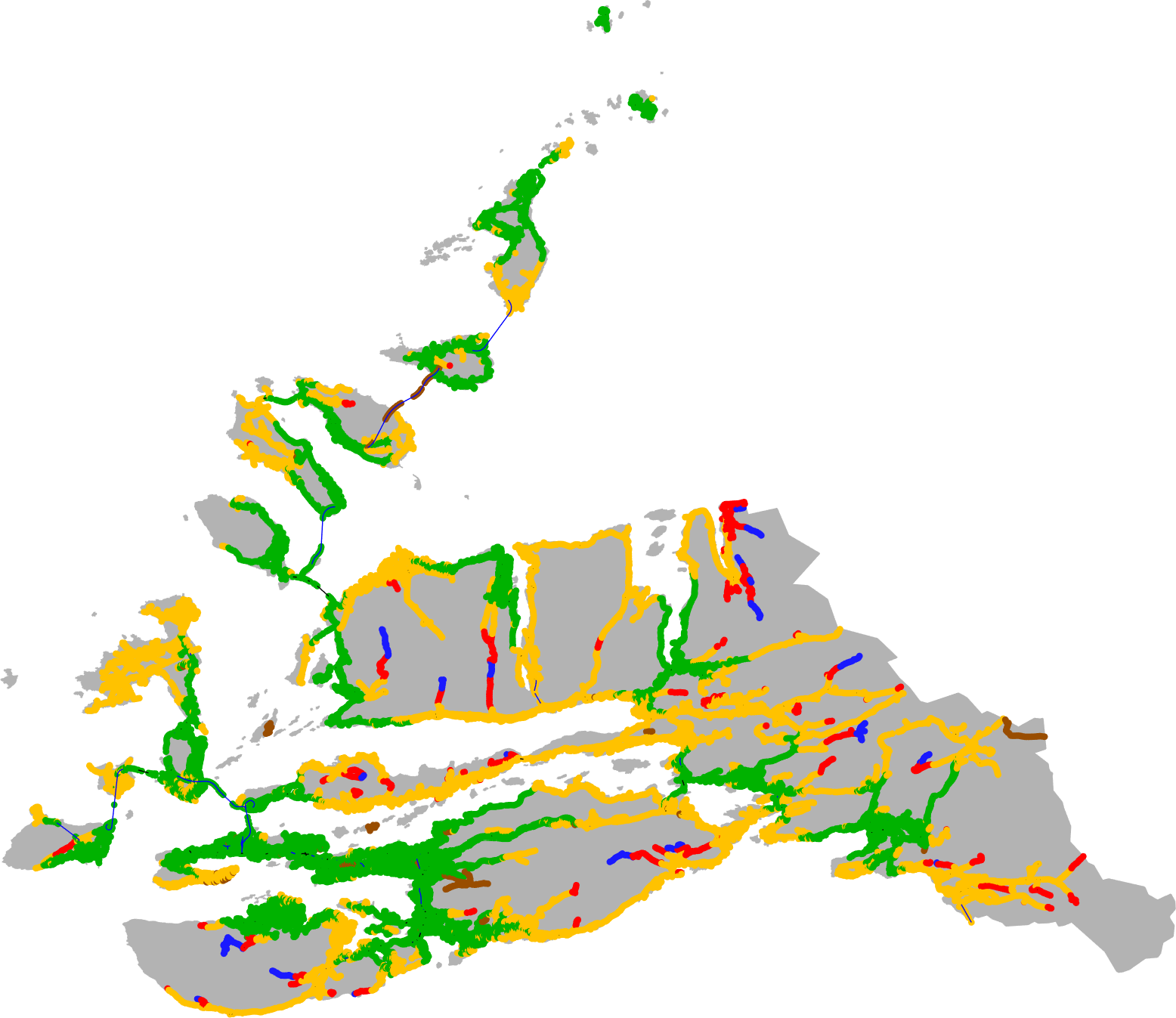}
   \hfill
   3\hspace{-1em}\includegraphics[width=0.38\textwidth]{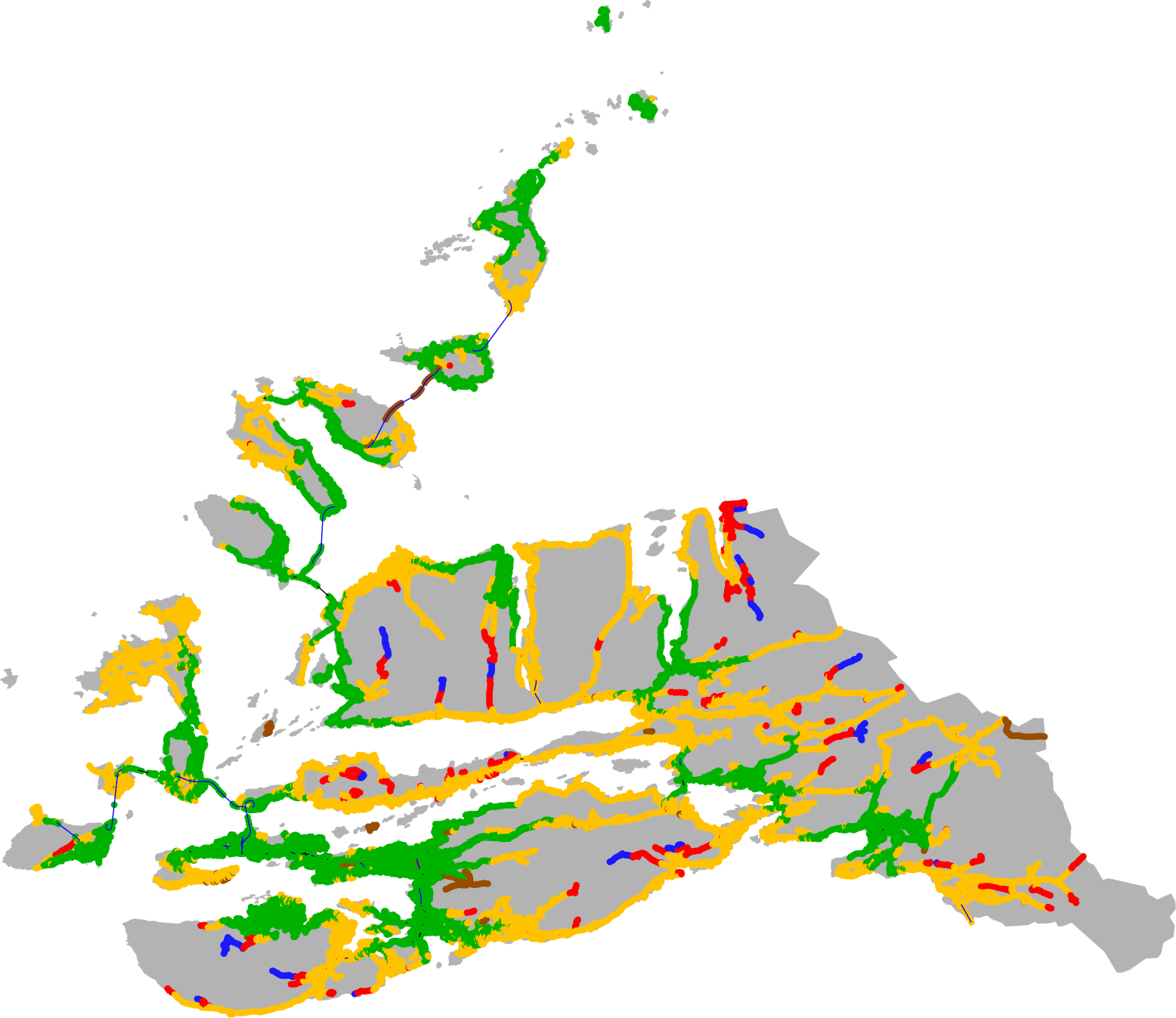} \\
   4\hspace{-1em}\includegraphics[width=0.38\textwidth]{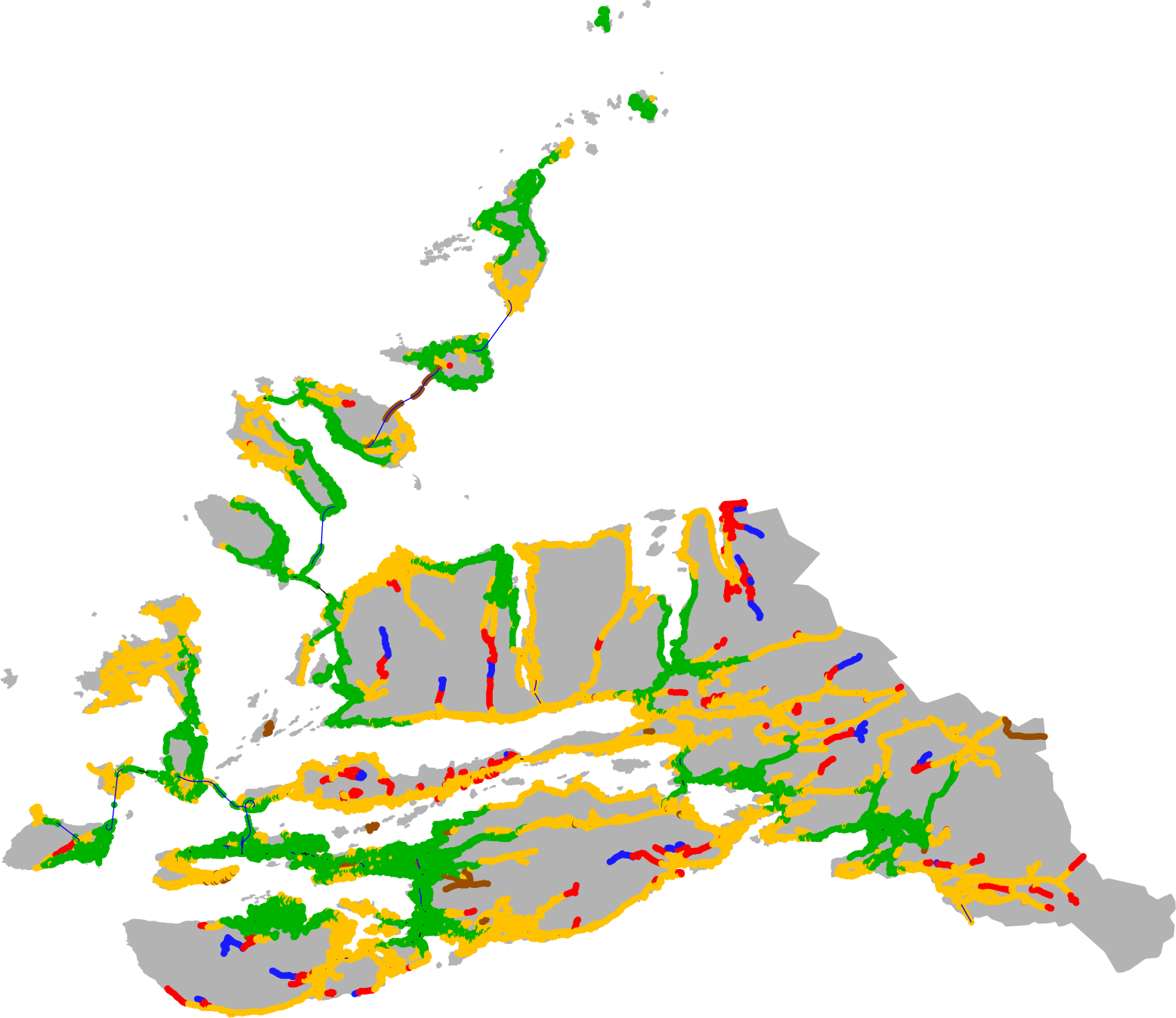}
   \hfill
   5\hspace{-1em}\includegraphics[width=0.38\textwidth]{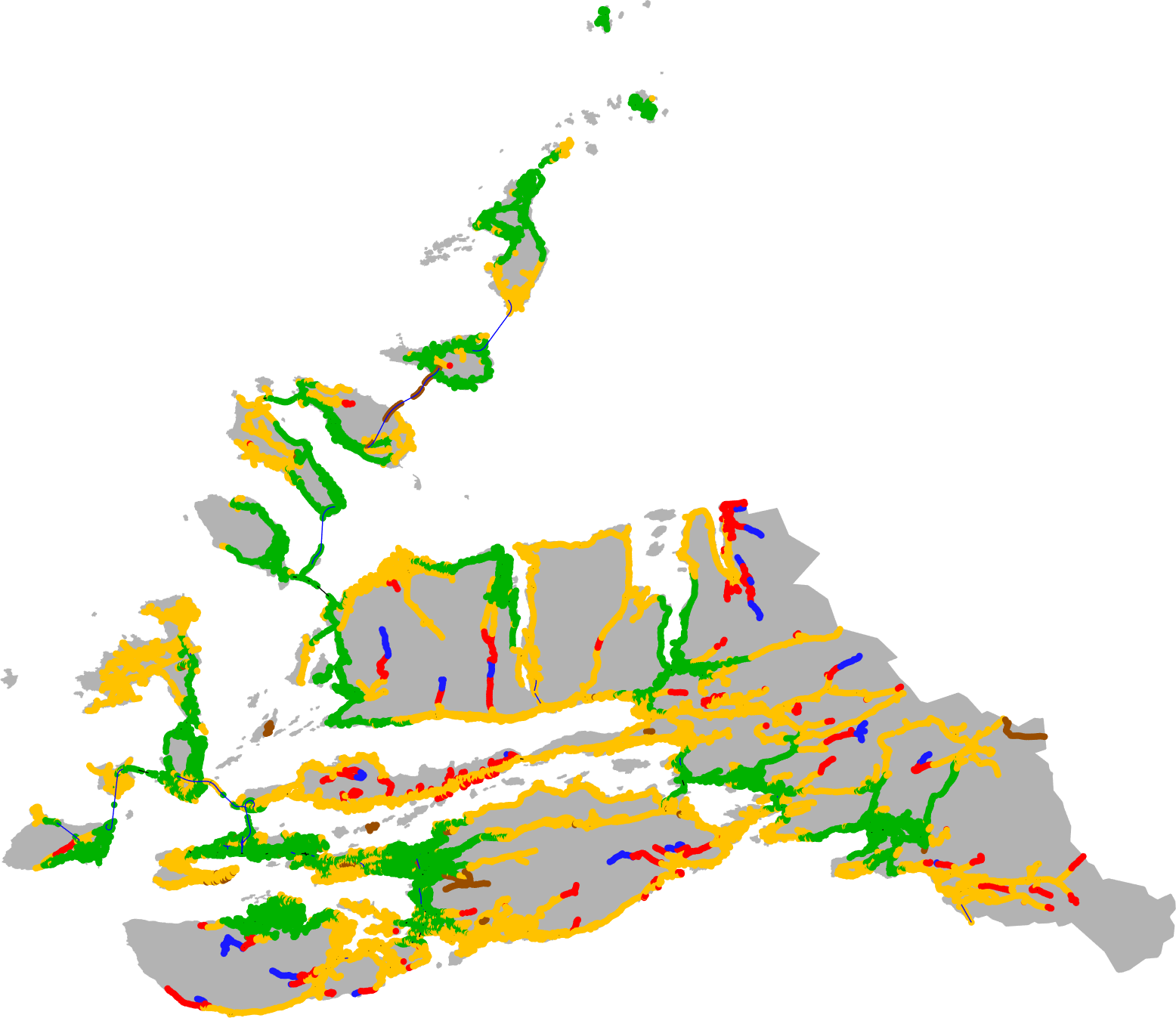}
   \caption{The effect on time of setting the full-time fire stations' call-out delay to 0 to 5 minutes, as labelled.
            The baseline model is 0 minutes.}
   \label{fig:full-time-callout-delays}
\end{figure}

\begin{figure}[p]
   \centering
   0\hspace{-1em}\includegraphics[width=0.38\textwidth]{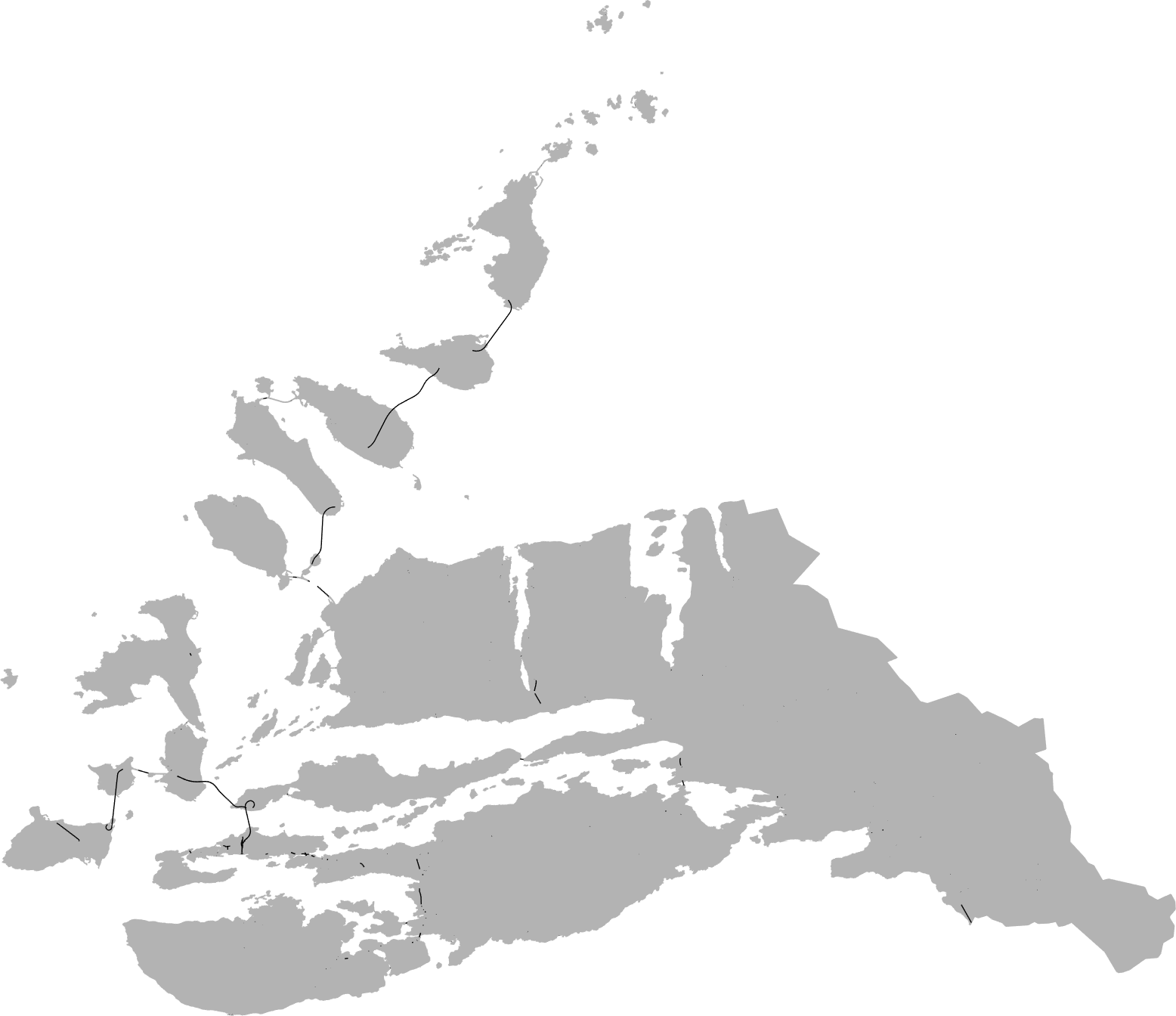}
   \hfill
   1\hspace{-1em}\includegraphics[width=0.38\textwidth]{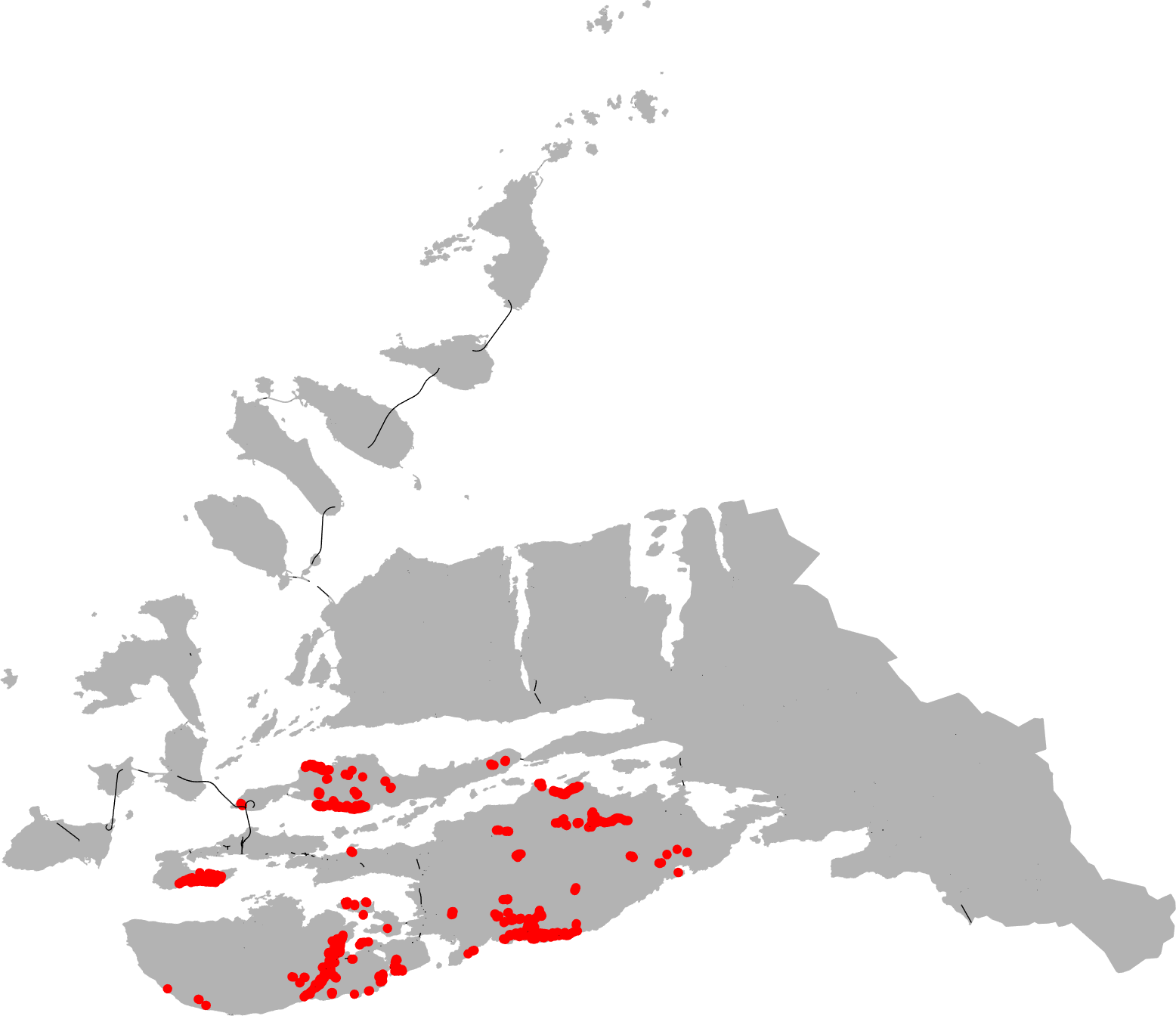} \\
   2\hspace{-1em}\includegraphics[width=0.38\textwidth]{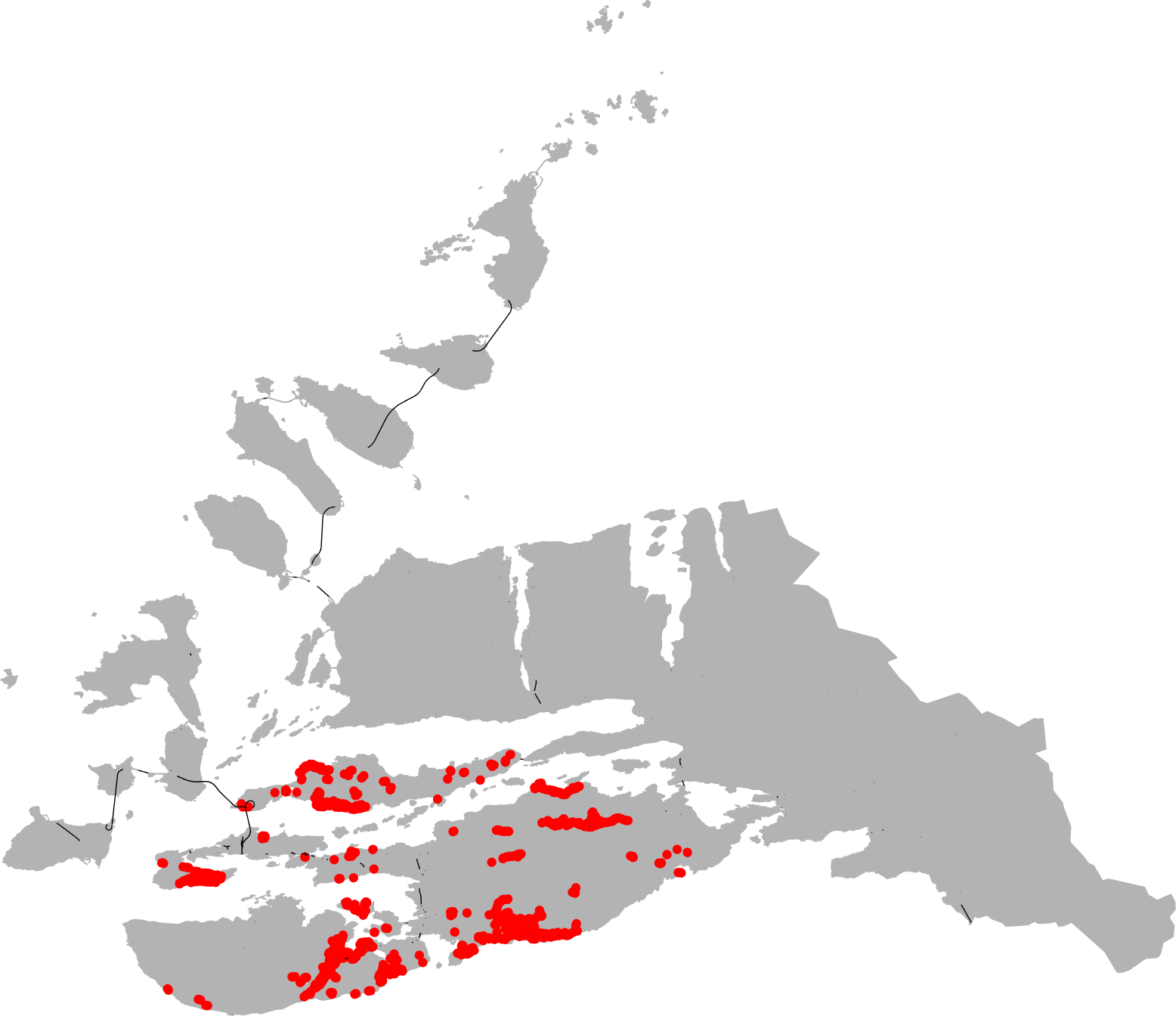}
   \hfill
   3\hspace{-1em}\includegraphics[width=0.38\textwidth]{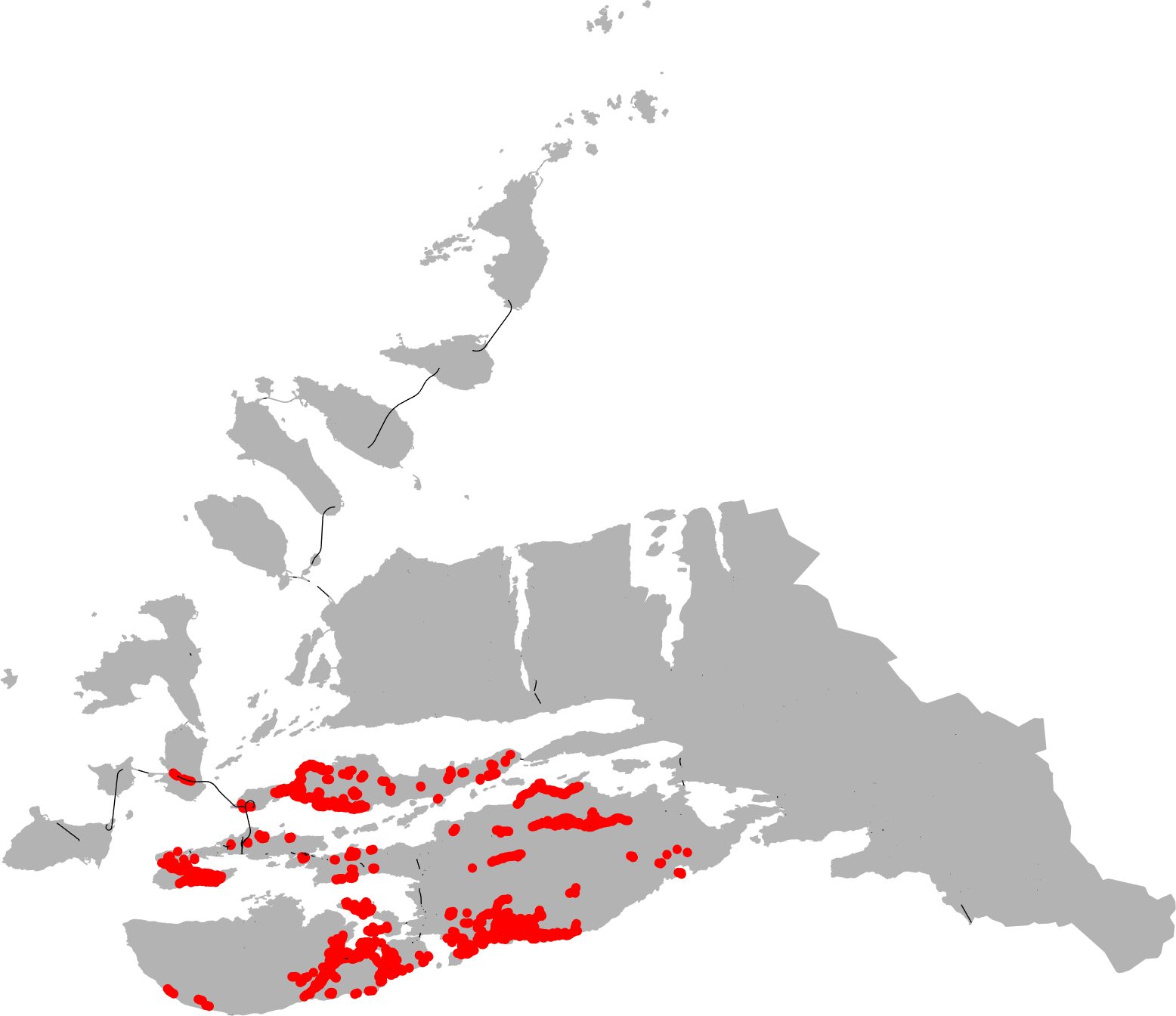} \\
   4\hspace{-1em}\includegraphics[width=0.38\textwidth]{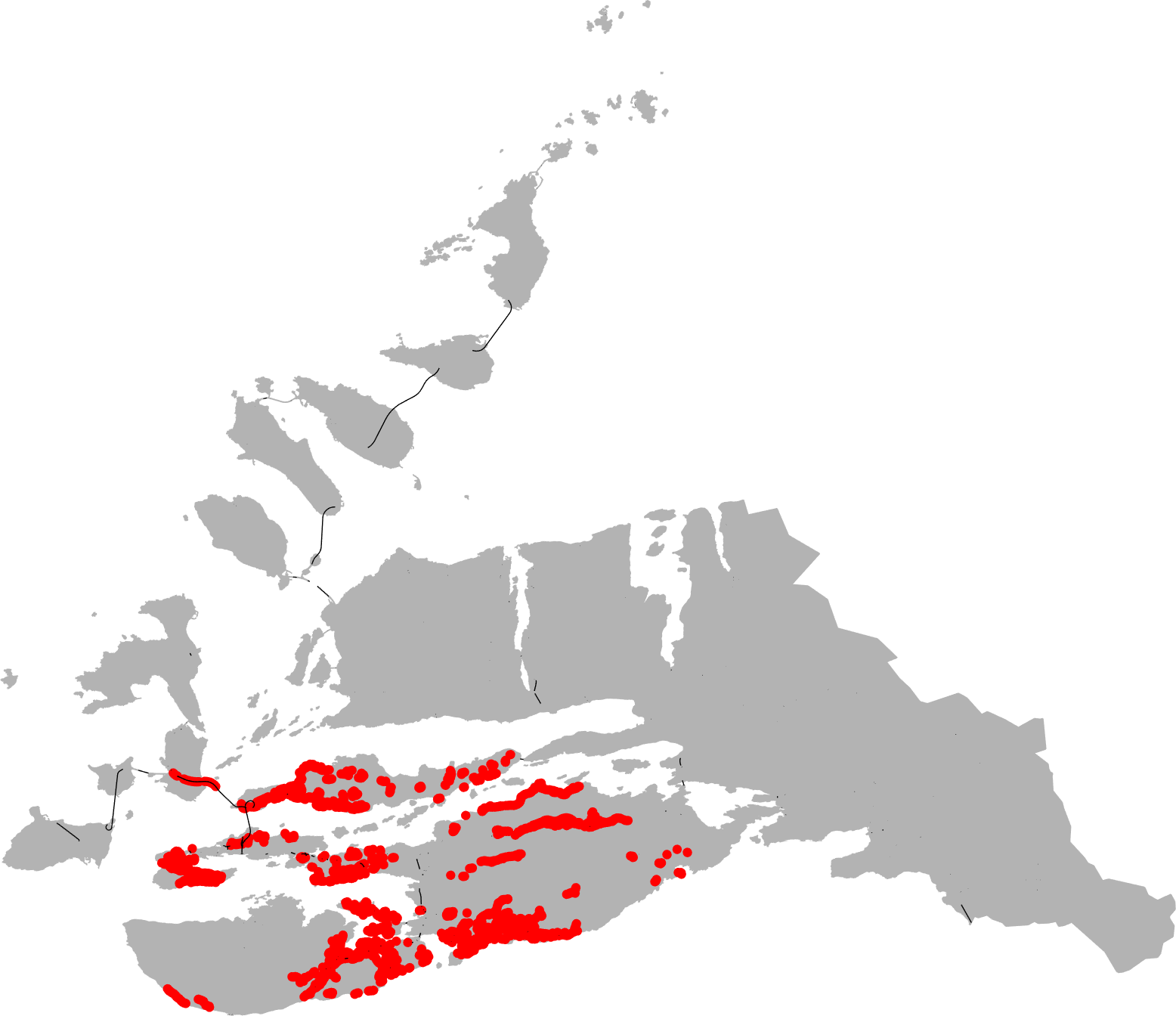}
   \hfill
   5\hspace{-1em}\includegraphics[width=0.38\textwidth]{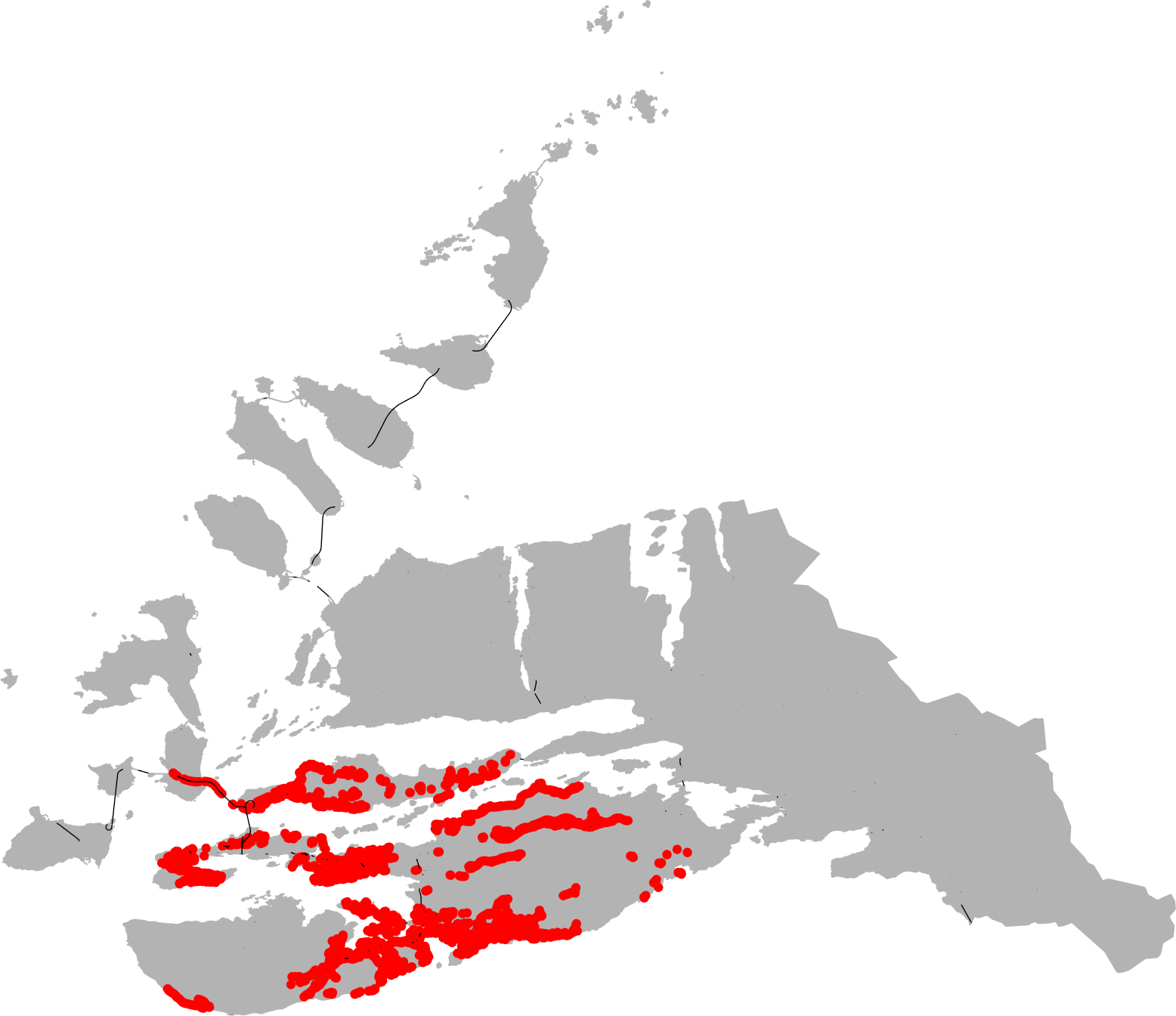}
   \caption{The difference maps associated with the heatmaps shown in figure \ref{fig:full-time-callout-delays}.}
   \label{fig:full-time-callout-diffs}
\end{figure}

\begin{figure}[p]
   \centering
   1\hspace{-1em}\includegraphics[width=0.38\textwidth]{fig_station_speedTime_scenario0}
   \hfill
   1.1\hspace{-1em}\includegraphics[width=0.38\textwidth]{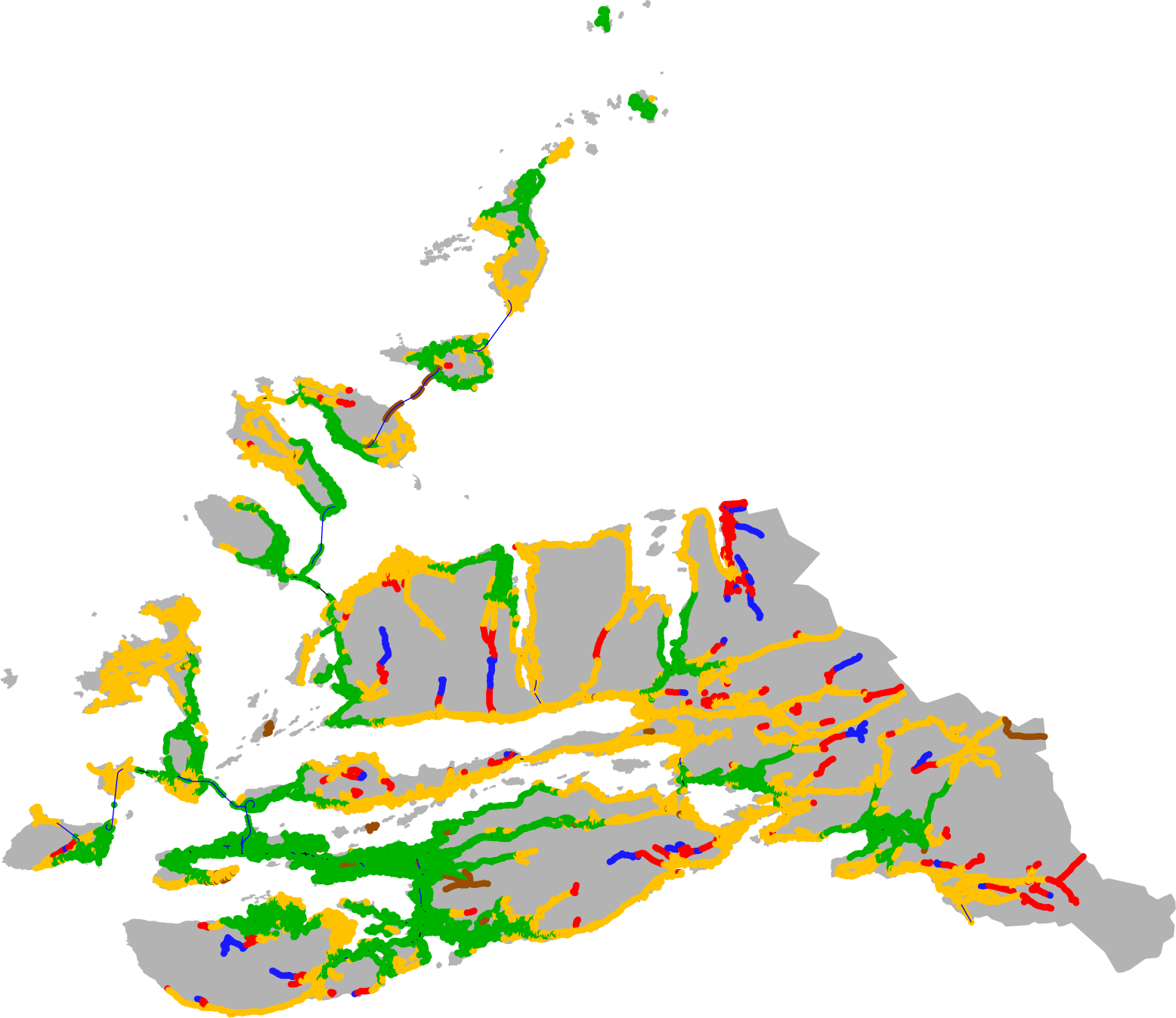} \\
   1.2\hspace{-1em}\includegraphics[width=0.38\textwidth]{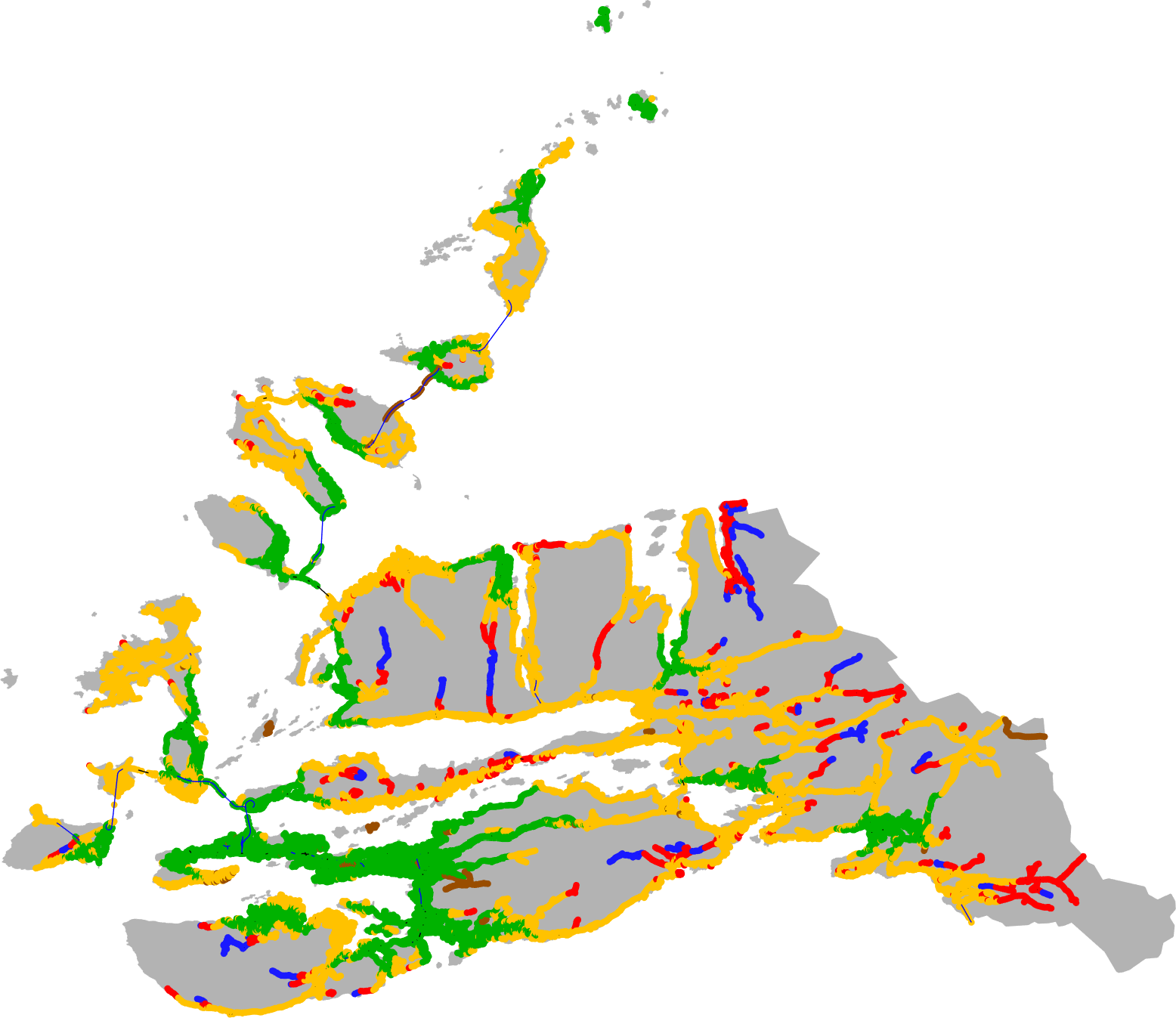}
   \hfill
   1.3\hspace{-1em}\includegraphics[width=0.38\textwidth]{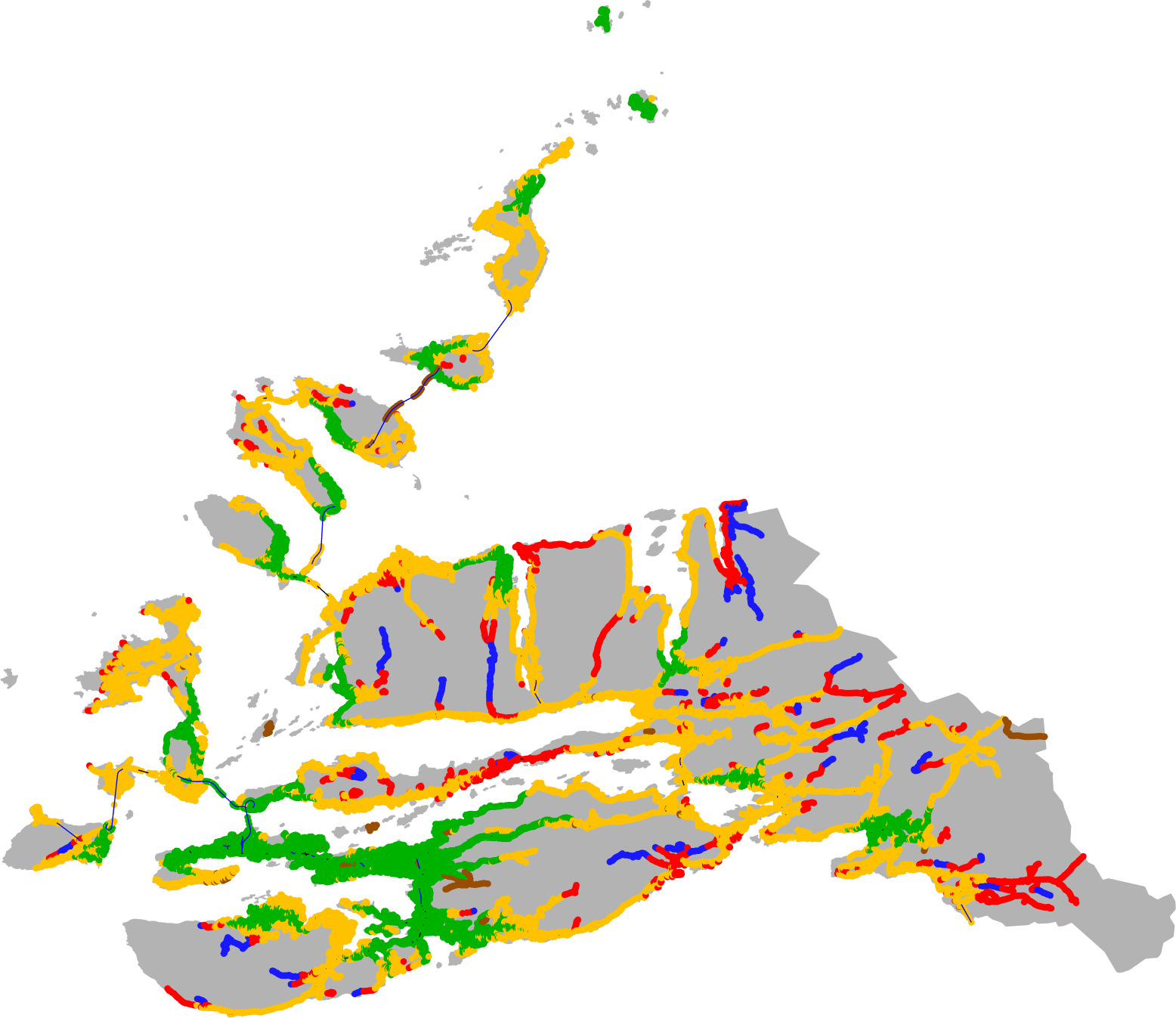} \\
   1.4\hspace{-1em}\includegraphics[width=0.38\textwidth]{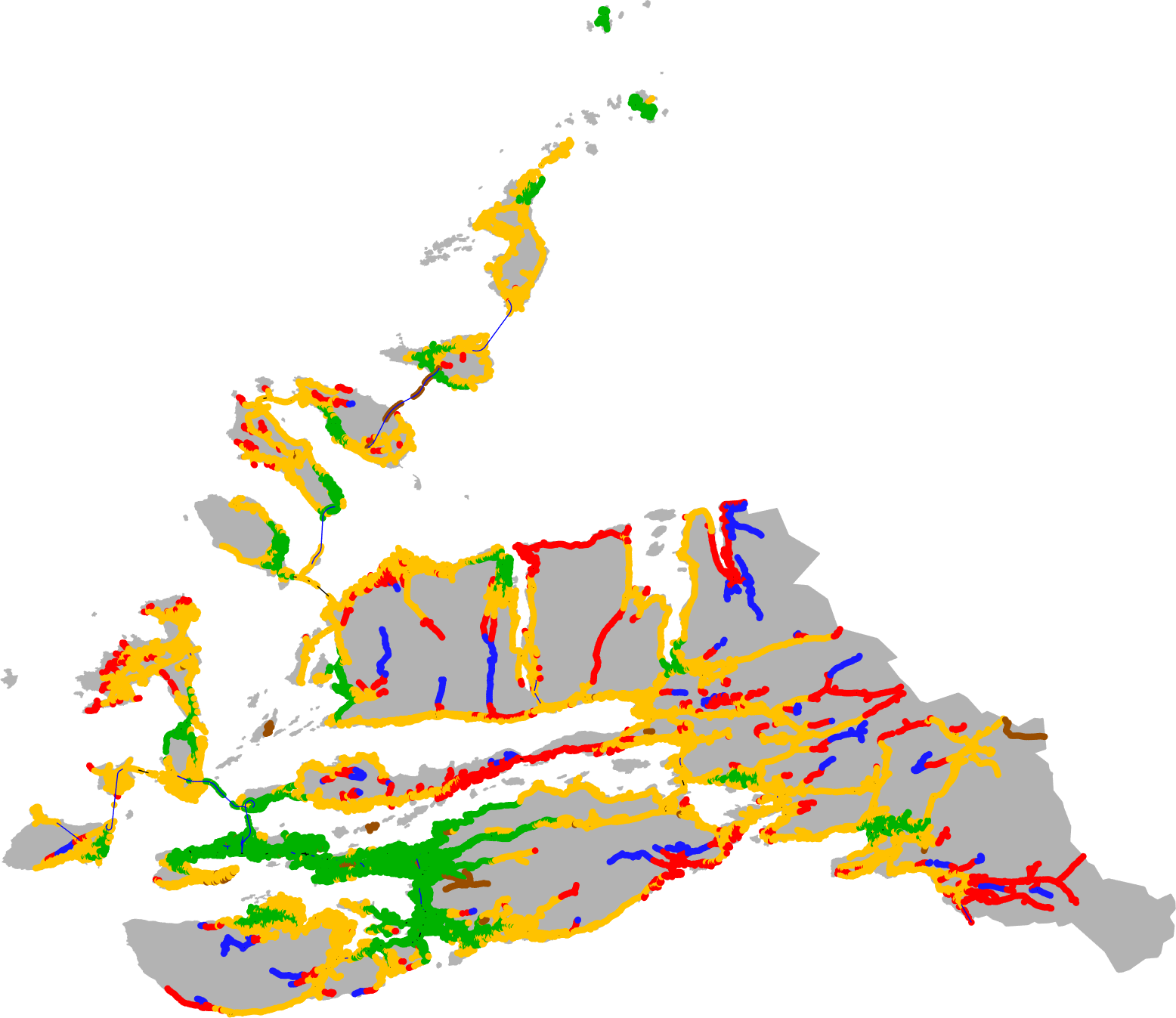}
   \hfill
   1.5\hspace{-1em}\includegraphics[width=0.38\textwidth]{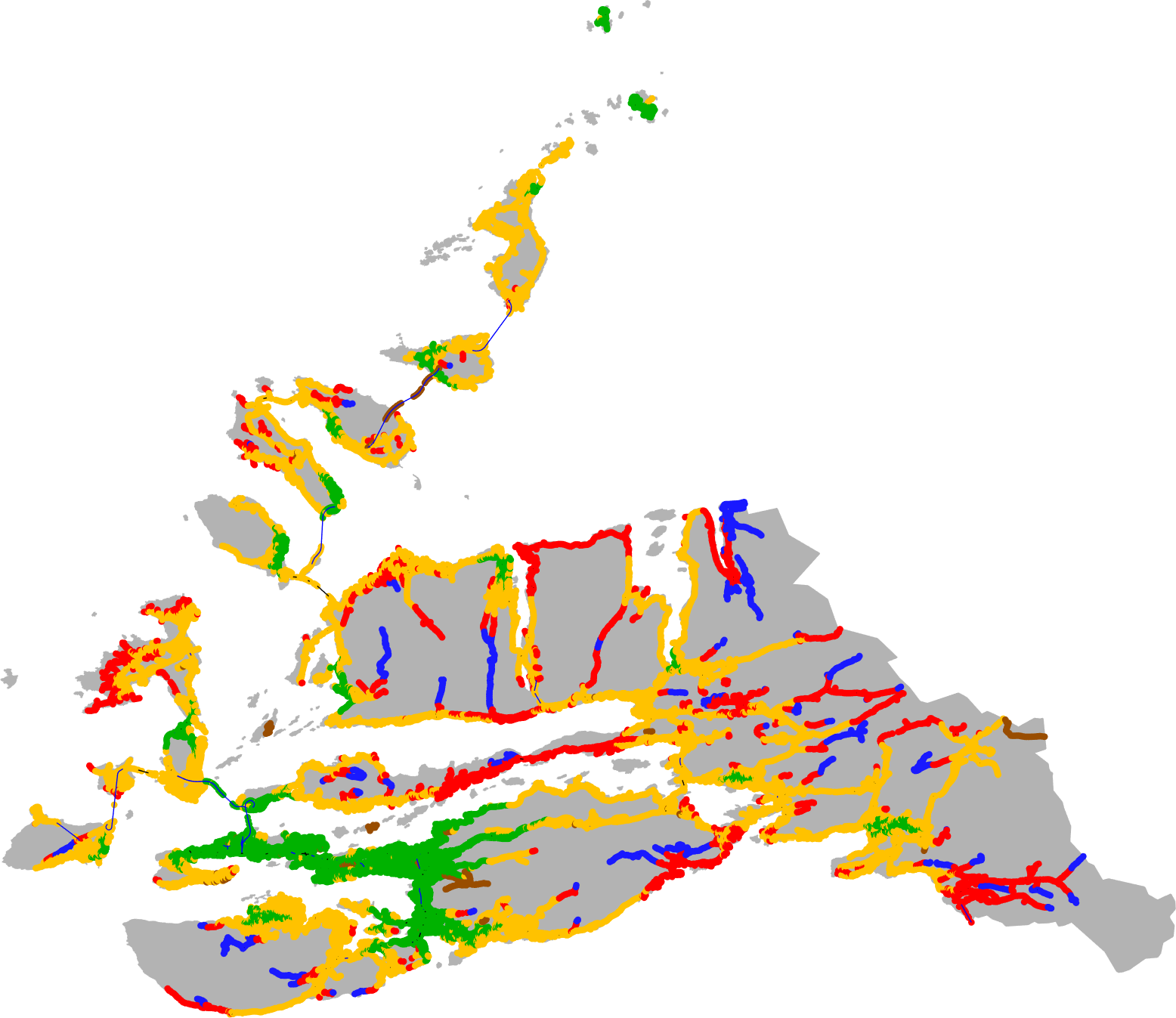}
   \caption{The effect on time of increasing the travel time by a factor, as labelled.
            The baseline model is the factor of 1.}
   \label{fig:traffic-time-delays}
\end{figure}

\begin{figure}[p]
   \centering
   1\hspace{-1em}\includegraphics[width=0.38\textwidth]{fig_diff_scenario00vs00}
   \hfill
   1.1\hspace{-1em}\includegraphics[width=0.38\textwidth]{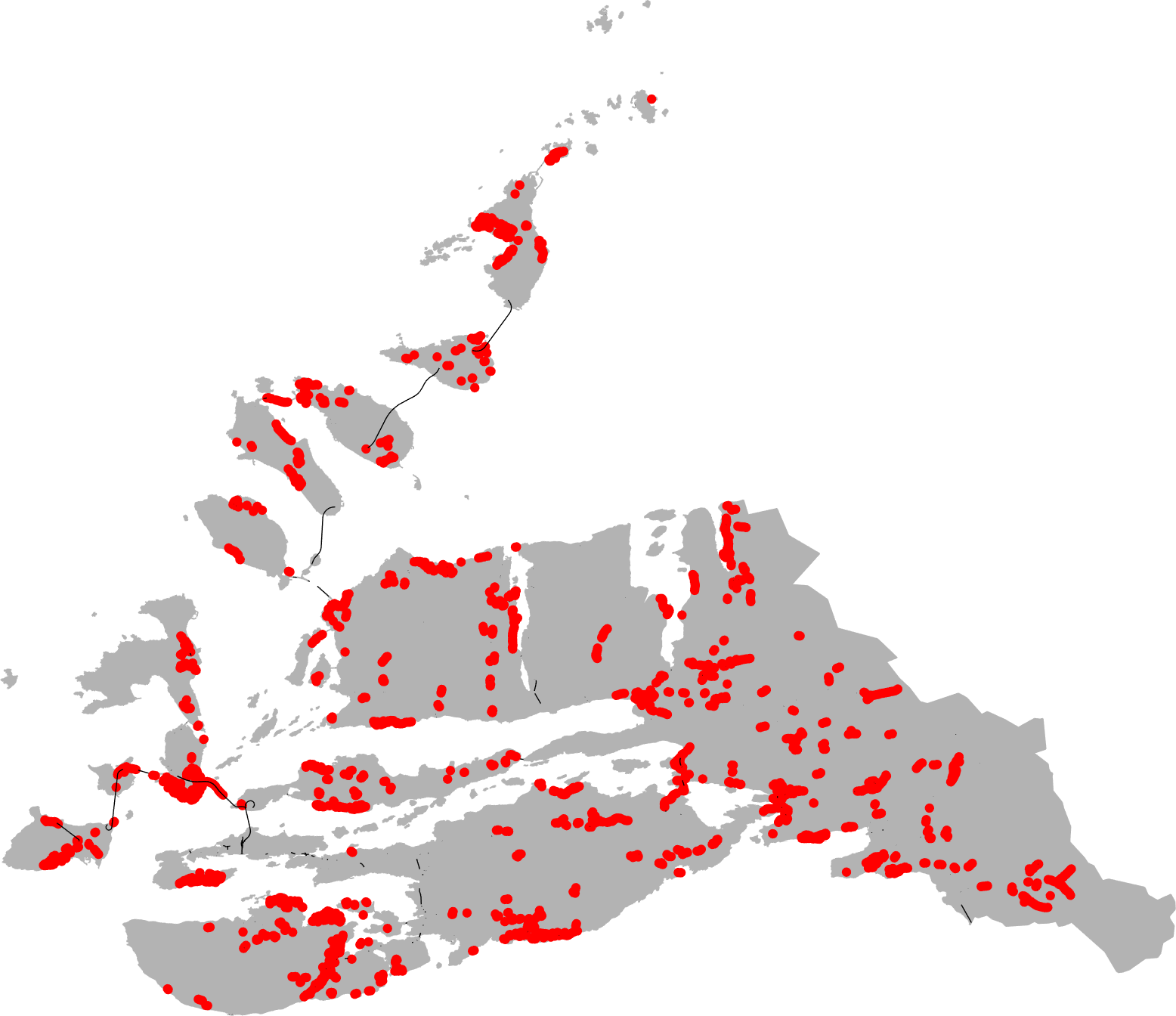} \\
   1.2\hspace{-1em}\includegraphics[width=0.38\textwidth]{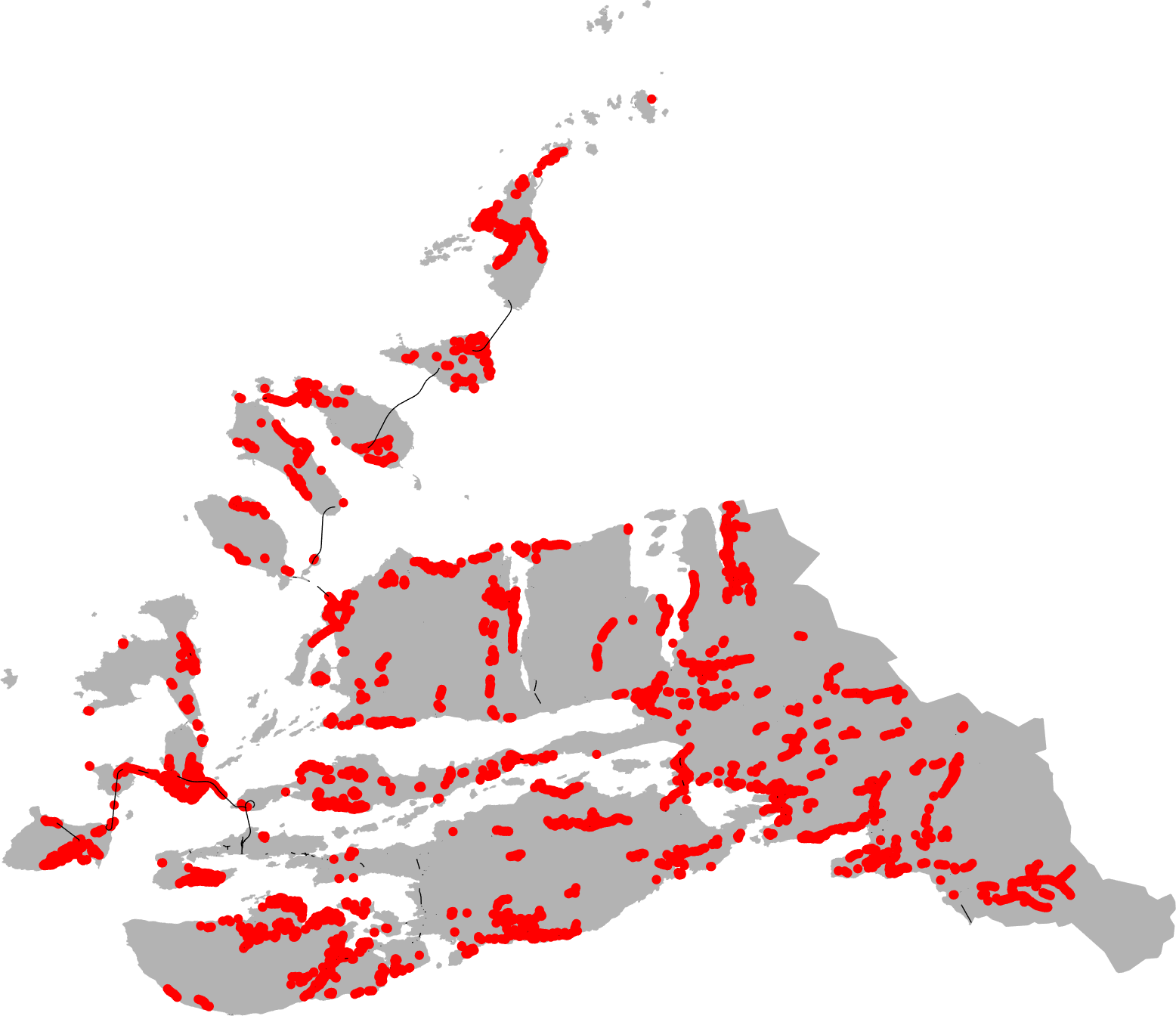}
   \hfill
   1.3\hspace{-1em}\includegraphics[width=0.38\textwidth]{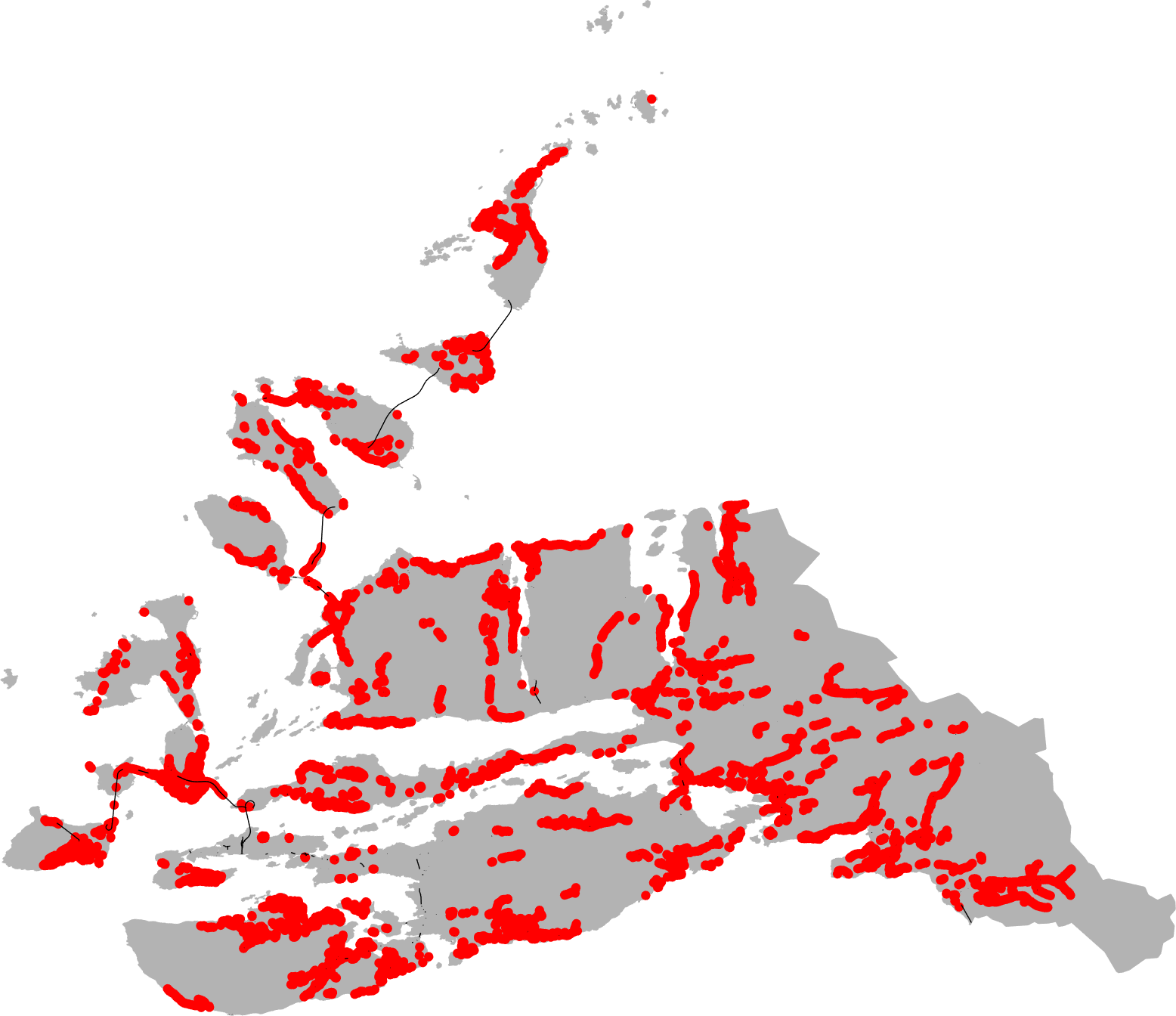} \\
   1.4\hspace{-1em}\includegraphics[width=0.38\textwidth]{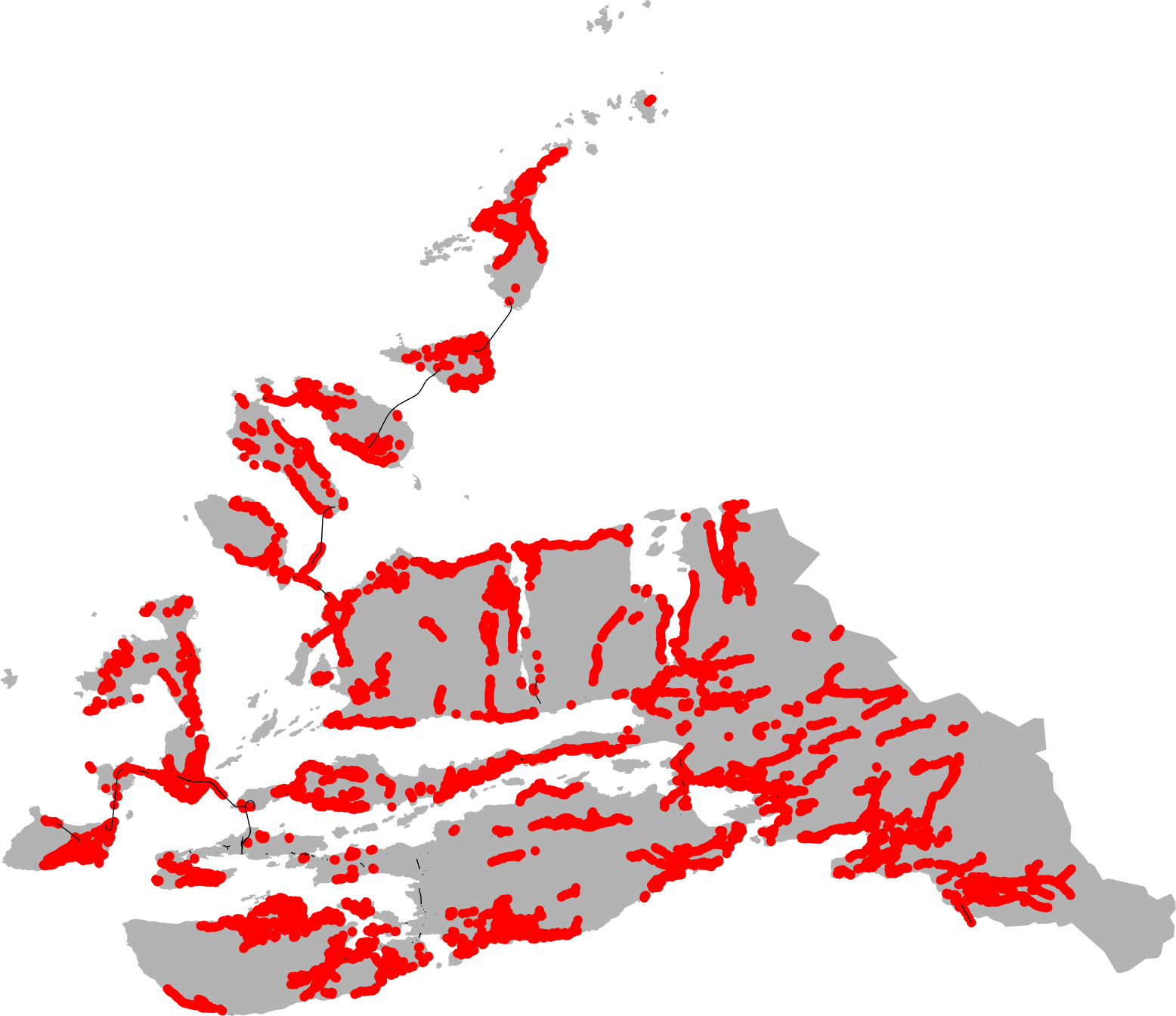}
   \hfill
   1.5\hspace{-1em}\includegraphics[width=0.38\textwidth]{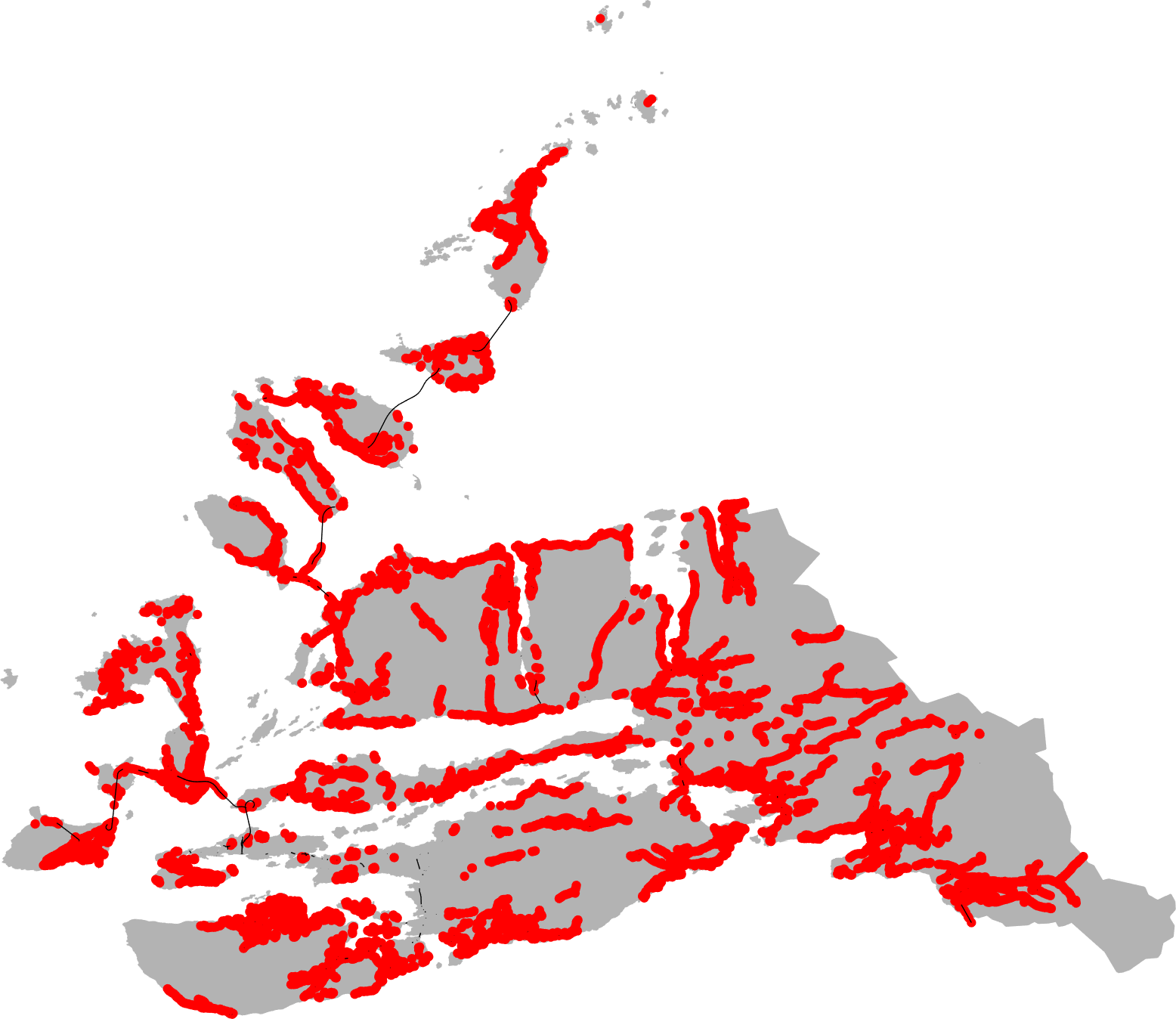}
   \caption{The difference maps associated with the heatmaps shown in figure \ref{fig:traffic-time-delays}.}
   \label{fig:traffic-time-diffs}
\end{figure}

However, there are a number of ``what if'' scenarios that can be tested without having to recalculate distances.  We have looked at the following cases:
\begin{enumerate}
\item Closing any one fire station.
      Example results are shown in figure \ref{fig:closing-one-station}.
\item Changing any one fire station from part-time to full-time, or vice versa.
      Example results are shown in figure \ref{fig:switching-one-station}.
\item Changing the delay between the alarm being raised and the fire truck leaving the station.
      This is currently set to 0 minutes for full-time fire stations, and 5 minutes for part-time stations.
      Results for part-time stations are shown in figures \ref{fig:part-time-callout-delays} and \ref{fig:part-time-callout-diffs}.
      Results for full-time stations are shown in figures \ref{fig:full-time-callout-delays} and \ref{fig:full-time-callout-diffs}.
\item Changing the overall speed at which the fire truck drives, as a factor applied to the maximum speed.
      Results are shown in figures \ref{fig:traffic-time-delays} and \ref{fig:traffic-time-diffs}.
\end{enumerate}
In each case the new time heatmap is displayed, along with the associated ``difference map'', which shows nodes that have decreased response times in green and increased response times in red, when compared with the benchmark case (see figure \ref{fig:baseline}) which has the following characteristics:
\begin{itemize}
\item All fire stations are open.
\item {\AA}lesund Hovedbrannstasjon and Spjelkavik Brannstasjon are full-time; all others are part-time.
\item The callout delay for full-time stations is 0 minutes; for part-time stations it is 5 minutes.
\item The speed at which a fire truck drives along a given road is the maximum speed registered for that road in OSM.
      If no maximum speed is registered for a given road then default speeds are used, as shown in table \ref{tab:highway-types}.
\end{itemize}

For each of these scenarios (including the baseline), we record the response time to each of the 54 locations that require a response time of less than 10 minutes.  For the baseline case, there are five locations for which the response time exceeds 10 minutes:
\begin{center}
   \begin{tabular}{lc}\toprule
   \textbf{location} & \textbf{response time (minutes)} \\
   \midrule
   Vigra sjukeheim og bukollektiv & 10.6 \\
   {\AA}lesund lufthavn Vigra     & 10.6 \\
   Fiskerstrand Verft A/S         & 10.4 \\
   Fredheim aldersboliger         & 14.6 \\
   Eidet omsorgssenter            & 10.5 \\
   \bottomrule
   \end{tabular}
\end{center}
Unfortunately, in all scenarios the minimal response time to Fredheim aldersboliger is $\ge$14.6 minutes.  This estimate is supported in the dataset of real response times, where there is one call to a location \SI{254}{\metre} from Fredheim aldersboliger, with a response time of 18.1 minutes, and one to a location \SI{329}{\metre} away with a response time of 16.7 minutes.

\newpage
\section{Comparison with real response times}
\label{sec:comparison}

\begin{figure}[b]
   \centering
   \subcaptionbox{\label{subfig:real-model-hist}}
      {\includegraphics[width=0.48\textwidth]{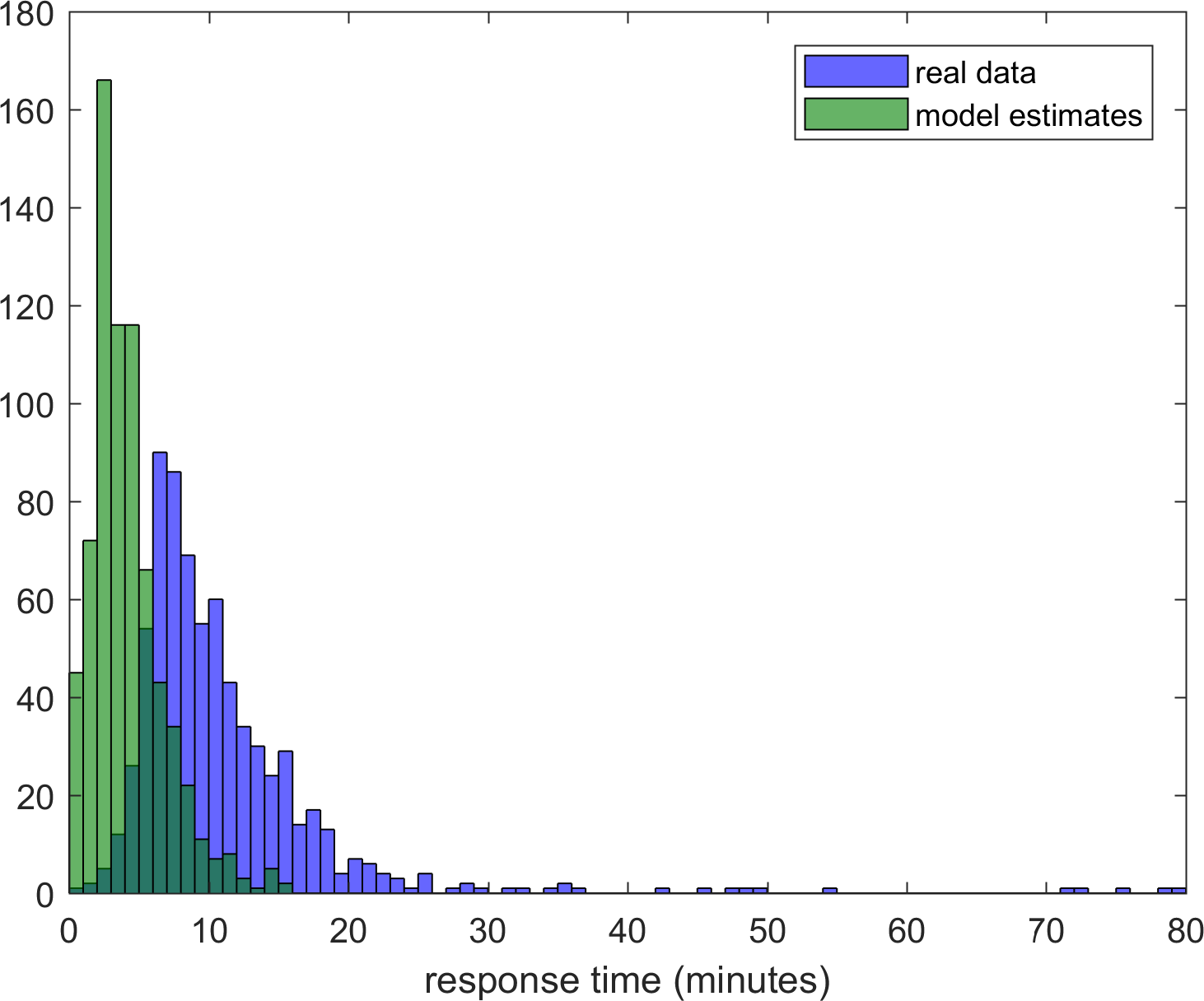}}
   \hfill
   \subcaptionbox{\label{subfig:real-model-gamfit}}
      {\includegraphics[width=0.48\textwidth]{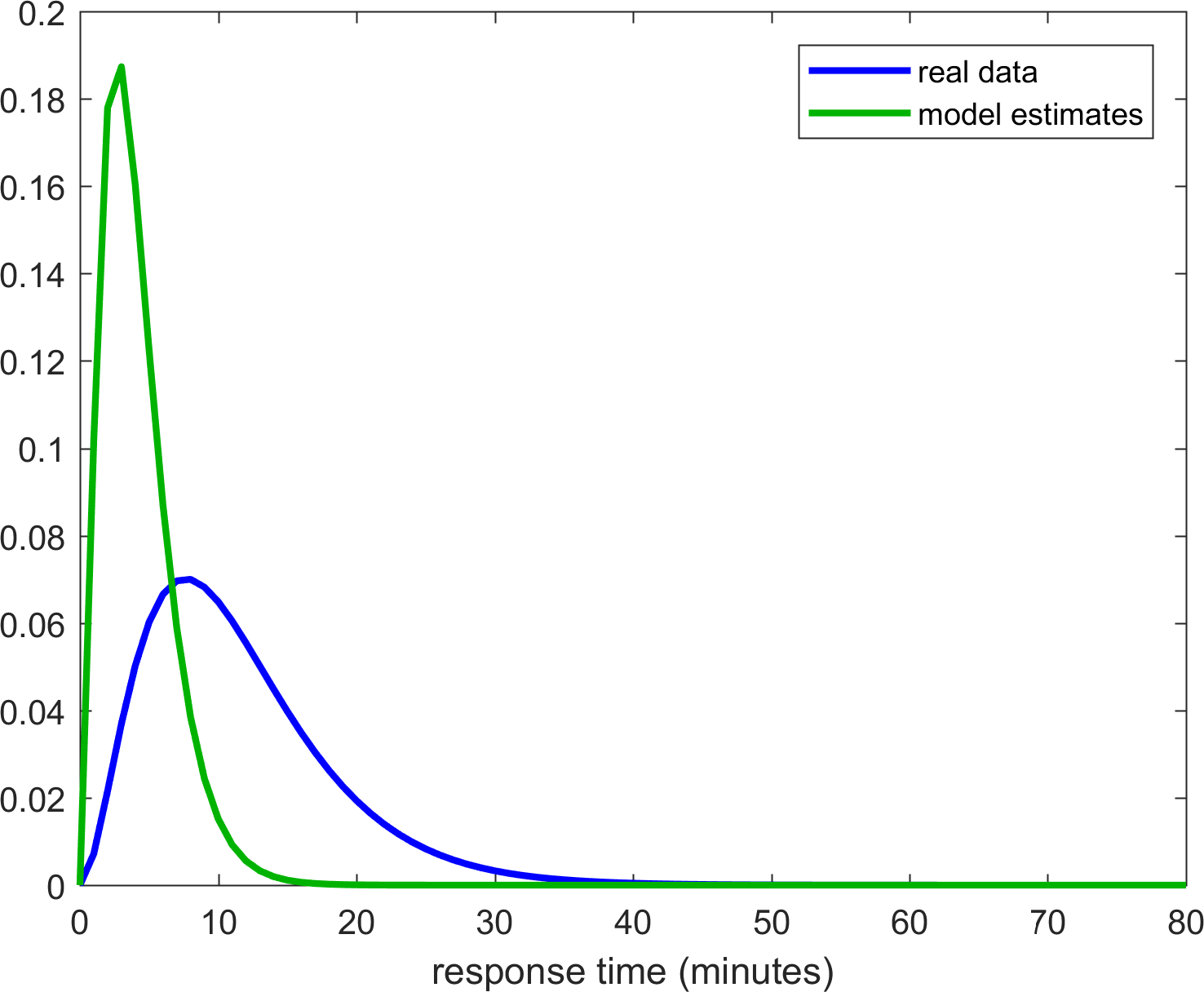}}
   \caption{Comparisons between real response times (in blue) and the baseline model's estimates (in green), showing, in
            \subref{subfig:real-model-hist}, a histogram, and in
            \subref{subfig:real-model-gamfit}, the gamma probability distributions
            fitted to the values.}
   \label{fig:real-model}
\end{figure}

We were provided with a dataset containing the real response times to 732 callouts reported as fires.  For each of these we located the nearest node in the road map; 10 were more than \SI{100}{\metre} from any node, so these were removed from the set, leaving 722 callouts.  For each of these nearest nodes, we calculated the estimated response time to that node according to our baseline model.  Figure \ref{subfig:real-model-hist} shows histograms comparing the real response times (blue) with those predicted by the model (green).  The values appear to be distributed according to a Gamma probability distribution.  Figure \subref{subfig:real-model-gamfit} compares the Gamma distributions fitted to the real and model response times.

\begin{figure}[tb]
   \centering
   \subcaptionbox{\label{subfig:real-model-hist-fixed}}
      {\includegraphics[width=0.48\textwidth]{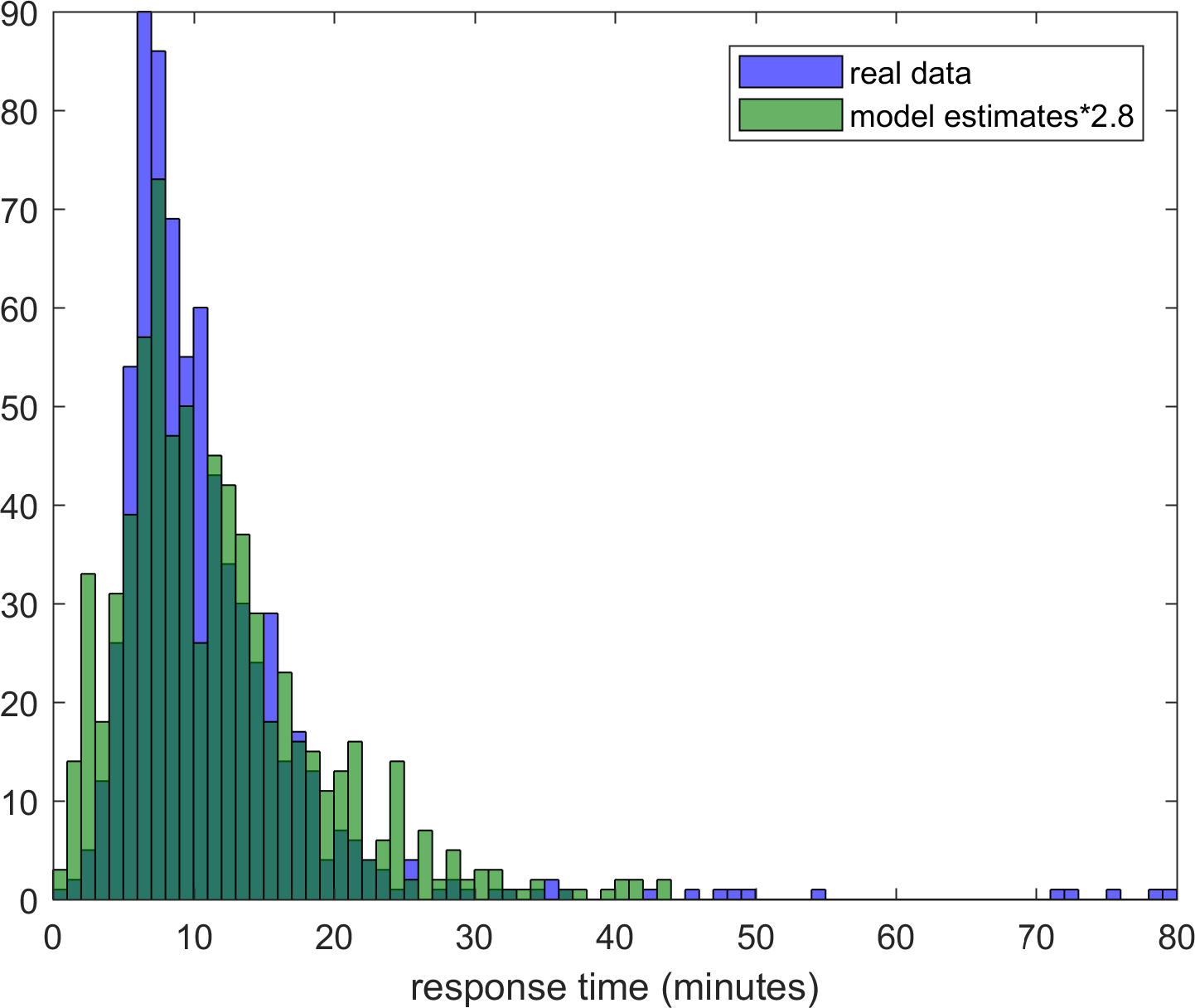}}
   \hfill
   \subcaptionbox{\label{subfig:real-model-gamfit-fixed}}
      {\includegraphics[width=0.48\textwidth]{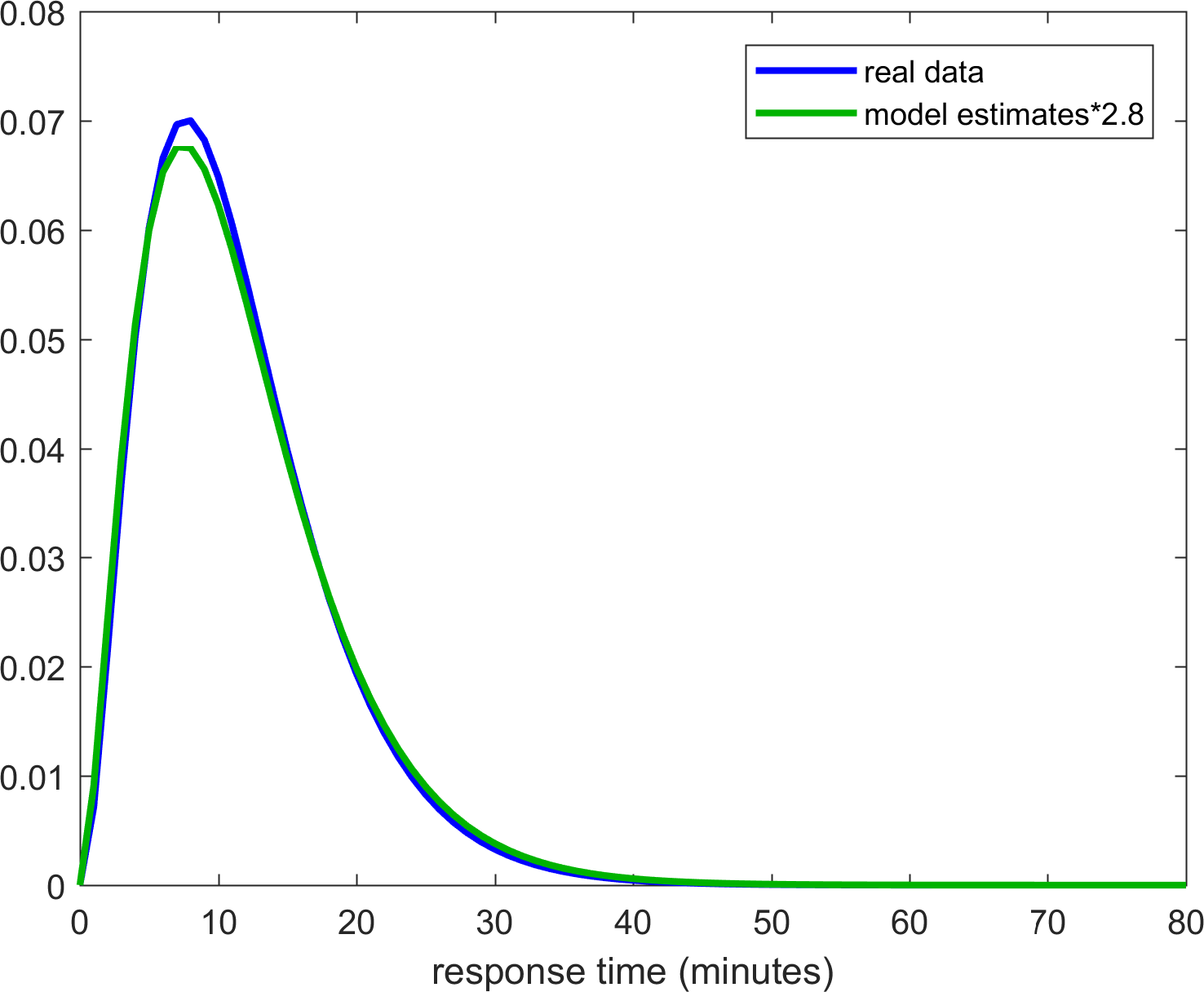}}
   \caption{The same comparisons between real and baseline model response times (as shown in figure \ref{fig:real-model}),
            but in this case the model response times have been multiplied by the factor of 2.8, making their distribution very
            similar to the real times.}
   \label{fig:real-model-fixed}
\end{figure}

By multiplying the model response times by the factor of 2.8 (empirically arrived at), the two distributions become very similar to one another (see figure \ref{fig:real-model-fixed}), suggesting that the driving speeds used in the model calculations significantly overestimate the true values.

\begin{figure}[p]
   \centering
   \subcaptionbox{original baseline model and response times\label{subfig:baseline-model-orig}}
      {\includegraphics[width=0.8\textwidth]{fig_station_speedTime_scenario0}}
   \\
   \subcaptionbox{baseline model with response times multipled by 2.8\label{subfig:baseline-model-factor}}
      {\includegraphics[width=0.8\textwidth,clip=true,trim=1pt 0pt 1pt 0pt]{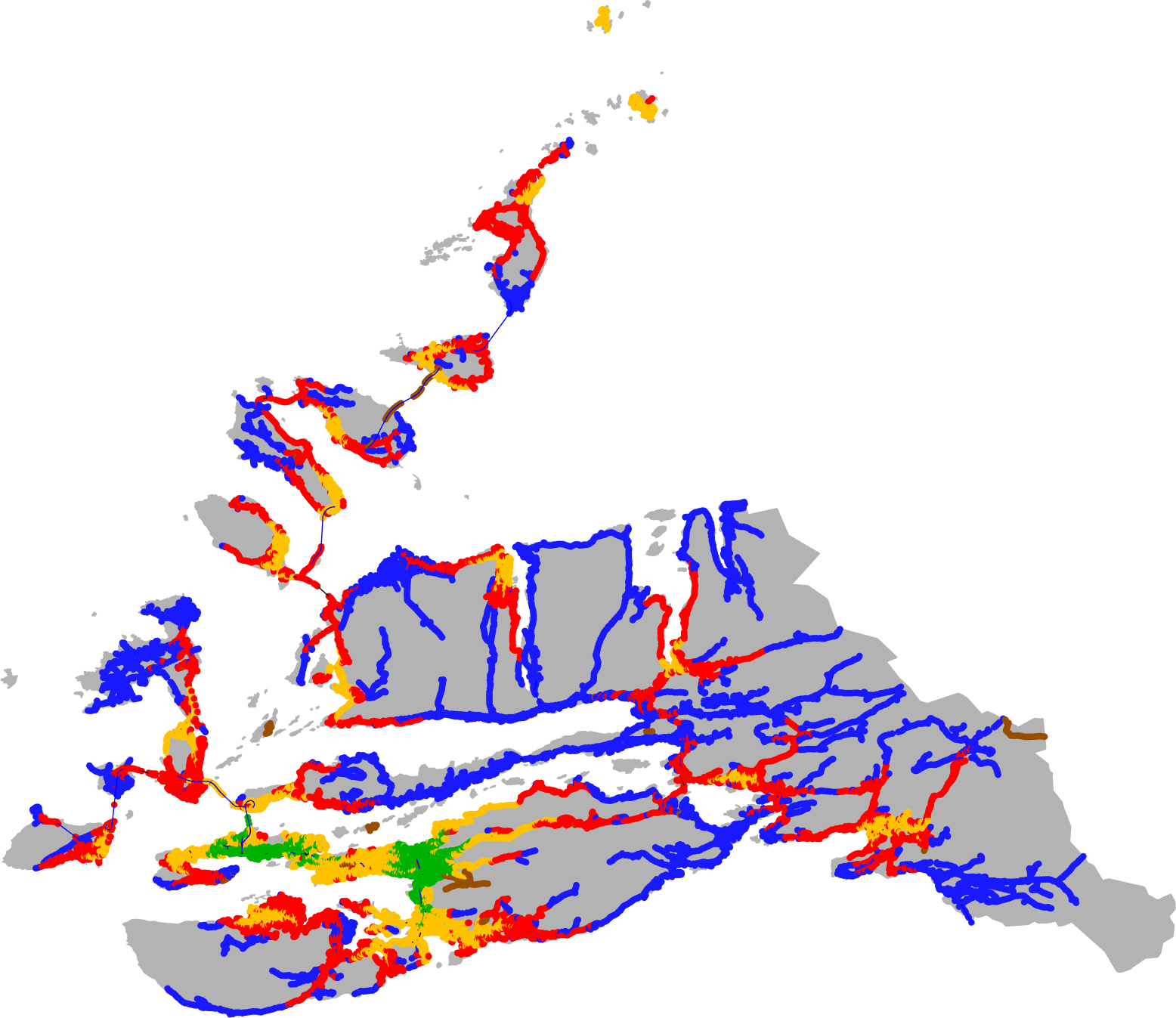}}
   \caption{Figure \subref{subfig:baseline-model-orig} shows the original baseline heatmap;
            \subref{subfig:baseline-model-factor} shows the same, but with the response times multiplied by the factor of 2.8.}
   \label{fig:cf-factor}
\end{figure}

Figure \ref{fig:cf-factor} compares the heatmap for the original baseline with that for the baseline model with response times multiplied by the factor of 2.8.  

\afterpage{\clearpage}

\newpage
\section{``What if'' scenarios, part 2}
\label{sec:scenarios-2}

\begin{figure}[h!]
   \centering
    \includegraphics[width=\textwidth]{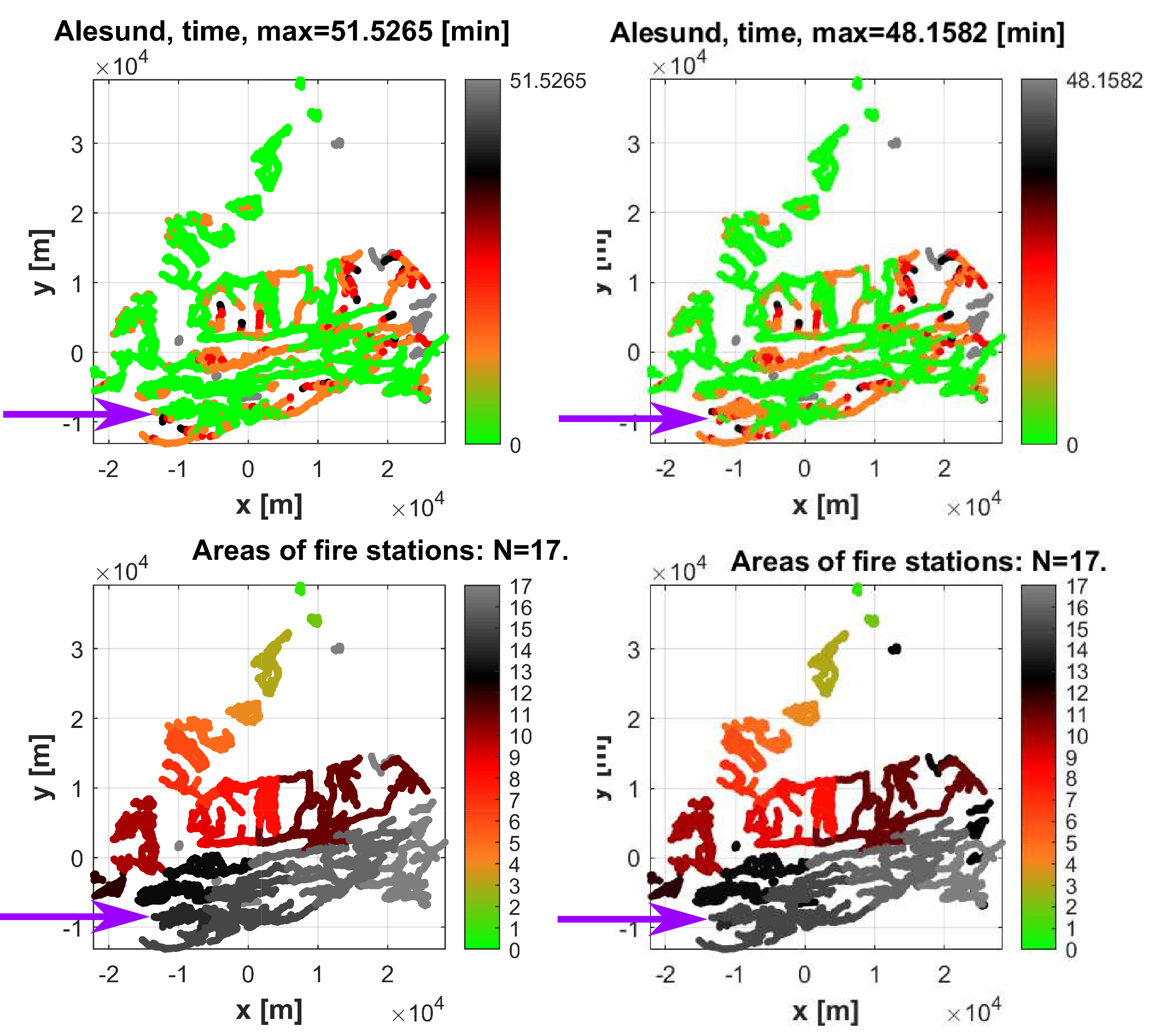}
    \caption{The comparisons of the response time heatmaps between the baseline and interactively modified a single fire station location.}
   \label{fig:interactive2}
\end{figure}

As mentioned above, calculating the distances for a new graph, where a single fire station location has been modified, takes roughly 3-5 minutes. It is possible to build the new heatmap interactively, so that a user can play with the {\AA}lesund map by picking a single fire station and placing it in a new location. The code would recalculate the new response times for all the nodes and display the results in the figure with the appropriate colormap. To achieve the interactive map capability for the code, we used the \texttt{ginput} commands of the MATLAB to extract the current location ($x$ and $y$ coordinates) of the mouse pointer. Then, any of the 17 full and/or part-time fire stations is selected. After, the new location for the selected fire station is chosen. The code selects the closest node for the selected new location and marks it as a new fire station location. The Dijkstra algorithm is run with the new fire station locations. Finally, the computation results are displayed by updating the figure with the new heatmap. Two examples are shown below. 
\begin{figure}[h!]
   \centering
    \includegraphics[width=\textwidth]{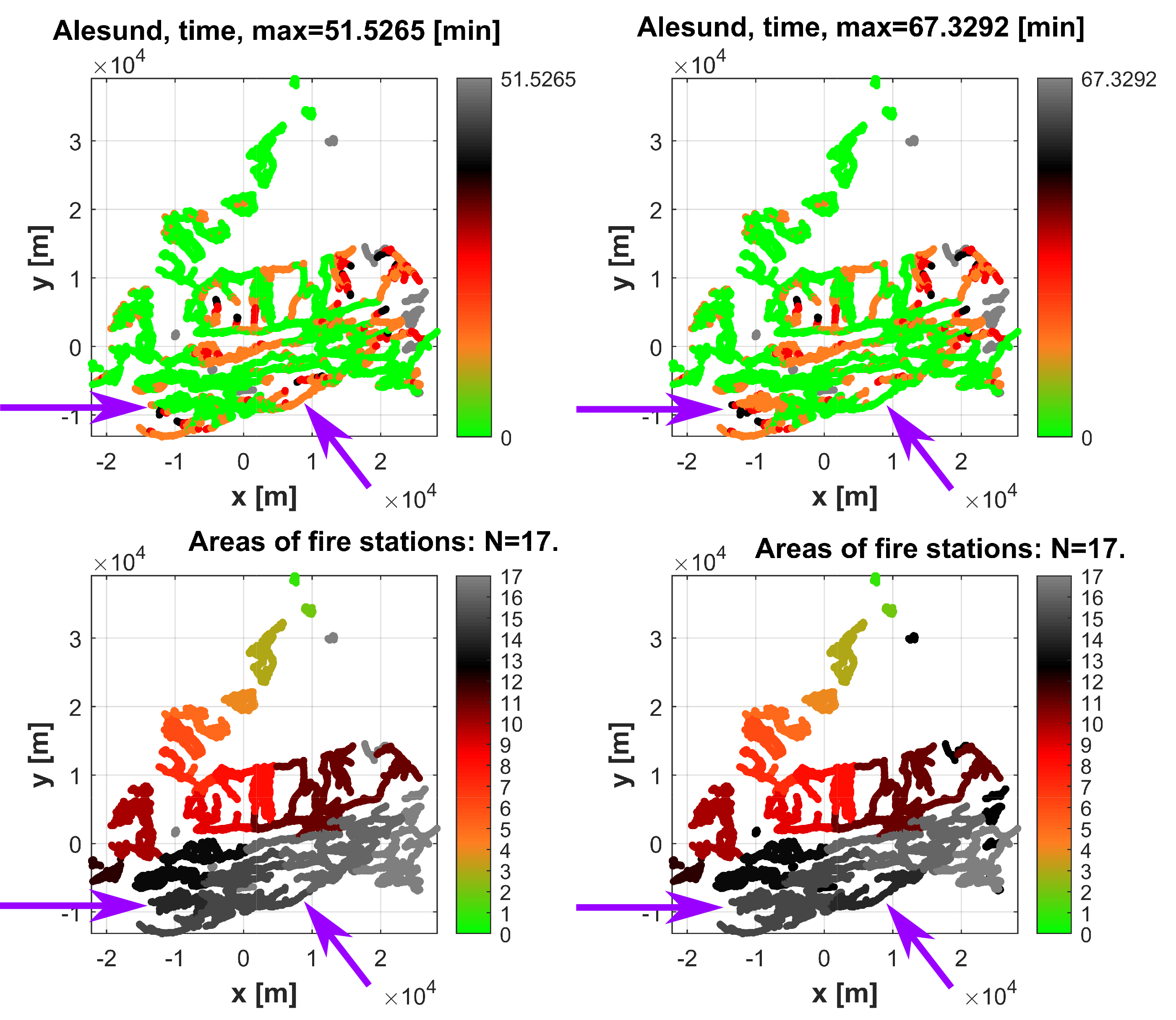}
    \caption{The comparisons of the response time heatmaps between the baseline and interactively modified a single fire station location.}
   \label{fig:interactive1}
\end{figure}
In the figure~\ref{fig:interactive1}, a fire station number 14 with longitude 6.2001542 and latitude 62.4403443 is selected and placed in the new location with longitude 6.5578 and latitude 62.4432. For this particular case, the computation time was 239.8 seconds. The selected fire station and its new location are indicated by violet arrows. The left subplots show the baseline case, while the right subplots show the updated results. The top subplots show the response time, while the bottom subplots show the areas of responsibility of each fire station (nodes are grouped in color where each color represents a zone of responsibility of the closest fire station). In this special case, the change of the fire station location leads to an increased maximum response time: from 51.5 minutes to 67.3 minutes. Thus, the possible new fire station location makes the response time worse and is not recommended. The next example, figure~\ref{fig:interactive2}, shows the case when the same fire station number 14 is selected and placed in the new location with longitude 6.1748 and latitude 62.4273. For this case, the computation time was 347 seconds. Similarly to the previous plot, the baseline and new fire station locations are indicated by the violet arrow. In this case, the maximum response time changes from 51.5 minutes to 48.2 minutes. In other words, the response time decreases and this new fire station location is recommended, because it improves the response time. These two figures demonstrate the interactive regime of the code where fire station locations can be modified manually and the corresponding changes in results can be analyzed visually or numerically. The same interactive code can be implemented for the changes made in the locations of critical objects. For example, if a hospital is moved to the new location. This can happen when an old hospital is closed and in turn a new hospital is built in the new place, which gives a net effect of hospital moving to the new location. 

This interactive code can be utilized to find the optimal placement of a fire station or of a new critical object in the map. To achieve this, in principle, all new locations (nodes) for an object being moved (fire station or a critical object) should be tried. Then, the best (optimal) heatmap should be selected. However, we anticipate that doing this by a brute force for all nodes would require large computational resources and time. By parallelizing the computations, vectorizing the Dijkstra's algorithm, and using the recursive codes, this problem can be alleviated somewhat. However, it would remain to be a significant obstacle. More efficient code to recompute the heatmap would be based on local recomputation of the response time of only the nodes affected by the change. However, this has not been explored and implemented yet. 

\newpage
\section{Optimum placement of a fire station, game theory approach}
\label{sec:game-theory}

Eventually, this section concerns a study of the optimum placement of a fire station by using a game theory approach. More precisely, we try to formulate a mathematical model for calculating an optimal distance of a fire station from all the fire locations (residential areas, nurseries, hospitals, schools etc.), so that in the case of emergency fire brigade can reach out the fire location in the minimal time.
For the convenience of readers, firstly we are going to explain a simple game theoretic model which deals with minimization of the cost of each players of a non-cooperative strategic game.

We begin our analysis by introducing several important notations and mathematical tools. Throughout this section, the abbreviation ``a.e." stands for ``almost everywhere". We assume that the non-cooperative strategic game  comprises $N$ players and the interval $[0,T]$ in $\mathbb{R}_{+}$ is called the planning horizon of players, where $T$ is the arbitrary time. We consider our non-cooperative strategic game is evolving over time $t\in [0,T]$. The dynamic behaviour of the non-cooperative strategic game is described by the strategy vector $x(t)\in L^{2}([0,T], \mathbb{R}^{n})$ of all the  players at the given moment $t\in [0,T]$. Let $x^{w}(t)\in L^{2}([0,T], \mathbb{R}^{n_{w}})$ be the strategy vector of the player $w$ and $x^{-w}(t)\in L^{2}([0,T], \mathbb{R}^{n-n_{w}})$ be the strategy vector of all players except the player $w$ at the given moment $t\in [0,T]$. Here $n=\sum\limits_{w =1}^{N} {n_{w}}$. To emphasize the strategy vector of the  player $w$, we write the strategy vector $x(t)$ of all the players as $x(t)=(x^{w}(t),x^{-w}(t))$.
This is just another way of writing the vector $x(t)=(x^{1}(t),x^{2}(t),\ldots,x^{w-1}(t),x^{w}(t),x^{w+1}(t),\ldots,x^{N}(t))\in L^{2} ([0,T],\mathbb{R}^{n})$. We recall that $L^{2} ([0,T],\mathbb{R}^{n})= L^{2} ([0,T],\mathbb{R}^{n_{w}}) \times L^{2} ([0,T],\mathbb{R}^{n-n_{w}})$. Moreover, in order to write the strategy vector $x$ is a function of the time parameter $t$, we write $x=x(t)$, and we use the same convention for the strategy of the player $w$ and the strategies of the decision variables of all the players except the player $w$. That is, we write $x^{w}=x^{w}(t)$ and $x^{-w}=x^{-w}(t)$, respectively. We consider a nonempty, closed and convex subset $K$ of $L^{2} ([0,T],\mathbb{R}^{n})$, and for any given vector $x^{-w}(t)$ of rival players, the nonempty, closed and convex feasible strategy set of the player $w$ is $K_{w}(x^{-w}(t)) \subset L^{2} ([0,T],\mathbb{R}^{n_{w}})$. We say that the strategy $x=x(t)$ is feasible if for all $w=1,2,\ldots,N$ and for all $t\in [0,T]$, we have $x^{w}(t)\in K_{w}(x^{-w}(t))$. Each player $w$ has an objective function which is known as the cost/loss function and which depends on both the player's own variable $x^{w}(t)$ and on the rival players' variables $x^{-w}(t)$. We interpret the total cost/loss function that the player $w$ incurs when the rival players have chosen the strategy $x^{-w}(t)$, $F^{w}: L^{2} ([0,T],\mathbb{R}^{n}) \to \mathbb{R}$, in terms of the integral
\(F^{w}(x(.)) = \displaystyle \int_{0}^{T} f^{w}(x^{w}(t),x^{-w}(t)) dt, \) where $f^{w}(x^{w}(t),x^{-w}(t))$ is a real-valued continuously differentiable function.\\

\noindent
Now, we define the dynamic generalized Nash equilibrium problem as follows:

\noindent

to find a feasible strategy vector $x=x(t)\in L^{2} ([0,T],\mathbb{R}^{n})$ such that for all the players $w=1,2,\ldots,N$, we have $x^{w}(t)\in K_{w}(x^{-w}(t))$ and
\begin{equation}\label{1.1}
	F^{w}(x(t))=F^{w}(x^{w}(t),x^{-w}(t))\leq F^{w}(p^{w}(t),x^{-w}(t)) \; ~\forall~ p^{w}(t)\in K_{w}(x^{-w}(t)).
\end{equation}

Now, we are going to motivate the model (\ref{1.1}) in the terms of optimum location of the fire station. In order to do so, we assume that there are $n$ fire locations (residential areas, nurseries, hospitals, schools, etc.) which we further call nodes. The considered time period for arriving fire brigades at the fire location is $[0,T]$. Further, $x^{w}(t)\in L^{2}([0,T], \mathbb{R}_{+})$ is the chosen distance of fire station from the fire location/node $w$ by the local authorities who are responsible for organizing and maintaining the fire station and ambulance networks in such a way that the response time does not exceed 10 minutes for hospitals and nursing homes, 20 minutes for residential areas and 30 minutes otherwise, $x^{-w}(t)\in L^{2}([0,T], \mathbb{R}^{n-1}_{+})$ is the chosen distance of fire station from the  all nodes  except the node $w$, and $x(t) \in L^{2}([0,T], \mathbb{R}_{+}^{n})$ is the chosen distance vector of fire station from the all fire locations/nodes by the authorities. Let's say, government imposes a maximum distance condition $X(t)$ on the fire station location from the all nodes (the fire locations). We have to also deal with traffics for arriving at the fire locations on time. Therefore, we consider $y^{w}(s)\in L^{2}([0,T], \mathbb{R}_{+}) $ is the traffic flow of the unit distance on the route between the fire station and the fire location $w$. Consequently, the traffic flow of the distance $x^{w}(t)$ is $x^{w}(t)y^{w}(t)$ on the route between the fire station and the fire location $w$. According to the traffic law, studied by \cite{dan,singh1,singh2} and the references therein, every feasible traffic flow has to satisfy the following capacity constraint
$$\eta^{w}(t)\leq x^{w}(t)y^{w}(t)\leq \theta^{w}(t),~\forall~w=1,2,\ldots,n,~\text{ a.e., in}~[0,T],$$
and the traffic conservation law/demand requirements
$$\sum\limits_{w=1}^{n} x^{w}(t)y^{w}(t)=\rho (t),~\text{ a.e., in}~[0,T], $$    
where $\eta^{w}(t), \theta^{w}(t) \in L^{2}([0,T], \mathbb{R}_{+})$ are the given bounds and the function $\rho(t)\in L^{2}([0,T], \mathbb{R}_{+})$  is the given demand. Now, we have the following nonempty feasible set for our optimum placement of a fire station model,
$$K=\left \{x(t)\in L^{2}([0,T], \mathbb{R}^{n}):~x^{w}(t)\leq X(t),~\eta^{w}(t)\leq x^{w}(t)y^{w}(t)\leq \theta^{w}(t),~\forall~w = 1,2,\ldots,n,\right.$$ $$\left.\text{and} \sum\limits_{w=1}^{n} x^{w}(t)y^{w}(t)=\rho (t)~\text{ a.e., in}~[0,T] \right \}.$$
For any given $x^{-w}(t)$ the nonempty, closed and convex feasible set for the chosen location of fire station to the fire location (node) $w$ is denoted by $K_{w}(x^{-w}(t))$, and defined as $$K_{w}(x^{-w}(t))=\{x^{w}(t)\in L^{2}([0,T], \mathbb{R}_{+}) :(x^{w}(t),x^{-w}(t))\in K\}.$$
We interpret the total response time function of the fire location/node $w$ from the fire station location is $F^{w}: L^{2} ([0,T],\mathbb{R}^{n}) \to \mathbb{R}$, is defined as in terms of the integral
\(F^{w}(x(.)) = \displaystyle \int_{0}^{T} f^{w}(x^{w}(t),x^{-w}(t)) dt, \) where $f^{w}(x^{w}(t),x^{-w}(t))$ is a real-valued continuously differentiable function.

\noindent
Now, the local authorities' aim is to calculate the distance vector of fire station from the all nodes $x(t)\in K$, for the given $x^{-w}(t)$ so that they can minimize the total response time, i.e., to solve the following optimization problem for all $w=1,2,\ldots,n$
\begin{equation}\label{1.2}
	\aligned
	&\underset{x^{w}(t)}\min \int_{0}^{T} f^{w}(x^{w}(t),x^{-w}(t)) dt\\
	&\text{subject~to}~ x^{w}(t)\in K_{w}(x^{-w}(t)).
	\endaligned
\end{equation}

\noindent
\textbf{Remark.}
	We have main concern that how to define the response time function $F^{w}(.)$. Indeed, at this stage it needs a deep literature survey regarding fire station works for defining a proper mathematical response time function. Nevertheless, we would like to define the response time function in the quadratic form, for instance,  \(F^{w}(x(.)) = \displaystyle \int_{0}^{T} (ax^{w}(t)+b (x^{w}(t))^{2}) dt, \) where $a$ and $b$ are any given real numbers. The reason behind this to adopt the solving methods and mathematical results of \cite{singh2} and the references therein for solving our optimum placement of a fire station model (\ref{1.2}).

\newpage
\section{Conclusion}
\label{sec:conclusion}

In this section we give a summary of the whole project and outline potential applications and further development.

First we created a map of the region of responsibility of {\AA}lesund Brannvesen based on publicly available data. The process of creating the map of the region of responsibility (the OpenStreetMap data, data selection and transformation, plotting nodes and edges, plotting fire stations, plotting critical objects) was described in Section \ref{sec:source-data} and Section \ref{sec:heatmaps}.

Then we used Dijkstra's algorithm to calculate minimal distances and time responses for each node in Section \ref{sec:resp_times}. We plotted areas of responsibility of the particular fire stations in one map, and finally the heat map which was then taken as the reference (baseline) model.

A number of ``what if'' scenarios was tested in Section \ref{sec:scenarios-1}. In particular, we focused on events like closing a fire station, changing characteristics of fire stations and changing maximum speed of the fire truck. We suggest that one of the next steps might be to test more ``what if'' scenarios including more demanding events like closing a road/bridge/tunnel, or a combination of more events. For example, we might expect that if the only bridge to Hessa island was closed, it would not be possible to fulfill the law requirement for critical objects on Hessa.

Comparison of results of our simulations to real data in Section \ref{sec:comparison} showed that histograms of time responses of both the model and real data have similar (Gamma distribution) shape. However, results of the simulations were more optimistic. Multiplying the model estimates by the factor 2.8 made the distributions practically identical. The same factor 2.8 was then used to generate a more realistic response time heat map in Figure \ref{subfig:baseline-model-factor}.

Implementation of interactivity into the heat map in MATLAB was described in Section \ref{sec:scenarios-2}. A potential application of the interactive heat map might be to find optimal covering of the region of responsibility by fire stations. To achieve this, further work is needed on algorithm optimisation (parallelisation, vectorisation, local recomputation etc.).

In addition, optimal placement of fire stations using the game theory was introduced in Section \ref{sec:game-theory}. The main unfinished task here is to define the response time function, preferably in the quadratic form.

We can conclude that the main goal of the project - to develop an interactive heat map of response times based on publicly available data - was achieved. Further development of ``what if'' scenarios and the optimal placement algorithm is possible and recommended.

\newpage\appendix
\section{Haversine formula}
\label{appendix:haversine}

The Haversine formula determines the great-circle distance between two points on a sphere given their longitudes and latitudes.

\begin{lstlisting}[style=custommatlab]
function [x,y] = std_lonlat2m (fix,lon,lat)
   % Uses the Haversine formula:
   % see www.movable-type.co.uk/scripts/latlong.html

   % convert the reference point from degrees to radians
   refLat  = fix(2)*pi/180;
   refLong = fix(1)*pi/180;
   % convert the given lon/lat from degrees to radians
   latRad = lat*pi/180;
   lonRad = lon*pi/180;

   % Earth's radius in metres (average of equatorial and polar)
   R = (6378137+6356752)/2;

   deltaLat  = latRad - refLat;
   deltaLong = lonRad - refLong;

   a = sin(deltaLat/2).^2
       + cos(refLat).*cos(latRad).*sin(deltaLong/2).^2;
   c = 2*atan2(sqrt(a),sqrt(1-a));
   distance = R*c;
   theta = atan2(sin(deltaLong).*cos(latRad),
           cos(refLat).*sin(latRad)-sin(refLat).*cos(latRad).*cos(deltaLong));
   % theta is clockwise from north;
   % Matlab uses anticlockwise from east
   thetaMat = 2*pi-theta + pi/2;
   % convert from polar coordinates back into Cartesian
   [x,y] = pol2cart(thetaMat,distance);
end
\end{lstlisting}

\newpage
\bibliographystyle{unsrt}
\bibliography{esgi156}

\end{document}